\documentclass[pre,singlecolumn,noshowpacs,preprintnumbers,amsmath,amssymb]{revtex4}

\usepackage{mathrsfs}
\usepackage[dvipdfmx]{graphicx}
\usepackage{dcolumn}
\usepackage{bm}
\usepackage{color}
\usepackage{comment}
\usepackage{enumitem}
\usepackage{ulem}
\usepackage{ascmac}

\newcommand{\mc}{\mathcal}

\newcommand{\tl}{\tilde}
\newcommand{\ra}{\rangle}
\newcommand{\la}{\langle}
\newcommand{\tb}{\textbf}

\newcommand{\eps}{\epsilon} 
\newcommand{\homg}{\hat{\omega}}

\newcommand{\hv}{\hat{v}}

\newcommand{\hxi}{\hat{\xi}}

\newcommand{\hz}{\hat{z}}

\newcommand{\heta}{\hat{\eta}}

\newcommand{\ve}{\varepsilon}

\newcommand{\mrG}{\mathrm{G}}
\newcommand{\mrP}{\mathrm{P}}

\newcommand{\mrT}{\mathrm{T}}

\newcommand{\mrss}{\mathrm{ss}}

\newcommand{\omg}{\omega}
\newcommand{\mcF}{\mathcal{F}}
\newcommand{\mcH}{\mathcal{H}}

\newcommand{\htt}{\hat{t}}

\newcommand{\hx}{\hat{x}}
\newcommand{\hnu}{\hat{\nu}}
\newcommand{\hN}{\hat{N}}
\newcommand{\Lpath}{\mathcal{L}_{\mathrm{path}}}

\begin{document}

\title{Field master equation theory of the self-excited Hawkes process}

\author{Kiyoshi Kanazawa$^{1}$ and Didier Sornette$^{2-4}$}

\affiliation{
	$^1$ Faculty of Engineering, Information and Systems, The University of Tsukuba, Tennodai, Tsukuba, Ibaraki 305-8573, Japan\\
	$^2$ ETH Zurich, Department of Management, Technology and Economics, Zurich, Switzerland\\
	$^3$ Tokyo Tech World Research Hub Initiative, Institute of Innovative Research, Tokyo Institute of Technology, Tokyo, Japan\\
	$^4$ Institute of Risk Analysis, Prediction and Management, Academy for Advanced Interdisciplinary Studies, Southern University of Science and Technology, Shenzhen, China
}
\date{\today}

\begin{abstract}
	A field theoretical framework is developed for the Hawkes self-excited point process with arbitrary memory kernels by embedding the original non-Markovian one-dimensional dynamics onto a Markovian infinite-dimensional one.
	The corresponding Langevin dynamics of the field variables is given by stochastic partial differential equations that are Markovian. 
	This is in contrast to the Hawkes process, which is non-Markovian (in general) by construction as a result of its (long) memory kernel.
	We derive the exact solutions of the Lagrange-Charpit equations for the hyperbolic master equations in the Laplace representation
	in the steady state, close to the critical point of the Hawkes process.
	The critical condition of the original Hawkes process is found to correspond to a transcritical bifurcation in the Lagrange-Charpit equations.
	We predict a power law scaling of the PDF of the intensities in an intermediate asymptotics regime, which crosses over
	to an asymptotic exponential function beyond a characteristic intensity that diverges as the critical condition is
	approached. 
	We also discuss the formal relationship between quantum field theories and our formulation.
	Our field theoretical framework provides a way to tackle complex generalisation of the Hawkes process, such as
	nonlinear Hawkes processes previously proposed to describe the multifractal properties of earthquake seismicity and of financial volatility.
\end{abstract}

\maketitle

\section{Introduction}

	The self-excited conditional Poisson process introduced by Hawkes \cite{Hawkes1,Hawkes2,Hawkes3} 
	has progressively been adopted as a useful first-order model of intermittent processes
	with time (and space) clustering, such as those occurring in seismicity and financial markets.
	The Hawkes process was first used and extended in statistical seismology and remains probably the most
	successful parsimonious description of earthquake statistics \cite{KK1981,KK1987,Ogata1988,Ogata1999,HelmsSor02,Shyametal2019}. 
	More recently, the Hawkes model has known a burst of interest in finance (see e.g. \cite{HawkesRev18} for a short review) as it was realised that
	some of the stochastic processes in financial markets can be well represented by
	this class of models \cite{FiliSor12}, for which the triggering and branching processes capture the herding nature
	of market participants (be they due to psychological or rational imitation of human traders or as a result of machine learning and adapting). 
	In field of financial economics, the Hawkes process has been successfully involved in issues as diverse as estimating
	the volatility at the level of transaction data, estimating the market stability \cite{FiliSor12,FiliSor15,WheatleyWeh2019},
	accounting for systemic risk contagion, devising optimal execution strategies or
	capturing the dynamics of the full order book \cite{Bacry1}. Another domain of intense use of the Hawkes model
	and its many variations is found in the field of social dynamics on the Internet, including instant messaging and blogging such on Twitter \cite{Zhao2015}
	as well as the dynamics of book sales \cite{SorDeschatres04}, video views \cite{CraneSor08}, success of movies \cite{EscobarSor08}, and so on.

	The present article, together with the joint-submission Letter~\cite{KKDS_PRL}, can be considered as a sequel complementing a series of papers
	devoted to the analysis of various statistical properties of the Hawkes process
	\cite{SaiSor2004,SaiHSor2005,SaiSor2006,SaiSorTimes2007,SaiSor2014}.
	These papers have extended the general theory of point processes \cite{DalayVere03}
	to obtain general results on the distributions of total number of events, total number of generations, and so on,
	in the limit of large time windows. Here, we consider the opposite limit of very small time windows,
	and characterise the distribution of ``intensities'', where the intensity $\nu(t)$ of the Hawkes process
	at time $t$ is defined as the probability per unit time that an event occurs (more precisely, $\nu(t) dt$
	is the probability that an event occurs between $t$ and $t+dt$).
	We propose a novel natural formulation of the Hawkes process
	in the form of a field theory of probability density functionals taking the form of a field master equation.
	This formulation is found to be ideally suited to investigate the hitherto ignored properties of the
	distribution Hawkes intensities, which we analyse in depth in a series of increasingly sophisticated
	forms of the memory kernel characterising how past events influence the triggering of future events.
	 
	This paper is organised as follows. 
	Section~\ref{sec:review} presents the Hawkes process in its original definition.
	In addition, we provide a comprehensive review of the previous literature 
	on non-Markovian stochastic processes of diffusive transport developed in traditional statistical physics.
	Historically, the analytical properties of the generalized Langevin equation (GLE) have been intensively studied 
	and we provide a brief review on the similarity and dissimilarity between the GLE and the Hawkes process 
	from the view point of the Markov embedding of the original non-Markovian one-dimensional dynamics onto a Markovian field dynamics.
	We then proceed to develop a stochastic Markovian partial differential equation equivalent to the Hawkes process in section~\ref{sec:ModelMasterEq}. This is done
	first for the case where the memory kernel is a single exponential, then made of two exponentials,
	an arbitrary finite number of exponentials and finally for general memory kernels. It is in section~\ref{sec:ModelMasterEq}
	that the general field master equations are derived. Section~\ref{sec:solutions} presents the analytical treatment
	and provides the solutions of the master equations, leading to the derivation of the probability density
	function of the Hawkes intensities for the various above mentioned forms of the memory kernel.
	We also discuss a formal relationship between quantum field theories and our formulation in Section~\ref{sec:Discussion_quantumFieldTheory}.
	Section~\ref{sec:conclusion} summarises and concludes by outlining future possible extensions of the formalism.
	These sections are complemented by seven appendices, in which the detailed 
	analytical derivations are provided.

\section{Model and literature review}\label{sec:review}

	\subsection{Notation}
		Firstly, we explain our mathematical notation for stochastic processes.  
		By convention, we denote stochastic variables with a hat symbol, such as $\hat{A}$, to distinguish them from
		the non-stochastic real numbers $A$, corresponding for instance to a specific realisation of the random variable. 
		The ensemble average of any stochastic variable $\hat{A}$ is also denoted by $\la\hat{A} \ra$. 		
		The probability density function (PDF) of any stochastic $\hat{A}$ is denoted by $P_t(\hat{A}(t)=A) = P_t(A)$.
		The PDF characterizes the probability that $\hat{A}(t) \in [A,A+dA)$ as $P_t(A)dA$.
		Using the notation of the PDF, the ensemble average reads $\la \hat{A}(t)\ra:=\int AP_t(A)dA$.

		We also make the following remark on notations used for our functional analysis. 
		For any function $z(x) \in \mathcal{S}_F$ defined for $x \in \bm{R}_+ = (0,\infty)$ with a function space $\mathcal{S}_F$, we can consider a functional $f[\{z(x)\}_{x\in \bm{R}_+}]$:
		i.e., $f: \mathcal{S}_F\to \bm{R}$. 
		Functionals in this paper are often abbreviated as $f[z]:=f[\{z(x)\}_x]$, with the square brackets emphasized to distinguish from ordinary functions.
		
	\subsection{Definition of the Hawkes conditional Poisson process}
	
		\begin{figure}
			\centering
			\includegraphics[width=75mm]{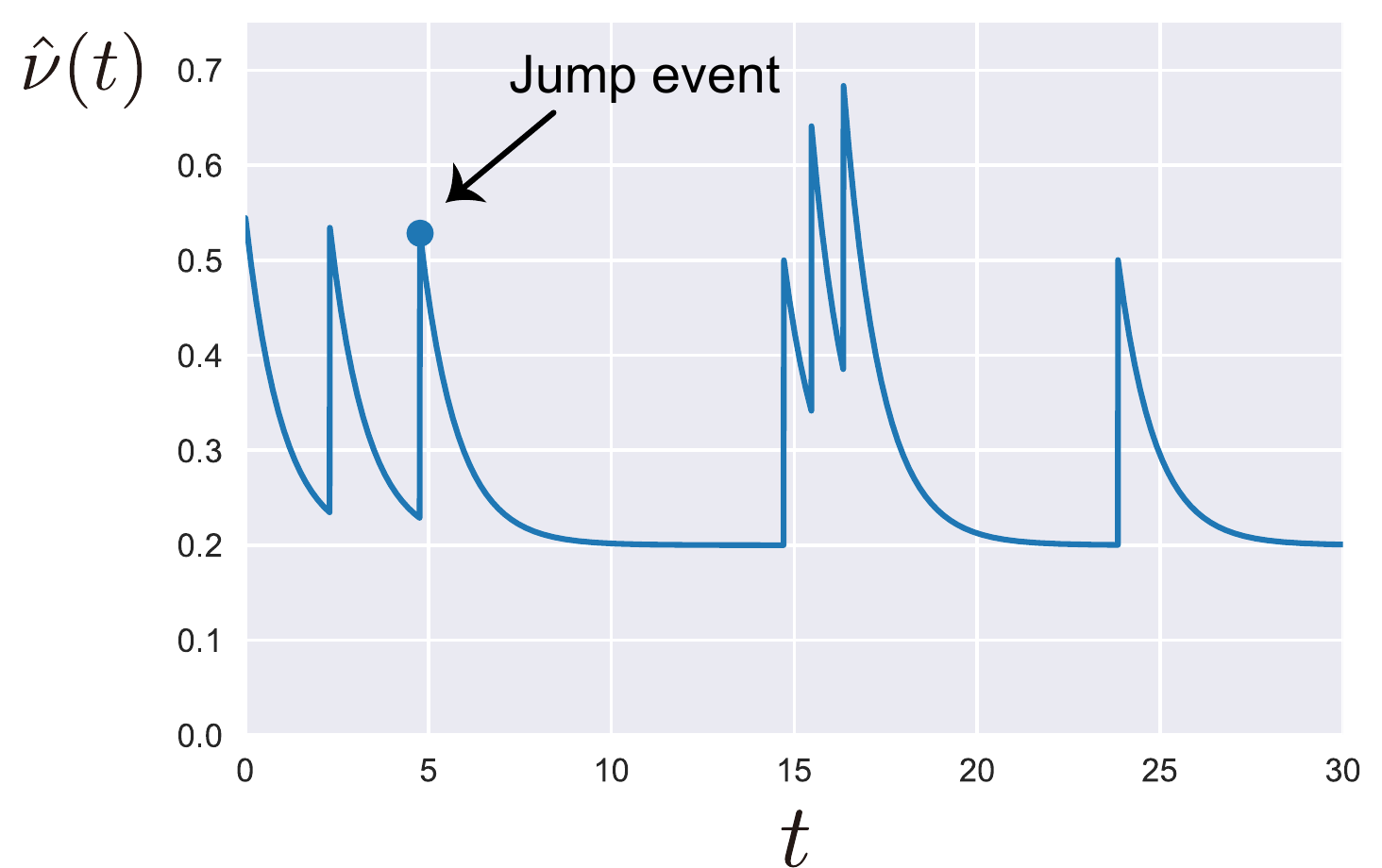}
			\caption{
						Schematic representation of the Hawkes process (\ref{def:Hawkes_general}) with an exponential kernel~\eqref{def:single_expon_kernel}. 
						The event occurring at a given time stamp is represented by a jump in the intensity $\hnu(t) $, the probability per unit time that a next event will occur.
					}
			\label{fig:trj_singleExpon}
		\end{figure}
		The Hawkes process is the simplest self-excited point process, which describes with a linear intensity function $\hnu$ how past
		events influence the triggering of future events. Its structure is particularly well-suited to address the general and important
		question occurring in many complex systems of disentangling the exogenous from the endogenous sources of observed activity.
		It has been and continues to be a very useful model in geophysical, social and financial systems. 
		
		The Hawkes process, as any other point process, deals with 
		events (``points'' along the time axis). The theory of point processes indeed considers events as being characterised by a time of 
		occurrence but vanishing duration (the duration of events is very small compared to the inter-event times).
		Thus, to a given event $i$ is associated a time $t_i$ of occurrence. In the case where one deals with spatial point processes,
		the event has also a position ${\vec r}_i$.  And a ``mark'' $m_i$ can be included to describe the event's size, or its ``fertility'', i.e. the
		average number of events it can trigger directly.
		
		The stochastic dynamics of the Hawkes process is defined as follows. 
		Let us introduce a state variable $\hnu$, called the intensity. 
		The intensity $\hnu$ is a statistical measure of the frequency of events per unit time (i.e., a shock occurs during $[t,t+dt)$ with the 			probability of $\hnu dt$). 
		In the Hawkes process, the intensity satisfies the following stochastic sum equation (see Fig.~\ref{fig:trj_singleExpon}): 
		\begin{align}
			\hnu(t) = \nu_0 + n\sum_{i=1}^{\hN(t)} h(t-\htt_i), \label{def:Hawkes_general}
		\end{align}
		where $\nu_0$ is the background intensity, $\{\htt_i\}_i$ represent the time series of events, $n$ is a positive number called the branching ratio, $h(t)$ is a normalized positive function (i.e., $\int_0^\infty h(t)dt=1$), 
		and $\hN(t)$ is the number of events during the interval $[0,t)$ (called ``counting process'').  
		One often refers to $\hnu(t)$ as a conditional intensity in the sense that,
		conditional on the realised sequence of $\hN(t)=k$ (with $k\geq 0$) events, the probability that the $(k+1)$th event 
		occurs during $[t,t+dt)$, such that $\htt_{k+1}\in [t,t+dt)$,  is given by $\hnu(t)dt$. 		
		The pulse (or memory) kernel $h(t)$ represents the non-Markovian influence of a given event, and is non-negative definite.

		The branching ratio $n$ is a very fundamental quantity, which 
		is the average number of events of first generation (``daughters'') triggered by a given event \cite{DalayVere03,HelmsSor02}.
		This definition results from the fact that the Hawkes model, as a consequence of the linear
		structure of its intensity (\ref{def:Hawkes_general}), can be mapped exactly onto a branching process,
		making unambiguous the concept of generations: more precisely, a given realisation of the Hawkes process can be represented by the 
		set of all possible tree combinations, each of them weighted by a certain probability derived from the intensity function
		\cite{ZhuangVere02}. The branching ratio is the control parameter separating three different regimes: (i) $n<1$: subcritical; 
		(ii) $n=1$: critical and (iii) $n>1$: super-critical or explosive (with a finite probability). The branching ratio $n$
		can be shown to be also the fraction of events that are endogenous, i.e., that have been triggered by previous events \cite{HelmsSor03}.

	\subsection{Review of non-Markovian stochastic processes in the framework of the generalized Langevin equation}
		This subsection provides the background of previous methods for non-Markovian stochastic processes in statistical physics, 
		by focusing on the diffusive dynamics of Brownian particles (e.g., see Refs.\cite{Hanggi1990} for detailed reviews). 
		While this class of physical stochastic processes exhibits dynamical characteristics that are quite different from those of the Hawkes processes, 
		our framework can been formally related to such standard theories (in particular for the Markov embedding techniques for non-Markovian processes).
		We thus offer a comprehensive review to prepare the reader to better understand our theoretical developments.
		This subsection is written in a self-contained way and is not needed to understand our main results; 
		readers only interested in our formulation can skip this section. 
	
		\subsubsection{Markovian Langevin equations}
			\paragraph*{Langevin equation.}
			In the context of statistical physics, non-Markovian stochastic processes have been studied from the viewpoint of diffusion processes~\cite{KuboB,ZwanzigB}. 
			One of the typical diffusive models is the Langevin equation,
			\begin{equation}
				M \frac{d\hv(t)}{dt} = -\frac{dU(\hx)}{d\hx} - \gamma \hv(t) + \heta(t), \>\>\> \frac{d\hx(t)}{dt} = \hv(t), \label{eq:Langevin_underdamped}
			\end{equation}
			where $\hv(t)$ and $\hx(t)$ are the velocity and the position of the Brownian particle, $M$ is its mass, $U(\hx)$ is the confining potential, $\gamma$ is the viscous friction coefficient, 
			and $\heta(t)$ represents the thermal fluctuation modelled by the zero-mean white Gaussian noise. 
			Equation~\eqref{eq:Langevin_underdamped} is one of the stochastic differential equations (SDE) first written down in the history of Physics. 
			
			The noise term embodying the presence of thermal fluctuation satisfies the fluctuation-dissipation relation (FDR)
			\begin{equation}
				\la \heta(t)\heta(t')\ra=2\gamma T \delta(t-t')
			\end{equation}
			where $T$ is the temperature. In this paper, the Boltzmann constant is taken unity: $k_B=1$.
			The FDR must hold for relaxation dynamics near equilibrium states because both viscous friction and thermal fluctuation come from the same thermal environment~\cite{ZwanzigB}. 
			
			The standard analytical solution to this Langevin equation can be obtained via
			the time-evolution equation for the joint PDF $P_t(v,x)$, which is given by the Fokker-Planck (FP) equation~\cite{GardinerB,VanKampenB}:
			\begin{equation}
				\frac{\partial P_t(v,x)}{\partial t} = \mathcal{L}_{\mathrm{FP}}P_t(v,x), \>\>\>
				\mathcal{L}_{\mathrm{FP}}:= - \frac{\partial}{\partial x}v
				 + \frac{\gamma}{M}\left[ \frac{\partial}{\partial v}\left(v+\frac{dU(x)}{dx}\right) + \frac{T}{M}\frac{\partial }{\partial v}\right].
				\label{eq:FP_equation}
			\end{equation}
			It is remarkable that the FP equation is always linear in terms of the PDF even if the Langevin equation has nonlinear terms in general. 
			While there is a one-to-one correspondence between the Langevin equation and the FP equation, the FP equation is often analyzed because the standard methods based on linear algebra is available, such as the eigenfunction expansion.
			
			The Langevin equation~\eqref{eq:Langevin_underdamped} can be interpreted as the equation of motion for a Brownian particle, 
			obtained after integrating out the many degrees of freedom of the original microscopic dynamics except for those of the Brownian particles. 
			For example, Eq.~\eqref{eq:Langevin_underdamped} can be systematically derived from the Hamiltonian dynamics of the Brownian particle surrounded by a dilute gas (see the kinetic framworks~\cite{McDonald,Spohn1980,VanKampenB} for examples).

			\paragraph*{Non-Markovian nature as a result of variable elimination.}
			Let us focus on the case of a harmonic potential $U(\hx)={1 \over 2}k\hx^2$ and eliminate the velocity~\cite{ZwanzigB}, 
			to change the system descriptions from $(\hv,\hx)$ to $\hx$. 
			By integrating out the velocity degree, we obtain the non-Markovian dynamics for the position $\hx(t)$: 
			\begin{equation}
				\frac{d\hx(t)}{dt} = -\int_{0}^{\infty}dsK(s)\hx(t-s) + \hat{\zeta}(t), \>\>\> \hat{\zeta}(t) := \frac{1}{M}\int_0^\infty ds e^{-\gamma s/M}\heta(t-s). \label{eq:Langevin_memory_overdamped}
			\end{equation}
			Here, the memory kernel $K(t):= ke^{-\gamma t/M}/M$ represents the retarded potential effect 
			and the noise term $\hat{\zeta}(t)$ is the colored Gaussian noise 
			with zero mean $\la \hat{\zeta}(t) \ra = 0$ and auto-correlation $\la \hat{\zeta}(t)\hat{\zeta}(t') \ra = (T/M)e^{-\gamma |t-t'|/M}$. 
			Remarkably, the non-Markovian nature has appeared as a result of variable elimination. 
			Here the non-Markovian version of the FDR holds $\la \hat{\zeta}(t)\hat{\zeta}(t') \ra = \la x^2\ra_{\rm{eq}} K(|t-t'|)$ with equilibrium average $\la x^2\ra_{\rm{eq}}:= k_{B}T/k$.

		\subsubsection{Generalized non-Markovian Langevin equation.}
			While the Markovian Langevin description~\eqref{eq:Langevin_underdamped} is reasonable for dilute thermal environments~\cite{LiReview2013,LiNature2010}, 
			such a Markovian description is not available for dense thermal environments, such as liquids~\cite{LiReview2013,Huang2011,Franosch2011}. 
			Indeed, Eq.~\eqref{eq:Langevin_underdamped} is not valid even for a Brownian particle in water, which is one of the most historically important cases.
			For such cases, the Langevin description must be modified to accommodate non-Markovian effects originating from hydrodynamic interactions in liquids. 
			The minimal model for such systems is given by the generalized Langevin equation (GLE)~\cite{KuboB,ZwanzigB}:
			\begin{equation}
				M\frac{d\hv(t)}{dt} = -\frac{dU(\hx)}{d\hx}-\int_0^\infty dsK(s)\hv(t-s) + \heta(t),\label{eq:generalized_Langevin}
			\end{equation}
			where $K(t)$ is the memory kernel for viscous friction and $\heta(t)$ is the thermal fluctuation modeled by a colored 
			Gaussian noise with zero mean $\la \heta(t)\ra=0$.
			Since both viscous friction and thermal fluctuation share the same origin, they are related to each other via the FDR for non-Markovian processes:
			\begin{equation}
				\la \heta(t)\heta(t')\ra = T K(|t-t'|).\label{eq:FDR_generalized}
			\end{equation}
			For typical three-dimensional liquid systems, the memory kernel has a long tail due to the hydrodynamic retardation effect, 
			such that $K(t) \propto t^{-3/2}$, which was verified by direct experiments in Refs.~\cite{LiReview2013,Huang2011,Franosch2011}. 
		
			The GLE~\eqref{eq:generalized_Langevin} can be derived from microscopic dynamics 
			by rearrangement of the Liouville operator by the method of the projection operators~\cite{Mori1965,ZwanzigB}.
			Assuming that the initial state of the system is sufficiently close to equilibrium and that a sufficient number of macroscopic variables are 
			accessible via experiments, 
			the projection operator formalism provides a microscopic foundation of the non-Markovian stochastic processes with intuitive physical interpretation. 
		
			The solution of the GLE~\eqref{eq:generalized_Langevin} is much harder to obtain than that of the Markovian Langevin equations due to its non-Markovian nature. 
			However, due to the linearity of the dynamics and the Gaussianity of the thermal fluctuations (while they are specific to the GLE equation~\eqref{eq:generalized_Langevin}),
			all the moments and correlation functions can be analytically calculated based on the Laplace transformation of the SDE~\cite{Morgado2002} and response functions can be expanded as sums of exponentials through the residue theorems~\cite{Bao2006}. 
			In addition, the recurrence methods of Ref.~\cite{Lee1982} are also available for this model. 

		\subsubsection{Origin of the non-Markovian property and Markov embedding}\label{subsubsec:orgn_non-Markovian}
			We have seen that the non-Markovian property appears as the result of variable elimination through the example of Eq.~\eqref{eq:Langevin_memory_overdamped}. 
			This shows that some non-Markovian systems with an exponential memory kernel can be converted back to a Markovian system by adding an auxiliary variable~\cite{ZwanzigB}. 
			This procedure is called Markov embedding~\cite{Goychuk2009,Kupferman2004} and can be generalized to transform the GLE~\eqref{eq:generalized_Langevin} into simultaneous Markovian SDEs 
			when the kernel is given by a sum of exponentials:
			\begin{equation}
				K(t) = \sum_{k=1}^K \kappa_ie^{-t/\tau_i} \>\>\>
				\Longrightarrow \>\>\>
				M\frac{d\hv(t)}{dt} = \sum_{k=1}^K \hat{u}_k(t), \>\>\> 
				\frac{d\hat{u}_k(t)}{dt} = -\frac{\hat{u}_k(t)}{\tau_k} - \kappa_k\hv(t) + \sqrt{\frac{2\kappa_kT}{\tau_k}}\hxi^{\mrG}_k(t)
				\label{eq:GLE_MarkovEmbedding_discrete}
			\end{equation}
			with independent standard Gaussian noises $\hxi^{\mrG}_k(t)$, satisfying $\la \hxi^{\mrG}(t)_k\ra=0$ and $\la \hxi^{\mrG}_k(t)\hxi^{\mrG}_j(t')\ra =\delta_{kj}\delta(t-t')$. 
			Here we have taken the free potential case $U(\hx)=0$ for simplicity. 
			We thus find that the exponential-sum memory case is Markovian by introducing the new system-variable set $\hat{\Gamma}:= (\hv, \hat{u}_1,\dots, \hat{u}_K)$,
			while it was non-Markovian in the original variable representation $\hv(t)$. 
			 
			In this sense, whether a system is regarded as Markovian or non-Markovian crucially depends on which variable set is taken for the description of the system.  
			This story is actually consistent with the projection operator formalism: 
			while the original Hamiltonian dynamics obeys the Markovian dynamics (i.e., the time-evolution of the phase-space distribution is given by the Liouville equation),
			the reduced dynamics of macroscopic variables obeys the non-Markovian Langevin dynamics due to integrating out irrelevant variables. 
			Furthermore, a systematic method of Markov embedding~\cite{Marchesoni1983} was proposed on the basis of the continued-fraction expansion~\cite{Mori1965}. 
			
		\subsubsection{Fokker-Planck descriptions of non-Markovian processes}
			Compared with Markovian stochastic processes, there are few mathematical techniques that can be applied to the Fokker-Plank (master) equations associated with general non-Markovian processes. 
			One of the formal methods is to use the time-convolution Fokker-Planck equation for macroscopic variables $\bm{x}$:
			$\partial P_t(\bm{x})/\partial t = \int_{0}^{t} ds\mathcal{L}_{\mathrm{GFP}}(\bm{x},s)P_{t-s}(\bm{x})$,
			which was derived from the projection operator formalism~\cite{ZwanzigB}. 
			Also, some specific class of non-Markovian SDEs can be formulated as a Fokker-Planck equation with time-dependent coefficients via the functional stochastic calculus~\cite{HanggiReview1995}. 
			While these equations are formally correct, they are not easy to exploit for practical calculations due to their genuine non-Markovian nature.
			
			One of the most powerful approaches to tackle GLE is to use Markov embedding, thus making the system Markovian. 
			This research direction was proposed by Ref.~\cite{Marchesoni1983} and there are a variety of ways to select the auxiliary variables. 
			By taking the selection in Eq.~\eqref{eq:GLE_MarkovEmbedding_discrete} according to Ref.~\cite{Goychuk2009}, we obtain
			the complete Fokker-Planck equation for the GLE with the exponential-sum memory $K(t)=\sum_{k=1}^K\kappa_ie^{-t/\tau_k}$ in the absence of potential $U(\hx)=0$,
			\begin{equation}
				\frac{\partial P_t(\Gamma)}{\partial t} = \sum_{k=1}^K\left[-\frac{\partial}{\partial v}\frac{u_k}{M}+\frac{\partial}{\partial u_k}\left(\frac{u_k}{\tau_k}+\kappa_kv\right) + 
				\frac{\kappa_kT}{\tau_k}\frac{\partial^2}{\partial u_k^2}\right]P_t(\Gamma)
				\label{eq:FP_GLE_discrete}
			\end{equation}
			for the extended phase point $\Gamma:= (v,u_1,\dots,u_K)$ and the corresponding PDF $P_t(\Gamma)$.
			
		\subsubsection{Field description for an infinite number of auxiliary variables}\label{sec:SPDE_GLE}
			There are two classes for Markov embedding: one requires a finite number of auxiliary variables (e.g., the GLE  with a finite discrete sum of exponential terms to describe the memory~\eqref{eq:GLE_MarkovEmbedding_discrete}),
			and the other one requires an infinite number of auxiliary variables. 	
			The latter class is essentially similar to the classical field theory of stochastic processes, represented by stochastic partial differential equations (SPDE). 
			Indeed, the continuous version of the Markov embedding~\eqref{eq:GLE_MarkovEmbedding_discrete} can be expressed in terms of SPDEs as follows.
			For the continuous decomposition
			\begin{equation}
				K(t) = \int_0^\infty dx \kappa(x)e^{-t/x},
			\end{equation}
			we obtain an equivalent Markov embedding representation in the absence of the potential $U(\hx)=0$,
			\begin{align}
				\frac{d\hv(t)}{dt} = \frac{1}{M}\int_0^\infty dxu(t,x), \>\>\> \frac{\partial \hat{u}(t,x)}{\partial t} = -\frac{\hat{u}(t,x)}{x} - \kappa(x)\hv(t) + \sqrt{\frac{2\kappa(x)T}{x}}\hxi^{\mrG}(x,t)~,
				\label{eq:GLE_MarkovEmbedding_continuous}
			\end{align}
			with the spatial white Gaussian noise term $\hxi^{\mrG}(t,x)$ satisfying $\la \hxi^{\mrG}(t,x)\ra$=0 and $\la \hxi^{\mrG}(t,x)\hxi^{\mrG}(t',x')\ra = \delta(x-x')\delta(t-t')$. 
			This is a simple equation in terms of the time derivative but it can be regarded as a SPDE, since $\hat{u}(t,x)$ is spatially distributed over the auxiliary field $x\in (0,\infty)$. 
			Indeed, this interpretation enables us to apply the functional calculus historically developed for the analytical solution of SPDEs~\cite{GardinerB}.
			
			In the domain of mathematics dealing with SPDEs, the Fokker-Planck description is based on the functional calculus for the field variable (called the functional Fokker-Planck equation in Ref.~\cite{GardinerB}). 
			The corresponding field Fokker-Planck equation can be derived by replacing the discrete sum with the functional derivative:
			\begin{equation}
				\frac{\partial P_t[\Gamma]}{\partial t} = \int_0^\infty dx \left[-\frac{\partial}{\partial v}\frac{u(x)}{M}+\frac{\delta}{\delta u(x)}\left(\frac{u(x)}{x}+\kappa(x)v\right) + \frac{\kappa(x)T}{x}\frac{\delta^2}{\delta u^2(x)}\right]P_t[\Gamma]
				\label{eq:field_master_GLE}
			\end{equation}
			for the extended phase point $\Gamma := (v, \{u(x)\}_{x\in (0,\infty)})$ and the corresponding probability functional $P_t[\Gamma]$. 
			
			We note that the mathematical precise definition of such Fokker-Planck description has not been established yet~\cite{GardinerB}.
			One can see a formal divergent term in the FP field equation, such as $[\delta/\delta u(x)]u(x) = \delta(0)$. 
			This divergence is irrelevant for physical observables as shown in Sec.~\ref{sec:quantumFieldTheoryGLE} since the SPDE is linear.
			But, general non-linear SPDEs can have divergences even for observables (see the example of the nonlinear stochastic reaction-diffusion model in Ref.~\cite{GardinerB}, Chapter 13.3.3).
			According to convention, the safest interpretation is that all the procedures are implicitly discrete and the continuous notation is regarded as a useful abbreviation of the discrete underlying model.
			By introducing the finite lattice constant $dx$ for the auxiliary variable $x$, we have the variable transformations $u_k \to u(x_k)dx$ and $\kappa_k \to \kappa(x_k)dx$.
			The functional derivative is then introduced as the formal limit of $\delta F[\Gamma]/\delta u(x_k) := \lim_{dx \to 0}(\partial F[\Gamma]/\partial u_k)/dx$ (see Ref.~\cite{GardinerB}, Chapter 13.1.1).
			In this sense, the derivation of the field Fokker-Planck equation~\eqref{eq:GLE_MarkovEmbedding_continuous} here follows this convention: 
			we first confirm that the discrete Markov embedding~\eqref{eq:GLE_MarkovEmbedding_discrete} works well and then generalize it to its general continuous version~\eqref{eq:GLE_MarkovEmbedding_continuous}.

		\subsubsection{Relation to the non-Markovian Hawkes processes}
			The above historical discussion of non-Markovian diffusive process provides a clear guideline for different classes of stochastic processes, including the Hawkes process as follows.
			The summary highlighting dissimilarities and similarities between the GLE and the Hawkes process is presented in Table~\ref{table:comparison_GLE_Hawkes}. 
			
			\paragraph{Dissimilarities.}
			The Hawkes process is an example of non-Markovian point processes, triggering finite-size jumps along a sample trajectory.
			This is in contrast to the non-Markovian Langevin equation, which is based on infinitesimal Gaussian noise to describe 
			the diffusive local transport, and thus does not include trajectory jumps.  
			This difference should be reflected in the form of the time-evolution equation of the PDF. 
			Indeed, the time-evolution of the PDF for the Hawkes process will be shown to obey the master equation (integro-differential equations), 
			while that for the Langevin dynamics obeys the Fokker-Planck equations (second-order derivative equations).
			
			Another dissimilarity comes from the fact that the Hawkes process is an out-of-equilibrium model typically describing a branching process requiring immigrants or background events to drive the whole sequence of events, 
			while the Langevin equations are near-equilibrium models in the sense that initial distributions of the thermal baths are characterized by small perturbations from the Gibbs distribution. 
			Note that the physical validity of the projection operator formalism is not guaranteed in general for out-of-equilibrium systems~\cite{ZwanzigB} characterized by non-Gibbs initial distributions.
			This conceptual difference is important because the FDR~\eqref{eq:FDR_generalized} is not necessarily assumed for the Hawkes processes. 
			Indeed, the master equation for the Hawkes process does not satisfy the detailed balance condition~\cite{GardinerB} (or the time-reversal symmetry) as shown later.
			
			We should stress that the word ``non-equilibrium'' is used here for systems driven by external forces, which leads to non-Gibbs initial distributions for thermal baths in contact with the system. While we did not review these cases in detail, there are various understandings of the FDR.
			In the historical context of stochastic processes, the FDR is defined as being closely associated with the time-reversal symmetry of the stochastic processes. 
			In this sense, one can mathematically prove that the Hawkes processes is actually ``out-of-equilibrium'', in the sense that the corresponding master equation does not satisfy the symmetry condition in Gardiner's textbook \cite{GardinerB}. 
			We note that such ``out-of-equilibrium" processes are physically reasonable in general non-equilibrium setups. 
			Indeed, the shot noise process~\cite{GardinerB} and the Levy flights dynamics~\cite{KlafterB} can be observed in out-of-equilibrium systems 
			(e.g., see Refs.~\cite{KanazawaPRL2015,KanazawaNature2020} for their statistical-physics derivation from microscopic dynamics), while they do not satisfy the detailed balance condition.
			In the historical context of the projection operators, the ``near-equilibrium condition'' is defined such that the initial distribution for the noise-space is characterized by a linear response around the Gibbs distribution. 
			In the context of fluctuation theorems~\cite{FTJarzynski2000}, the FDR is derived from the assumptions of (1) time-reversal symmetry of the microscopic dynamics and
			 (2) validity of the Gibbs distribution for the thermal bath in contact with the target system. The common understanding is that the noise source (i.e. the thermal bath in contact with the systems) is characterised by a distribution close to the Gibbs distribution. 
			 Therefore, there should not be any confusion when considering the Hawkes process for which the FDR is irrelevant.
			
			\paragraph{Similarities.}
			The GLE~\eqref{eq:generalized_Langevin} can be mapped onto a Markovian process~\eqref{eq:GLE_MarkovEmbedding_continuous} by adding a sufficient number of auxiliary variables.
			As we show below, the same Markov embedding technique is available even for the non-Markovian Hawkes processes~\eqref{def:Hawkes_general}, by adding a sufficient number of auxiliary variables.
			Since the memory kernel can be a continuous sum of exponential kernels, the most general description of the Hawkes process should be based on an infinite number of auxiliary variables.
			As discussed above, such systems are typically described as a classical field theory driven by stochastic terms  
			and thus the dynamics of the original Hawkes process can be finally mapped onto an SPDE~\eqref{eq:SPDE_generalHawkes} and the field master equations~\eqref{eq:master_gen_functional}. 
			
			\begin{table}
				\begin{tabular}{|c|c|c|c|}
					\hline
					Model & GLE (exponential memory) & GLE (general memory) & Hawkes process (general memory) \\ \hline \hline
					Character & Diffusive transport & Diffusive transport & Point process \\
					Fundamental equation & Fokker-Planck equation~\eqref{eq:FP_GLE_discrete} & Field Fokker-Planck equation~\eqref{eq:field_master_GLE} & Field master equation~\eqref{eq:master_gen_functional} \\
					Typical systems & Near equilibrium & Near equilibrium & Out-of-equilibrium\rule[1mm]{0mm}{2mm}\\
					Phenomena & Relaxation & Relaxation & Critical bursts \\
					FDR & Yes & Yes  & No \\
					Non-Markovian representation & SDE~\eqref{eq:generalized_Langevin} & SDE~\eqref{eq:generalized_Langevin} & SDE~\eqref{def:Hawkes_general} \\
					Markovian representation  & SDE~\eqref{eq:GLE_MarkovEmbedding_discrete} & SPDE~\eqref{eq:GLE_MarkovEmbedding_continuous} & SPDE~\eqref{eq:SPDE_generalHawkes} \\
					\# of auxiliary variables & finite & inifinite (field description) & infinite (field description) \\ \hline
				\end{tabular}
				\caption	{
								Summary table to compare the generalized Langevin equation and the Hawkes process. 
								Here FDR stands for the fluctuation-dissipation relation. 
						}
				\label{table:comparison_GLE_Hawkes}
			\end{table}

\section{Master equations}\label{sec:ModelMasterEq}			
	We now formulate the master equation for the model~\eqref{def:Hawkes_general} and provide its asymptotic solution around the critical point. 
	
	\subsection{Markov embedding: introduction of auxiliary variables}
		As discussed in Sec.~\ref{subsubsec:orgn_non-Markovian}, non-Markovian properties in many stochastic systems often arise as a result of variable eliminations. 
		This suggests that, reciprocally, it might be possible to map a non-Markovian system onto a Markovian one by adding auxiliary variables (i.e., Markov embedding). 
		While the Hawkes process~\eqref{def:Hawkes_general} is non-Markovian in its original representation only based on the intensity $\hnu(t)$, 
		it can be also transformed onto a Markovian process by selecting an appropriate set of system variables.  
		In this section, we formulating such a Markov embedding procedure to derive the corresponding master equations.
		
	\subsection{The single exponential kernel case}
	
		\subsubsection{Mapping to Markovian dynamics}
		
			Before developing the general framework for arbitrary memory kernel $h(t)$, 
			we consider the simplest case of an exponential memory kernel:
			\begin{equation}
				h(t) = \frac{1}{\tau}e^{-t/\tau},\label{def:single_expon_kernel}
			\end{equation}
			satisfying the normalization $\int_0^\infty h(t)dt=1$.
			The decay time $\tau$ quantifies how long an event can typically trigger events in the future. 
			This special case is Markovian as discussed in Refs.~\cite{Oakes1975,Dassios2011},
			because the lack of memory of exponential distributions ensures that the number of events after time $t$ defines a Markov process in continuous time~\cite{Knopoff1997}.
								
\begin{figure}
				\centering
				\includegraphics[width=140mm]{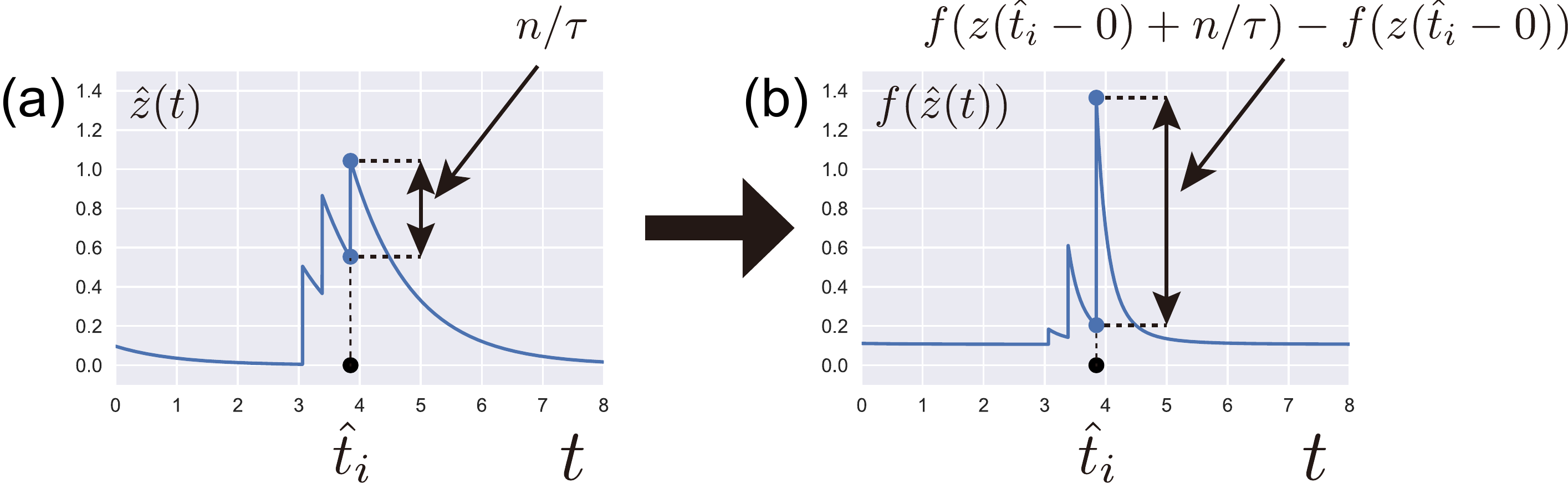}
				\caption{
							(a)~Schematic representation of a typical trajectory for $\hz(t)$ defined by (\ref{hygqfqtgqb}).
							A jump of size $n/\tau$ may occur in the time interval $[t,t+dt)$ with probability $\hnu(t)dt$. 
							(b)~Let us consider an arbitrary function $f(\hz)$. At the time $t=\htt_i$ of the jump, there is a corresponding jump in the trajectory of $f(\hz)$, 
							which is characterized by $df(\hz(t)):= f(\hz(\htt_i-0)+n/\tau)-f(\hz(\htt_i-0))$. 
							The plots shown here are based on a numerical simulation with the following parameters: $\tau=1$, $n=0.5$, $\nu_0=0.1$ and $f(z)=\exp[5(z-1)]+0.1$. 
						}
				\label{fig:trj_singleExpon_f}
			\end{figure}

			As shown in the example~\eqref{eq:Langevin_memory_overdamped}, some non-Markovian processes can be mapped onto a Markovian stochastic system if the memory function is exponential. Here we show that, likewise,
			this single exponential case~\eqref{def:single_expon_kernel} can be mapped onto an SDE driven by a state-dependent Markovian Poisson noise. 
			By decomposing the intensity as 
			\begin{equation}
				\hz:= \hnu-\nu_0,  \label{hygqfqtgqb}
			\end{equation}
			let us consider the Langevin dynamics
			\begin{equation}
				\frac{d\hz}{dt} = -\frac{1}{\tau}\hz + \frac{n}{\tau} \hxi^{\mrP}_{\hnu} \label{eq:Markov_exp_pulse}
			\end{equation}	
			with a state-dependent Poisson noise $\hxi^{\mrP}_{\hnu}$ with intensity given by $\hnu=\hz+\nu_0$ and
			initial condition $\hz(0)=0$. The introduction of $\hz$ is similar to the trick proposed 
			in \cite{BouchaudTradebook2018} for an efficient estimation of the maximum likelihood of the Hawkes process.
			By expressing the state-dependent Poisson noise as
			\begin{equation}
				\hxi^{\mrP}_{\hnu}(t) = \sum_{i=1}^{\hN(t)} \delta(t-\htt_i),  \label{hwtrgfq}
			\end{equation}
			we obtain the formal solution of equation (\ref{eq:Markov_exp_pulse})
			\begin{equation}
				\hnu(t) = \nu_0+\hz(t) = \nu_0 + \frac{n}{\tau}\int_0^t dt'e^{-(t-t')/\tau} \hxi^{\mrP}_{\hnu}(t') 
				= \nu_0 + n\sum_{i=0}^{\hN(t)} h(t-\htt_i).
			\end{equation}
			This solution shows that the SDE~\eqref{eq:Markov_exp_pulse} is equivalent to the Hawkes process~\eqref{def:Hawkes_general}. 					
			Equation~\eqref{eq:Markov_exp_pulse} together with (\ref{hwtrgfq}) is therefore a short hand notation for
			\begin{equation}
				\hz(t+dt) - \hz(t) = \begin{cases}
							-\frac{1}{\tau}\hz(t)dt & (\mbox{No jump during $[t,t+dt)$; probability} = 1-\hnu(t)dt)\\
							\frac{n}{\tau} & (\mbox{Jump in $[t,t+dt)$; probability} = \hnu(t)dt)
						\end{cases}
						\label{jehygbqgb}
			\end{equation}
			for the probabilistic time evolution during $[t,t+dt)$ (see Fig.~\ref{fig:trj_singleExpon_f}a for a schematic representation). Note that 
			the event probability explicitly depends on $\hnu(t)$, which reflects the endogenous nature of the Hawkes process.
			This is the first example of the Markov embedding of the Hawkes process. Note that this procedure corresponds to Eq.~\eqref{eq:GLE_MarkovEmbedding_discrete} for the case of the GLE with $K=1$.
			
		\subsubsection{Master equation}
			By introducing equation~\eqref{eq:Markov_exp_pulse} together with (\ref{hwtrgfq}), we have transformed
			a non-Markovian point process into a Markovian SDE. This allows us to derive the corresponding master equation 
			for the probability density function (PDF) $P_t(z)$ of the excess intensity $\hz$ (\ref{hygqfqtgqb}),
			\begin{equation}
				\frac{\partial P_t(z)}{\partial t} = \frac{1}{\tau}\frac{\partial }{\partial z}zP_t(z) + 
				\Big[(\nu_0+z-n/\tau)P_t(z-n/\tau)-(\nu_0+z) P_t(z)\Big], 
				\label{eq:master_exp}
			\end{equation}
			with the boundary condition
			\begin{equation}
				P_t(z)\Big|_{z=0} = 0
			\end{equation}
			$P_t(z)dz$ is thus the probability that $\hz(t)$ takes a value in the interval $\hz(t)\in [z,z+dz)$ at time $t$. 
			We note that this master equation after Markov embedding corresponds to the FP equation~\eqref{eq:FP_GLE_discrete} for the case of the GLE with $K=1$.
			
			The master equation (\ref{eq:master_exp}) is derived as follows. Let us consider an arbitrary function $f(\hz)$. 
			Using (\ref{jehygbqgb}), its time evolution during $[t,t+dt)$ is given by
			\begin{equation}
				f(\hz(t+dt)) - f(\hz(t)) = 	\begin{cases}
													-\frac{\hz(t)}{\tau}\frac{\partial f(\hz(t))}{\partial \hz}dt & (\mbox{Probability} = 1-\hnu(t)dt) \\
													f(\hz(t)+n/\tau) - f(\hz(t)) & (\mbox{Probability} = \hnu(t)dt)
												\end{cases},
			\end{equation}
			as schematically illustrated in Fig.~\ref{fig:trj_singleExpon_f}b.
			By taking the ensemble average of both sides, we obtain
			\begin{align}
				dt\int dz \frac{\partial P_t(z)}{\partial t}f(z) &= \int dz P_t(z)\left[-\frac{z}{\tau}\frac{\partial f(z)}{\partial z}dt + (z+\nu_0)\{f(z+n/\tau)-f(z)\} dt\right], \notag\\
				\Longrightarrow \int dz \frac{\partial P_t(z)}{\partial t}f(z) &= \int dz f(z)\left[\frac{1}{\tau}\frac{\partial }{\partial z}z P_t(z) +  \{(\nu_0+z-n/\tau)P(z-n/\tau)-(\nu_0+z) P(z)\}\right]. 
				\label{eq:derivation_master_exp}
			\end{align}
			This result (\ref{eq:derivation_master_exp}) is obtained by (i) using the identity
			\begin{equation}
				\left<f(\hz(t+dt))-f(\hz(t))\right> = \int dz [P_{t+dt}(z)-P_t(z)]f(z) = dt\int dz\frac{\partial P_t(z)}{\partial t}f(z),
			\end{equation}
			(ii) by performing a partial integration of the first term in the right-hand side of Eq.~\eqref{eq:derivation_master_exp}, and 
			(iii) by introducing the change of variable $z\to z-n/\tau$ for the second term. 
			Since (\ref{eq:derivation_master_exp})  is an identity holding for arbitrary $f(z)$, the integrants of the lelf-hand-side
			and right-hand-side must be equal for arbitrary $f(z)$, which yields the master equation~\eqref{eq:master_exp}. 
			
			Note that the above derivation of the master equation is not restricted to the exponential shape of the memory kernel.
			We are going to use the same derivation in the more complex examples discussed below.
			We also note that the master equation~\eqref{eq:master_exp} does not satisfy the detailed balance condition of Ref.~\cite{GardinerB}, which reflects the fact that the Hawkes process is a model for out-of-equilibrium systems.

	\subsection{Discrete sum of exponential kernels}\label{sec:discrete_sum_masterEq}
		\subsubsection{Mapping to Markovian dynamics}
			The above formulation can be readily generalized to the case of a memory kernel
			expressed as a discrete sum of exponential functions:
			\begin{equation}
				h(t) = \frac{1}{n}\sum_{k=1}^K \frac{n_k}{\tau_k}e^{-t/\tau_k}.
			\label{yjtyhnwrgbqbq}
			\end{equation}
			In this case, each coefficient $n_k$ quantifies the contribution of the $k$-th exponential with memory length $\tau_k$
			to the branching ratio $n = \sum_{k=1}^K n_k$, satisfying the normalization $\int_0^\infty h(t)dt=1$. 
			We note that this representation (\ref{yjtyhnwrgbqbq}) is quite general, as it can approximate well 
			the case of a power-law kernel with cut-off  up to a constant \cite{Hardimanetal2013}.
			
\begin{figure}
				\centering
				\includegraphics[width=140mm]{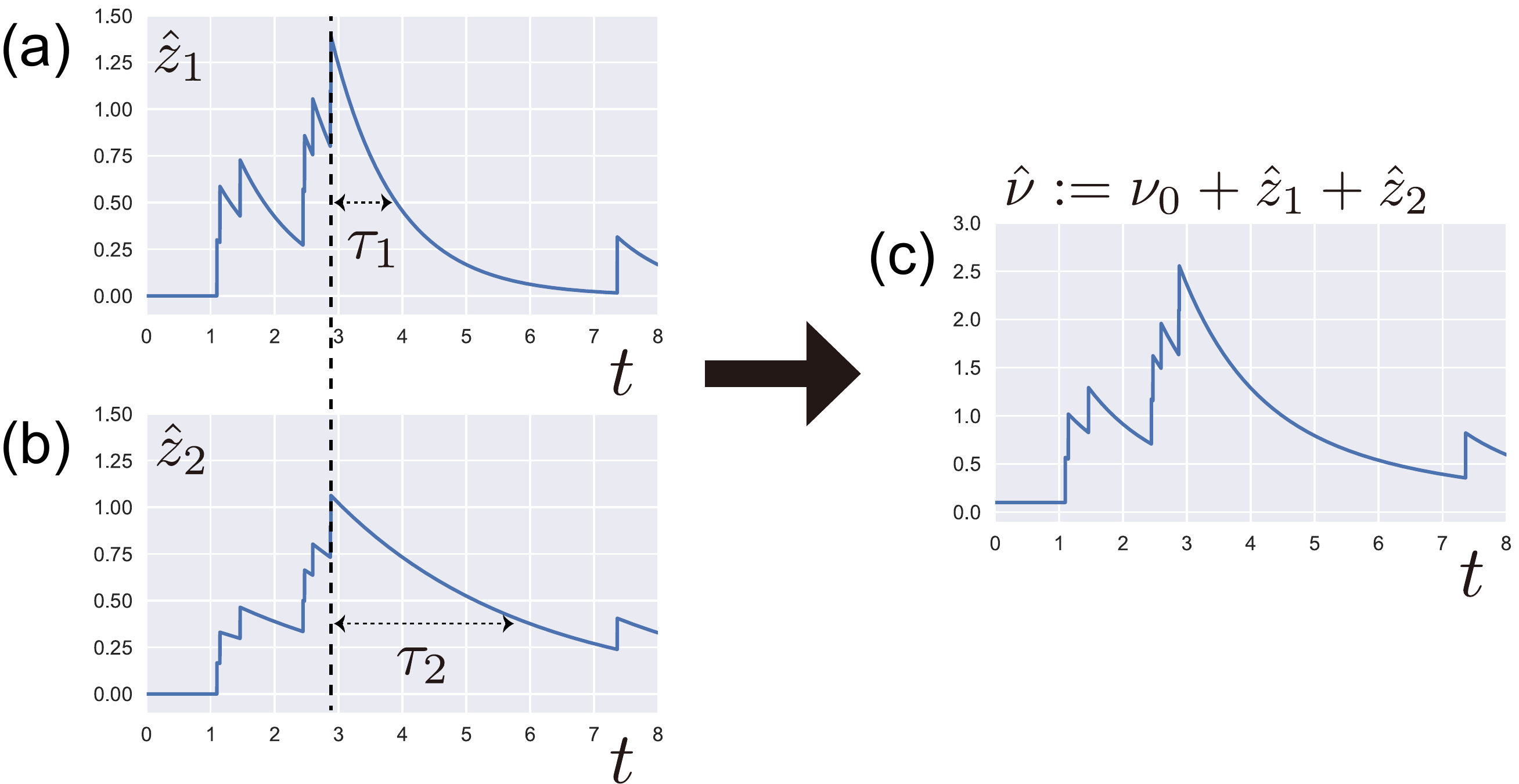}
				\caption{Case where the memory kernel is the sum of two exponentials. Panels (a) and (b) show
				the schematic trajectories of the two excess intensities $\hz_1$ and $\hz_2$ 
				and panel (c) that of the resulting total intensity $\hnu:=\nu0+\hz_1+\hz_2$.
				The parameters are $K=2$, $\tau_1=1$, $n_1=0.3$, $\tau_2=3$, $n_2=0.5$ (and thus $n=0.8$) and $\nu_0=0.1$.
						}
				\label{fig:trj_twoExpon}
			\end{figure}

			Harris suggested the intuitive notion that it is possible to map this case to a Markovian dynamics
			if the state of the system at time $t$ is made to include the list of the ages of all events \cite{Harris1963}. 
			The problem is that this conceptual approach is
			unworkable in practice due to the exorbitant size of the required information.
			By introducing an auxiliary age pyramid process, Ref.~\cite{Boumezoued2016} identified 
			some key components to add to the Hawkes process and its intensity to make the dynamics Markovian.
			Here, in order to map model~\eqref{def:Hawkes_general} onto a Markovian stochastic process,
			we propose a more straightforward Markov embedding approach, which generalised the previous case of a single 
			exponential memory function.
			We decompose the intensity into a sum of $K$ excess intensities $\{z_k\}_{k=1}^K$ as follows:
			\begin{equation}
				\hnu(t) = \nu_0 + \sum_{k=1}^K \hz_k(t)~.\label{eq:sum_expon_discrete}
			\end{equation}
			Each excess intensity $\hz_k$ is the solution of a Langevin equation driven by a state-dependent 
			Markovian Poisson shot noise 
			\begin{equation}
				\frac{d\hz_k}{dt} = -\frac{\hz_k}{\tau_k} + \frac{n_k}{\tau_k}\hxi^{\mrP}_{\hnu}~. 
				\label{eq:SDE_general_superposition_discrete}
			\end{equation}
			Note that the same state-dependent Poisson noise $\hxi^{\mrP}_{\hnu}(t)$ defined by expression (\ref{hwtrgfq}) acts on
			the Langevin equation for each excess intensity $\{\hz_k\}_{k=1,\dots,K}$. In other words,
			each shock event impacts simultaneous the trajectories for all excess intensities $\{\hz_k\}_{k=1,\dots,K}$ 
			(see the vertical broken line in Fig.~\ref{fig:trj_twoExpon}a and \ref{fig:trj_twoExpon}b and the resulting trajectory of $\hnu(t)$ in Fig.~\ref{fig:trj_twoExpon}c).
			Note that this Markov embedding corresponds to Eq.~\eqref{eq:GLE_MarkovEmbedding_discrete} for the case of the GLE.
			
		\subsubsection{Master equation}
			As the set of SDEs for $\hat{\bm{z}}:= (\hz_1,\hz_2,\dots,\hz_K)^{\mrT}$ are standard Markovian stochastic processes,
			we obtain the corresponding master equation:
			\begin{equation}
				\frac{\partial P_t(\bm{z})}{\partial t} = \sum_{k=1}^K\frac{\partial }{\partial z_k}\frac{z_k}{\tau_k}P_t(\bm{z}) 
				+ \left[ \left\{\nu_0+\sum_{k=1}^K (z_k- n_k/\tau_k)\right\}P_t(\bm{z}-\bm{h}) - \left\{\nu_0+\sum_{k=1}^K z_k \right\}P_t(\bm{z})  \right].
				\label{eq:master_n_expon}
			\end{equation}
			The jump-size vector is given by $\bm{h} := (n_1/\tau_1, n_2/\tau_2,\dots, n_K/\tau_K)^{\mrT}$. 
			The PDF $P_t(\bm{z})$ obeys the following boundary condition 
			\begin{equation}
				P_t(\bm{z})\Big|_{\bm{z} \in \partial \bm{R}^{K}_+} = 0~,  \label{eq:boundary_condition_n_expon}
			\end{equation}
			on the boundary $\partial \bm{R}^{K}_+:= \{\bm{z} | z_k = 0 \mbox{ for some }k \}$. 
			This equation (\ref{eq:master_n_expon}) can be derived following the procedure used for the single exponential 
			case that led us to the master equation \eqref{eq:master_exp} (see Appenedix~\ref{sec:master_eq_n_expon} for an explicit derivation). 
			Note that this master equation corresponds to the FP equation~\eqref{eq:FP_GLE_discrete} for the case of the GLE.

		\subsubsection{Laplace representation of the master equation}
			
			The master equation (\ref{eq:master_n_expon}) takes a simplified form under the Laplace representation,
			\begin{equation}
				\tl{P}_t(\bm{s}) := \mathcal{L}_{K}[P_t(\bm{z});\bm{s}]~,   \label{def:Laplace_PDF_n_gen}
			\end{equation}
			where the Laplace transformation in the $K$ dimensional space is defined by
			\begin{equation}
				\mathcal{L}_{K}[f(\bm{z}); \bm{s}] := \int_0^\infty d\bm{z} e^{-\bm{s}\cdot \bm{z}}f(\bm{z}) 
			\end{equation}
			with volume element $d\bm{z}:= \prod_{k=1}^K dz_k$.  The wave vector $\bm{s} := (s_1, \dots, s_K)^{\mrT}$
			is the conjugate of the excess intensity vector $\bm{z} := (z_1, \dots, z_K)^{\mrT}$.
			
			The Laplace representation of the master equation~\eqref{eq:master_n_expon} is then given by
			\begin{equation}
				\frac{\partial \tl{P}_t(\bm{s})}{\partial t} = -\sum_{k=1}^K\frac{s_k}{\tau_k}\frac{\partial \tl{P}_t(\bm{s})}{\partial s_k} 
				+ \left(e^{-\bm{h}\cdot \bm{s}}-1\right) \left(\nu_0-\sum_{k=1}^K\frac{\partial }{\partial s_k}\right)\tl{P}_t(\bm{s}). \label{eq:master_n_expone_Laplace}
			\end{equation}
			Then, the Laplace representation~\eqref{def:Laplace_PDF_n_gen} of $P_t(\bm{z})$, which is the solution of 
			(\ref{eq:master_n_expone_Laplace}), allows us to obtain the Laplace representation
			$\tl{Q}_{t}(s)$ of the intensity PDF $P_t(\nu)$ according to
			\begin{equation}
				\tl{Q}_{t}(s) := \mathcal{L}_{1}[P_t(\nu); s] = \left< e^{-s(\nu_0+\sum_{k=1}^K\hz_k)}\right> 
				= e^{-\nu_0s} \tl{P}_t\left(\bm{s}=(s,s,\dots,)^{\mrT}\right)~.
			\end{equation}

	\subsection{General kernels}
	
		\subsubsection{Mapping to Markovian dynamics}
			The above formulation can be generalized to general forms of the memory kernel. Let us decompose 
			the kernel as a continuous superposition of exponential kernels,
			\begin{equation}
				h(t) = \frac{1}{n}\int_0^\infty \frac{n(x)}{x}e^{-t/x}dx~, \>\>\>
				n= \int_0^\infty n(x)dx~,
				\label{eq:continuous_decomposition_kernel}
			\end{equation}
			where we have introduced the set of continuous auxiliary variables $x \in \bm{R}_+$. 
			This decomposition satisfies the normalization condition $\int_0^\infty h(t)dt=1$.
			Here we use the notation $x$ for these auxiliary variables to emphasise the formal connection with the usual field theory of classical stochastic  (or quantum) systems. Indeed, the observables will be defined over the ``field''  $x \in \bm{R}_+$.
			The function $n(x)$ quantifies the contribution of the 
			$x$-th exponential with memory length $x$ to the branching ratio.
			We can then interpret $n(x)/n$ as a normalised distribution of time scales present in the memory kernel
			of the Hawkes process. As we show below, an important condition for solvability will be the existence of its first-order moment
			\begin{equation}
				\frac{\alpha}{n} :=  \langle \tau \rangle:= \int_0^\infty x \frac{n(x)}{n} dx < \infty~.   \label{rhr2bg2}
			\end{equation}
			This condition (\ref{rhr2bg2}) means that $n(x)$ should decay faster than $1/x^2$ at large $x$'s.
			Hence, the representation (\ref{eq:continuous_decomposition_kernel}) implies that the memory kernel
			has to decay at large times faster than $1/t^2$. This covers situations where the variance of the time
			scales embedded in the memory kernel diverges. But this excluded the cases $h(t) \sim 1/t^{1+\theta}$
			with $0<\theta<1$ that are relevant to the Omori law for earthquakes \cite{SorDeschatres04,CraneSor08} 
			and to the response to social shocks \cite{SaiSor2004}.
			This case $0<\theta<1$  for which $\alpha$ diverges needs to be treated separately and this is beyond
			the content of the present work.
			
			We then decompose the intensity of the Hawkes process as a continuous sum of excess intensities $\hz(t,x)$
			\begin{equation}
				\hnu(t) = \nu_0 + \int_0^\infty dx \hz(t,x)~.
			\end{equation}
			Each excess intensities $\hz_t(\tau)$ is the solution of the following dynamical equation
			\begin{equation}
				\frac{\partial \hz(t,x)}{\partial t} = - \frac{\hz(t,x)}{x} + \heta(t,x), \>\>\> \heta(t,x):= \frac{n(x)}{x}\hxi^{\mrP}_{\hnu},  \label{eq:SPDE_generalHawkes}
			\end{equation}
			where, as for the previous case of a discrete sum of exponentials, 
			the same state-dependent Poisson noise $\hxi^{\mrP}_{\hnu}(t)$ defined by expression (\ref{hwtrgfq}) acts on
			the Langevin equation for each excess intensity $\hz(t,x)$. 
			While Eq.~\eqref{eq:SPDE_generalHawkes} is a simple equation only related to the time derivative, this equation can be regarded as an SPDE since $\hz(t,x)$ is distributed over the auxiliary variable field $x\in (0,\infty)$.
			Furthermore, this interpretation allows us to use functional calculus, historically developed for the solution of SPDEs (such as the stochastic reaction-diffusion equations~\cite{GardinerB}). 
			Note that this SPDE corresponds to Eq.~\eqref{eq:GLE_MarkovEmbedding_continuous} for the case of the GLE.

			The set of SPDEs~\eqref{eq:SPDE_generalHawkes} expresses the fact that the 
			continuous field of excess intensity $\{\hz(t,x)\}_{x\in \bm{R}_+}$ tends to relax to zero,
			but they are intermittently simultaneously shocked by the shared shot noise term $\hxi^{\mrP}_{\hnu}$,
			with a $x$-dependent jump size $n(x)/x$. 
			This is in contrast to the SPDE representation~\eqref{eq:GLE_MarkovEmbedding_continuous} for the GLE~\eqref{eq:generalized_Langevin}, 
			where the noise term at $x$ has no correlation with that at another point $x'$: $\la \hxi^{\mrG}(t,x)\hxi^{\mrG}(t,x')\ra = 0$ for $x\neq x'$. 
			In a more conventional expression, we can rewrite the long-range nature of the spatial correlation in $\heta(t,x)$ as 
			\begin{equation}
				\la \heta(t,x)\heta(t,x')\ra_{\mrss} = \mathcal{K}(x,x')\delta (t-t'), \>\>\> \mathcal{K}(x,x') = \frac{n(x)n(x')}{xx'}\la  \hnu \ra_{\mrss}.
			\end{equation}
			If $K(x,x')$ was a $\delta$ function, the SPDE could be regarded as a non-interacting infinite variable systems. 
			This long-range nature means that all the excess intensity $\hz(t,x)$ at different points are strongly correlated through this noise term, 
			even though the SPDE~\eqref{eq:SPDE_generalHawkes} has no spatial derivatives.

		\subsubsection{Field master equation}
		
			The master equation corresponding to the SDE~\eqref{eq:SPDE_generalHawkes} can be derived by following the
			same procedure presented for the simple exponential case and for the discrete sum of exponentials.
			There is however a technical difference since the state of the system is now specified by the 
			continuous field variable $\{\hz(t,x)\}_{x \in \bm{R}_+}$. 
			Thus, the probability density function is replaced with the probability density functional 
			$P[\{\hz(t,x)=z(x)\}_{x \in \bm{R}_+}] = P_t[\{z(x)\}_{x \in \bm{R}_+}]$.
			In other words, the probability that the system state is in the state specified by $\{z(x)\}_{x \in \bm{R}_+}$ 
			at time $t$ is characterized by $P_t[\{z(x)\}_{x \in \bm{R}_+}]\mathcal{D}z$ with functional integral volume element $\mathcal{D}z$.
			
			We use the notational convention that any mapping with square bracket $A[\{f(x)\}_{x\in \bm{R}_+}]$ indicates
			that the map $A$ is a functional of $\{f(x)\}_{x \in \bm{R}_+}$.
			In addition, we sometimes abbreviate the functional $P_t[\{z(x)\}_{x \in \bm{R}_+}]$ by $P_t[z]:= P_t[\{z(x)\}_{x \in \bm{R}_+}]$
			for the sake of brevety.
			 
			The presence of a continuous field variable leads to several technical issues, such as in the correct application of the Laplace transform. 
			The functional Laplace transformation $\Lpath$ of an arbitrary functional $f[z]$ is defined by a functional integration (i.e., a path integral): 
			\begin{equation}
				\Lpath \big[f[z]; s\big] := \int \mathcal{D}z e^{-\int_0^\infty dx s(x)z(x)}f[z]  ~.   \label{etujeyhgq}
			\end{equation}
			This allows us to define the Laplace representation of the probability density functional by
			\begin{equation}
				\tl{P}_t[s] := \Lpath \big[P_t[z]; s\big]\label{def:prob_functional_Laplace}
			\end{equation}
			for an arbitrary nonnegative function $\{s(\tau)\}_{\tau\in \bm{R}_+}$. 
			
			As the natural extension of Eq.~\eqref{eq:master_n_expon}, the master equation for the probability density functional is given by
			\begin{equation}
				\frac{\partial P_t[z]}{\partial t} = \int_0^\infty dx\frac{\delta }{\delta z(x)}\frac{z(x)}{x}P_t[z] + 
				\Bigg[ \left\{\nu_0+\int_0^\infty (z- n/x)dx \right\}P_t[z-n/x] - \left\{\nu_0+\int_0^{\infty} z dx \right\}P_t[z] \Bigg]\label{eq:master_gen_functional}
			\end{equation}
			with the boundary condition 
			\begin{equation}
				P_t[z]\Big|_{z \in \partial \bm{R}^{\infty}_+} = 0
			\end{equation}
			where the boundary of the function space $\partial \bm{R}^{\infty}_+:= \{z | z(x) = 0 \mbox{ for some }x \in [0, \infty) \}$.
			This field master equation after Markov embedding corresponds to the FP field equation~\eqref{eq:GLE_MarkovEmbedding_continuous} for the GLE.
			
			\paragraph*{Interpretation.}
			As discussed in Sec.~\ref{sec:SPDE_GLE}, one of the safest interpretations is to regard this functional description 
			as the formal continuous limit from the discrete description in Sec.~\ref{sec:discrete_sum_masterEq}.
			By introducing a finite lattice interval $dx>0$ and rewriting $\tau_k \to x_k$, $\hz_k(t) \to \hz(t,x_k)dx$, and $n_k \to n(x_k)dx$, Eqs.~{\eqref{yjtyhnwrgbqbq} - \eqref{eq:SDE_general_superposition_discrete}} can be rewritten as
			\begin{equation}
				h(t) = \frac{1}{n}\sum_{k=1}^K \frac{n(x_k)}{x_k} e^{-t/x_k}dx, \>\>\>
				\hnu(t) = \nu_0 + \sum_{k=1}^K \hz(t,x_k)dx, \>\>\> 
				\frac{\partial z(t,x_k)}{\partial t} = - \frac{\hz(t,x_k)}{x_k} + \frac{n(x_k)}{x_k}\hxi^{\mrP}_{\hnu}(t).
			\end{equation}
			The master equation~\eqref{eq:master_n_expon} is rewritten as
			\begin{equation}
				\frac{\partial P_t(\bm{z})}{\partial t} = \sum_{k=1}^Kdx \left[ \frac{1}{dx}\frac{\partial }{\partial z(x_k)} \right] \frac{z(x_k)}{x_k}P_t(\bm{z}) 
				+ \left\{\nu_0+\sum_{k=1}^K (z(x_k)- n(x_k)/x_k)dx \right\}P_t(\bm{z}-\bm{h}) - \left\{\nu_0+\sum_{k=1}^K z(x_k)dx \right\}P_t(\bm{z}). 
			\end{equation}
			According to the convention in Ref.~\cite{GardinerB}, we take a formal limit $K\to \infty$ and $dx \to 0$ to apply the formal replacement
			\begin{equation}
				\int_0^\infty dx [...] := \lim_{dx \to 0}\sum_{k=1}^K dx[...], \>\>\> \frac{\delta }{\delta z(x)} := \lim_{dx\to 0}\frac{1}{dx}\frac{\partial }{\partial z(x_k)}.
			\end{equation}
			The master equation~\eqref{eq:master_gen_functional} for the field variables $\{\hz(t,x)\}_{x \in (0,\infty)}$ is thus derived. 
			While this derivation is based on a formal limit of a discrete description, Eq.~\eqref{eq:master_gen_functional} can be also derived by direct continuous operations based on the functional Taylor expansions 
			(See Appendix.~\ref{sec:master_eq_gen} for the detailed derivation based on functional calculus).
			
			We remark that the SPDE~\eqref{eq:SPDE_generalHawkes} is linear and serious divergence problems did not appear
			at least for our main results on physical observables. 
			In addition, our final results have been confirmed to be robust for both discrete and continuous cases as shown later. 
			This strategy is consistent with the safe prescription suggested in Ref.~\cite{GardinerB}:
			in the beginning, the functional descriptions for field variables should be based on a discrete formulation.
			The continuous description should be introduced afterward as a formal limit of zero-lattice intervals. 
			In this sense, our analysis has successfully avoided the delicate divergence problems on general SPDEs. 
			
		\subsubsection{Laplace representation of the master equation}
		
			In the functional Laplace representation~\eqref{def:prob_functional_Laplace}, 
			the master equation (\ref{eq:master_gen_functional}) takes the following simple first-order functional differential equation
			\begin{equation}
				\frac{\partial \tl{P}_t[s]}{\partial t} = \nu_0\left(e^{-\int_{0}^\infty dx' s(x')n(x')/x'} -1\right)\tl{P}_t[s] 
				-\int_0^\infty dx \left(e^{-\int_{0}^\infty dx' s(x')n(x')/x'} -1 + \frac{s(x)}{x}\right)\frac{\delta \tl{P}_t[s]}{\delta s(x)}~.\label{eq:master_gen_functional_Laplace}
			\end{equation}

		\subsubsection{General formulation }
		
			All the above forms of the memory kernel can be unified by remarking that 
			the variable transformation~\eqref{eq:continuous_decomposition_kernel} is equivalent to a Laplace transform, since it can be rewritten as
			\begin{align}
				h(t) = \frac{1}{n}\int_0^\infty \frac{1}{s}n\left(\frac{1}{s}\right)e^{-st}ds = \frac{1}{n}\mathcal{L}_1\left[\frac{1}{s}n\left(\frac{1}{s}\right); t\right]
				\>\>\>\> \Longleftrightarrow \>\>\>\>
				\frac{n(x)}{n} = \frac{1}{x}\mathcal{L}^{-1}_1\left[h(t); s\right]\bigg|_{s=1/x}.
			\end{align}
			This allows us to reformulate the several examples discussed above in a unified way presented in 
			Table~\ref{table_examples_general}. 
				
			\begin{table}
				\centering
				\begin{tabular}{|c|c|c|}
					\hline 
					Case  & $\displaystyle h(t)$ & $\displaystyle n(x)$ \\
					\hline \hline
					Single exponential kernel & $\displaystyle \frac{1}{\tau_1}e^{-t/\tau_1}$ & $\displaystyle n_1\delta (x-\tau_1)$\rule[-3mm]{0mm}{8mm} \\ \hline
					Discrete superposition of exponential kernel & $\displaystyle \frac{1}{n}\sum_{k=1}^K\frac{n_k}{\tau_k}e^{-t/\tau_k}$ & $\displaystyle \sum_{k=1}^Kn_k\delta (x-\tau_k)$\rule[-5mm]{0mm}{12mm}  \\ \hline
					Power-law kernel $(\beta \geq 0)$ & $\displaystyle \frac{1}{\tau^*} \frac{\beta}{(1+t/\tau^*)^{\beta+1}}$ & $\displaystyle \frac{n}{x}\left(\frac{\tau^*}{x}\right)^{\beta}\frac{e^{-\tau^*/x}}{\Gamma(\beta)}$\rule[-4mm]{0mm}{10mm} \\
					\hline
				\end{tabular}
				\caption{
							Examples of various memory kernel $h(t)$ and corresponding $n(x)$ 
							defined in expression (\ref{eq:continuous_decomposition_kernel}). 
						}
				\label{table_examples_general}
			\end{table}


\section{Solution}\label{sec:solutions}

	In section~\ref{sec:ModelMasterEq}, we have derived the master equations and their Laplace representations for the Hawkes processes 
	with arbitrary memory kernels.  
	Remarkably, the Laplace representations are first-order partial (functional) differential equations. 
	Because first-order partial (functional) differential equations can be formally solved by the method of characteristics (see Appendix~\ref{sec:app:method_of_characterisics} for a brief review), 
	various analytical properties of the Hawkes process can be studied in details. 
	
	In this section, we present novel properties of the Hawkes process unearthed from the solution of the master equations 
	by the method of characteristics. In particular, we focus on the behavior of the PDF of the steady-state intensity near the critical point $n=1$. 
	Under the condition of the existence of the first-order moment (\ref{rhr2bg2}), 
	an asymptotic analysis of the master equations shows that
	the PDF $P_{\mrss}(\nu):=\lim_{t\to \infty} P_t(\nu)$ exhibits a power-law behavior with a non-universal exponent: 
	\begin{equation}
		P_{\mrss}(\nu) \propto {1 \over \nu^{1-2\nu_0\alpha}}~,~~~~~{\rm with}~\alpha =  n  \langle \tau \rangle ~(\ref{rhr2bg2})
		 \label{eq:main_finding_power-law_gen}  
	\end{equation}
	for large $\nu$, up to an exponential truncation, which is pushed towards $\nu \to \infty$ as $n \to 1$. 
	As the tail exponent is smaller than $1$, the steady-state PDF $P_{\mrss}(\nu)$ is not renormalizable without the exponential cutoff. 
	However, the characteristic scale of the exponential tail diverges as the system approaches the critical point $n=1$, 
	and the power-law tail (\ref{eq:main_finding_power-law_gen}) can be observed over many orders 
	of magnitude of the intensity for near-critical systems, as we illustrate below.
	
	The parameter $\alpha = n  \langle \tau \rangle$ entering in the expression of tail exponent in 
	expression (\ref{eq:main_finding_power-law_gen}) has been defined by expression (\ref{rhr2bg2}).
	Since $\nu_0$ is the background intensity of the Hawkes intensity as defined in (\ref{def:Hawkes_general}),
	the exponent of $P_{\mrss}(\nu)$ depends on $\nu_0\alpha = n \nu_0 \langle \tau \rangle$, which is 
	$n$ times the average number of background events (or immigrants) occurring during a time equal 
	to the average time scale $\langle \tau \rangle$ of the memory kernel. Thus, the larger the memory $\langle \tau \rangle$,
	the larger the background intensity $\nu_0$ and the larger the branching ratio $n$, the smaller is the exponent $1-2\nu_0\alpha$.
	Note that $1-2\nu_0\alpha$ can even turn negative for $\nu_0\alpha > 1/2$, which corresponds to 
	a non-monotonous PDF $P_{\mrss}(\nu)$, which first grows according to the power law (\ref{eq:main_finding_power-law_gen})
	before decaying exponentially at very large $\nu$'s.
	
	In simple terms, the PDF (\ref{eq:main_finding_power-law_gen}) describes the distribution of the number $\nu dt$ of
	events in the limit of infinitely small time windows $[t, t+dt]$. We should contrast this limit to the other previously studied
	limit of infinitely large and finite but very large time windows. Standard results of branching processes (of which the 
	Hawkes model is a subset) give the total number of events generated by a given triggering event  (see Ref.~\cite{SaiHSor2005} for
	a detailed derivation). In equation (\ref{def:Hawkes_general}), this corresponds to counting all the events over 
	an infinitely large time window that are triggered by a single source event $\nu_0 =\delta(t)$ occurring at the origin of time.
	Ref.~\cite{SaiSor2006} has studied the distribution of ``seismic rates'' in the limit of large time windows which,
	in our current formulation, corresponds to the distribution of $N(t) := \int_t^{t+T} \nu(\tau) d\tau$, in the limit of large $T$'s.
	The corresponding probability density distributions are totally different from (\ref{eq:main_finding_power-law_gen}),
	which corresponds to the other limit $T \to 0$. 
	
	We derive our main result (\ref{eq:main_finding_power-law_gen}) first for the single exponential form of
	the memory kernel, then for the discrete sum of exponentials and then for the general case.

	\subsection{Single exponential kernel}
	
	As the first example, we focus on the single exponential kernel~\eqref{def:single_expon_kernel}. 
	While this special case is analytically tractable without the need to refer to the master equation approach~\cite{Dassios2011}, 
	we nevertheless derive its exact solution via the master equation approach, because the methodology 
	will be readily generalized to the more complex cases. 		
		
		\subsubsection{Steady state solution}
		
		Let us first study the steady solution of the PDF $P_{\mrss}(\nu)$. 
		By setting $K=1$ in Eq.~\eqref{eq:master_n_expone_Laplace}, we obtain the expression of the Laplace
		transform of the steady state $\tl{P}_{\mrss}(s):=\int_{0}^\infty d\nu e^{-s\nu}P_{\mrss}(z)$ of the master equation 
		(\ref{eq:master_exp}) in the form of a first-order ordinary differential equation
		\begin{equation}
			\left(e^{-ns/\tau}-1+\frac{s}{\tau}\right)\frac{d\tl{P}_{ss}(s)}{ds} = \nu_0 \left(e^{-ns/\tau}-1\right)\tl{P}_{ss}(s). \label{eq:maseter_single_expon_steady}
		\end{equation}
		
		By solving this equation, we obtain the exact steady solution below the critical point $n<1$,
		\begin{equation}
			\log \tl{Q}_{\mrss}(s) = -s\nu_0 + \log \tl{P}_{\mrss}(s) = -\frac{\nu_0}{\tau}\int_0^s \frac{sds}{e^{-ns/\tau}-1+s/\tau}\label{eq:exact_solution_single_expon_steady}
		\end{equation}
		with the renormalization condition 
		\begin{equation}
			\int_0^\infty P_{\mrss}(\nu) = \tl{Q}_{\mrss}(s=0) = 1.
		\end{equation}
			
			\paragraph{Near the critical point.}
				Let us evaluate the asymptotic behavior of $\tl{Q}_{\mrss}(s)$ for large $\nu$ 
				by assuming that the system is in the near-critical state, such that
				\begin{equation}
					\ve := 1-n \ll 1.
				\end{equation}
				By performing an expansion in the small parameter $\ve$, we obtain an asymptotic formula for small $s$ (large $\nu$), 
				\begin{equation}
					\log \tl{Q}_{\mrss}(s) \simeq -\frac{\nu_0}{\tau}\int_{0}^s \frac{ds}{\eps/\tau + s/2\tau^2} 
					= -2\nu_0 \tau\log\left(1+\frac{s}{2\tau \eps}\right),
				\end{equation}
				which implies a power-law behavior with a non-universal exponent, up to an exponential truncation: 
				\begin{equation}
					P_{\mrss}(\nu) \propto \nu^{-1+2\nu_0 \tau}~ e^{-2 \tau \eps \nu} ~.    \label{eq:power-law_single_expon_steady}
				\end{equation}
				for large $\nu$.
				This is a special case of expression~\eqref{eq:main_finding_power-law_gen} obtained for $n \to 1$ and
				$\langle \tau \rangle = \tau$.
				
				It is remarkable that the power-law exponent is less than one, and thus the PDF is not renormalizable without the exponential truncation. 
				This means that the power-law scaling actually corresponds to an intermediate asymptotics of the PDF, 
				according to the classification of Barenblatt~\cite{Barenblatt}.
				In addition, while this scaling can be regarded as a heavy ``tail" for $2\nu_0 \tau<1$, 
				the exponent can be negative when $2\nu_0 \tau>1$ (i.e., the PDF is a power-law increasing function until
				the exponential tail takes over and ensure the normalisation of the PDF). 
				
				The characteristic scale of the exponential truncation is  defined by
				\begin{equation}
					\nu_{\mathrm{cut}} := \frac{1}{2\tau \ve} =  \frac{1}{2\tau (1-n)}~,
				\end{equation}
				which diverges as the system approaches to the critical condition $n=1$. 
				This means that (i) if the system is in a near-critical state $\ve \ll 1$ and (ii) the background intensity is sufficiently 
				small $\nu_0<1/(2\tau)$, one can actually observe the power-law intermediate asymptotics for a wide 
				range $\nu \ll \nu_{\mathrm{cut}} = O(\ve^{-1})$, up to the exponential truncation.

			\paragraph{Numerical verification.}
				
				\begin{figure}
					\centering
					\includegraphics[width=180mm]{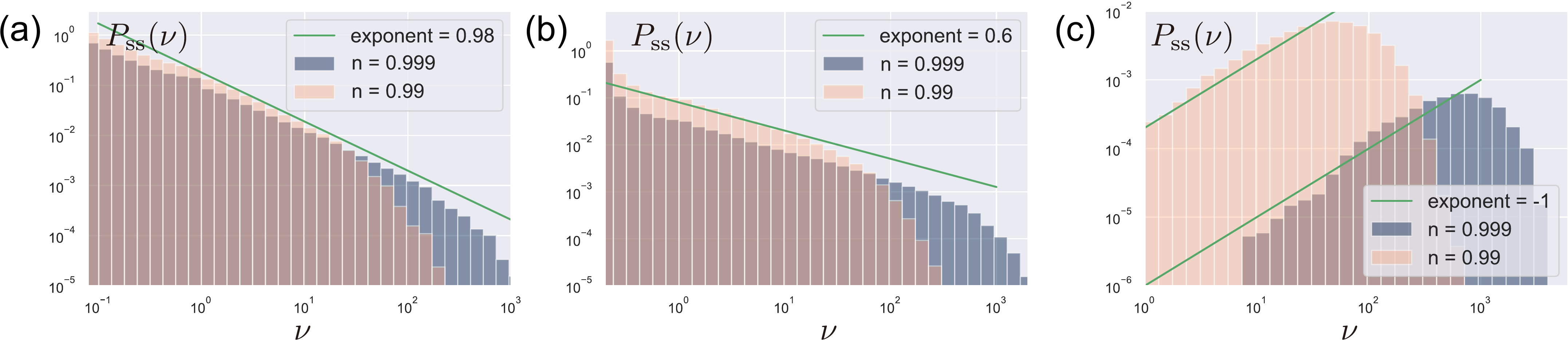}
					\caption{
						Numerical evaluation of the steady state PDF of the intensity $\hnu$ for the following parameter sets 
						near the critical point:  $n=0.999$ (blue bars) and $n=0.99$ (red bars). 
						The theoretical power law is shown by the green straight line.
								(a) Background intensity $\nu_0=0.01$, relaxation time $\tau=1$, leading to the power-law exponent $0.98$.
								(b) $\nu_0=0.2$, $\tau=1$, leading to the power-law exponent $0.6$. 
								(c) $\nu_0=1.0$, $\tau=1$, leading to the negative (i.e. growing) power-law exponent $-1.0$.
						For all simulations, the sampling time interval and total sampling time 
						are $dt=0.001$ and $T_{\mathrm{tot}}=10000$ with the initial condition $\hz(0)=0$.
						The initial 10\% of the sample trajectories were discarded from the statistics. 
							}
					\label{fig:Simulation_PDFs_SingleExpon}
				\end{figure}
				
				We now present numerical confirmations of our theoretical prediction, in particular for the
				intermediate asymptotics as shown in Fig.~\ref{fig:Simulation_PDFs_SingleExpon}. 
				There is a well-developed literature on the numerical simulation of Hawkes process \cite{Harte2010,DassiosZhao13}.
				We have used an established simulation package for Python called ``tick" (version 0.6.0.0) for 32-threads parallel computing.
				The total simulation time was $10^4$ seconds and the sampling time interval was $0.001$ second. 
				We note that the initial 10\% of the sampled trajectories were discorded for initialization. 
				
				For small background intensity $\nu_0\ll 1/\tau$, we obtain an approximate universal exponent $-1$. 
				For $2\nu_0 \tau <1$, we observe a decaying power law of exponent $1-2\nu_0\tau$, 
				while we observe a growing power law for $2\nu_0 \tau >1$. 
				The power-law intermediate asymptotics is truncated by the exponential function, as predicted and also 
				discussed in Ref.~\cite{BouchaudTradebook2018} (albeit with the error of missing the $1$ in the exponent
				and thus failing to describe the intermediate asymptotics), 
				ensuring the normalization of the PDF of the Hawkes intensities.

		\subsubsection{Time-dependent solution \label{htgb12fv1}}
			
			We now present the exact solution of the time-dependent master equation. 
			In the Laplace representation, the dynamics of the PDF of the intensities is given by the following first-order PDE, 
			\begin{equation}
				\frac{\partial \tl{P}_t(s)}{\partial t} + \left(e^{-ns/\tau}-1+\frac{s}{\tau}\right)\frac{\partial \tl{P}_t(s)}{\partial s} = \nu_0 \left(e^{-ns/\tau}-1\right)\tl{P}_t(s).  \label{eq:time_dependent_master_Laplace_one_expon}
			\end{equation}
			This equation can be solved by the method of characteristics (see Appendix~\ref{sec:app:method_of_characterisics} for a brief review).
			The corresponding Lagrange-Charpit equations are given by
			\begin{align}
				\frac{ds}{dt} = e^{-ns/\tau}-1+\frac{s}{\tau}, \>\>\>  ~~~~~~~~
				\frac{d\Phi}{dt} = \nu_o\left(e^{-ns/\tau}-1 \right),~~~{\rm with}~~ \Phi := \log \tl{P}~. 
			\end{align}
			These equations can be solved explicitly,
			\begin{align}
				t = \mcF(s) + C_1, \>\>\>\> ~~~~~~
				\Phi = \nu_0s - \frac{\nu_0}{\tau}\int_0^s \frac{s'ds'}{e^{-ns'/\tau}-1+s'/\tau} +C_2
			\end{align}
			with
			\begin{equation}
				\mcF(s) := \int_{s_0}^s\frac{ds'}{e^{-ns'/\tau}-1+s'/\tau} + C_1.
				\label{yi,m5i74jhnb2wg}
			\end{equation}
			$C_1$ and $C_2$ are constants of integration and $s_0$ is a positive constant 
			chosen to satisfy several convenient properties discussed below. 
			
			\paragraph{Summary of the properties of $\mcF$.}
			We present several analytical properties of $\mcF(s)$ (see Appendix~\ref{app:sec:mcF_characters_proof} for their proof):
			\begin{screen}
				\begin{enumerate}[label=($\alpha$\arabic*)]
					\item $\mcF(s)$ is a monotonically increasing function by choosing $s_0>0$ appropriately. 
					\item The inverse function $\mcF^{-1}(s)$ can be defined uniquely. 
				\end{enumerate}
			\end{screen}
			In addition, for the sub-critical case $n<1$, the following properties hold true:
			\begin{screen}
				\begin{enumerate}[label=($\alpha$\arabic*)]
					\setcounter{enumi}{2}
					\item $s_0$ can be set to any positive value. 
					\item $\lim_{s\to +0}\mc{F}(s) = -\infty$. 
					\item $\lim_{s\to \infty}\mc{F}(s) = +\infty$.
					\item $\mcF(s)$ can take all real values: $\mcF(s)\in (-\infty,\infty)$ for $s>0$. 
				\end{enumerate}
			\end{screen}
				
			\paragraph{Regularization of $\mcF(s)$.}
				In the following, we assume the sub-critical condition $n<1$. It is useful to decompose $\mcF(s)$ into regular and singular parts: 
				\begin{equation}
					\mathcal{F}(s) = \int_{s_0}^{s} \frac{ds}{e^{-ns/\tau}-1+s/\tau} \underbrace{- \int_{s_0}^s \frac{ds}{(1-n)s/\tau} + \frac{\tau}{1-n}\log\frac{s}{s_0}}_{\mbox{totally zero as an identity}}
					= \underbrace{\frac{\tau}{1-n}\int_{s_0}^{s} ds\frac{1-e^{-ns/\tau}-ns/\tau}{s(e^{-ns/\tau}-1+s/\tau)}}_{\mbox{regular part}}
					+ \underbrace{\frac{\tau}{1-n}\log \frac{s}{s_0}}_{\mbox{singular part}},
				\end{equation}
				where the regular part is well-defined even for $s_0 \to 0$. 
				This expression is useful since the divergent factor in $\mcF(s)$ can be renormalized into the integral constant $C_1$, such that
				\begin{equation}
					C_1 -\frac{\tau}{1-n}\log s_0 \to C_1.
				\end{equation} 
				We then take the formal limit $s_0\to 0$ and use the following regularized expression for the sub-critical condition $n<1$,  
				\begin{equation}
					\mathcal{F}(s) = \mathcal{F}_{\rm R}(s) + \frac{\tau}{1-n}\log s, \>\>\> \mathcal{F}_{\rm R}(s) := \frac{\tau}{1-n}\int_{0}^{s} ds\frac{1-e^{-ns/\tau}-ns/\tau}{s(e^{-ns/\tau}-1+s/\tau)},\label{eq:F_transform_under_critical}
				\end{equation}
				where the $s_0$-dependence is removed as the result of the renormalization. The regular part has no singularity at $s\simeq 0$, 
				and reads $\mcF_{\rm R}(s) \simeq -n^2s/\{2\tau(1-n)\}$.
			
			\paragraph{Explicit solution.}
				Building on the above, we now provide the solution of the master equation~\eqref{eq:time_dependent_master_Laplace_one_expon}. According to the method of characteristics (see Appendix.~\ref{sec:app:method_of_characterisics} for a brief review), the general solution is given by
				\begin{equation}
					C_2 = \mcH(C_1)
				\end{equation}
				with an arbitrary function $\mcH(\cdot)$. 
				The time-dependent solution $\log \tl{Q}_t(s) = \log \tl{P}_t(s) -\nu_0s$ is thus given by
				\begin{equation}
					\log \tl{Q}_t(s) = -\frac{\nu_0}{\tau}\int_{0}^s \frac{sds}{e^{-ns/\tau}-1+s/\tau} + \mathcal{H}\left(t - \mathcal{F}(s)\right). \label{eq:solution_time_dependent_single_expon}
				\end{equation}
				The function $\mcH(\cdot)$ is determined by the initial condition. 
				
				Let us assume that the initial PDF and its Laplace representation are given by $P_{t=0}(\nu)$ and $\tl{P}_{t=0}(s)$, respectively. 
				Then, we obtain
				\begin{equation}
					\mcH(-\mcF(s)) = \log \tl{P}_{t=0}(s) + \frac{\nu_0}{\tau}\int_{0}^s \frac{sds}{e^{-ns/\tau}-1+s/\tau}
				\end{equation}
				or equivalently,
				\begin{equation}
					\mathcal{H}(x) = \log \tl{P}_{t=0}\left(S(x)\right) + \frac{\nu_0}{\tau}\int_0^{S(x)} \frac{ sds}{e^{-ns/\tau}-1+s/\tau}, \>\>\>
					S(x) = \mathcal{F}^{-1}(-x).
				\end{equation}
				
				Note that the time-dependent solution~\eqref{eq:solution_time_dependent_single_expon} is consistent with the steady 
				solution~\eqref{eq:exact_solution_single_expon_steady}
				\begin{equation}
					\lim_{t\to \infty}\tl{Q}_t(s) = \tl{Q}_{\mrss}(s),
				\end{equation}
				since $\lim_{x\to +\infty}\mcH(x)=0$ (see Appendix.~\ref{app:sec:convergence_to_steady} for the proof). 
				We also note that, from the time-dependent solution ~\eqref{eq:solution_time_dependent_single_expon}, we can derive the 
				dynamics of the intensity $\hnu(t)$ for finite $t$ as 
				\begin{equation}
					\la \hnu(t)\ra = \nu_{\rm ini}e^{-(1-n)t/\tau} + \nu_0\frac{1-e^{-nt/\tau}}{1-n} \label{eq:avg_nu_single_expon_time_dependent}
				\end{equation}
				with the initial condition $\hnu(0)=\nu_{\rm ini}$ (see Appendix.~\ref{app:sec:average_nu_for_finite_time_single_expon} 
				for the derivation). This expression (\ref{eq:avg_nu_single_expon_time_dependent}) shows that the mean
				intensity converges at long times $t \to +\infty$ to $\la \hnu(t)\ra \to \nu_0 / (1-n)$, which is a well-known 
				result \cite{DalayVere03,HelmsSor02}.  Expression (\ref{eq:avg_nu_single_expon_time_dependent}) 
				also shows that an initial impulse decays exponentially with a renormalised time decay $\tau/(1-n)$, 
				which is also consistent with previous reports~\cite{Escobar2015}. This diverging time scale $\tau/(1-n)$, 
				as $n \to 1$, reflects the occurrence of all the generations of triggered events that renormalise the 
				``bare'' memory function into a ``dressed'' memory kernel with much longer memory.

			\paragraph{Asymptotic relaxation dynamics for large $t$.}
				The time-dependent asymptotic solution is given for large $t$ by
				\begin{equation}
					\log \tl{P}_t(s) \simeq -\frac{\nu_0}{\tau}\int_{S(t,s)}^s \frac{sds}{e^{-ns/\tau}-1+s/\tau} +\log \tl{P}_{t=0} \left(S(t,s)\right),\>\>\>
					S(t,s) = s\exp\left[-\frac{1-n}{\tau}\left(t-\mathcal{F}_{\rm R}(s)\right)\right],\label{eq:asymptotic_relaxation_single_expon}
				\end{equation}
				assuming that $t\gg \mcF(s)$ for a given $s$. 
				As a corollary of this formula, an asymptotic prediction for the distribution conditional on the initial intensity is given by the following formula
				\begin{equation}
					P(\nu, t| \nu_{\rm ini}, t=0) = \mathcal{L}^{-1}_{1}\left[\tl{P}_t(s); t\right], \>\>\> \log \tl{P}_{t=0}(s) = -\nu_{\rm ini}s
				\end{equation}
				for large $t$.
				Note that asymptotic convergence of these formula is not uniform in terms of $s$;
				indeed, the convergence of the Laplace representation for large $s$ is slower than that for small $s$. 
			
		\subsubsection{Another derivation of the power law exponent: linear stability analysis of the Lagrange-Charpit equation}\label{sec:single_expon_power-law_stability_analysis}
			We have presented both the steady state and time-dependent solutions of the master equations, based on exact or asymptotic methods. 
			While these formulations are already clear, here we revisit the power-law bahavior~\eqref{eq:power-law_single_expon_steady} 
			of the steady state PDF, 
			and present another derivation based on the linear stability analysis of the Lagrange-Charpit equation, which has the advantage
			of being generalisable to memory kernels defined as superposition of exponential functions.
			Indeed, while the derivation based on the exact solution~\eqref{eq:exact_solution_single_expon_steady} is clear and powerful, 
			it is not easy to extend this kind of calculation to general cases, such as superposition of exponential kernels. 
			In contrast, the derivation that we now present can be extended to arbitrary forms of the memory kernel of the Hawkes processes,
			as will be shown later. Moreover, we have found additional distinct derivations of \eqref{eq:power-law_single_expon_steady} 
			and we refer the interested reader to Appendix.~\ref{app:sec:various_derivation_power_law_single_expon}. 
		
			While the steady state master equation~\eqref{eq:maseter_single_expon_steady} is a ordinary differential equation which can be solved exactly, 
			let us consider its corresponding Lagrange-Charpit equations, 
			\begin{equation}
				\frac{ds}{dl} = -e^{-ns/\tau}+1-\frac{s}{\tau}, \>\>\>
				\frac{d}{dl}\log \tl{P}_{\mrss} = -\nu_0\left(e^{-ns/\tau}-1\right)
				\label{ryjty3hqtbq}
			\end{equation}
			where we introduce the parameter $l$ of the characteristic curve. 
			These equations can be regarded as describing a ``dynamical system" in terms of the auxiliary ``time" $l$. 
			This formulation is useful because the well-developed theory of dynamical systems is applicable even to more general cases as shown later. 
			
			\begin{figure}
				\centering
				\includegraphics[width=120mm]{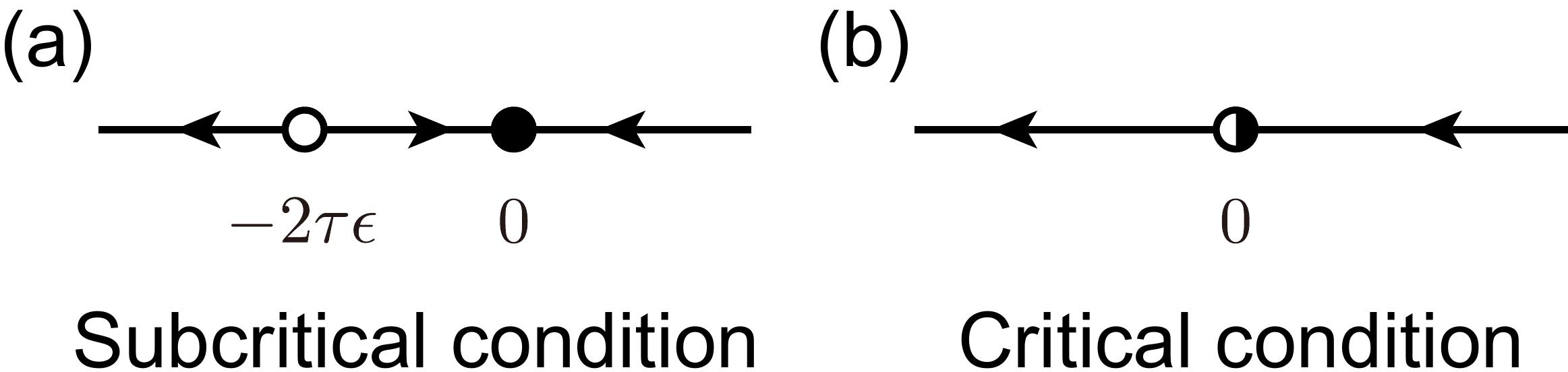}
				\caption{
							Schematic illustration of the vector field $V(s):=ds/dt=-e^{-ns/\tau}+1-s/\tau$ along the $s$ dimension as a function 
							of ``time'' $l$.
							(a)~Near critical condition $\eps:=1-n \ll 1$, two fixed points exist at $s=0$ (attractor) and $s\simeq -2\tau\eps$ (repeller). 
							(b)~At the critical condition $n=1$, the repeller merges with the attractor, which corresponds to a transcritical bifurcation. 
							}
				\label{fig:phaseSpace1D}
			\end{figure}
			
			\paragraph{Sub-critical condition $n<1$.}
			Let us first focus on the sub-critical case $n<1$ 
			and consider the expansion of equations (\ref{ryjty3hqtbq}) around $s=0$, which leads to 
			\begin{equation}
				\frac{ds}{dl} \simeq -\frac{1-n}{\tau}s - \frac{n^2s^2}{2\tau^2} + \dots, \>\>\> \frac{d}{dl}\log \tl{P}_{\mrss} \simeq \frac{n\nu_0}{\tau}s + \dots.
			\end{equation}
			
			The corresponding flow of this effective dynamical system $s(l)$ along the $s$ axis
			as a function of ``time'' $l$ is illustrated in Fig.~\ref{fig:phaseSpace1D}.
			Near the critical condition $\eps = 1-n\ll 1$, this ``dynamical system" has two fixed points $V(s)=0$ at 
			\begin{equation}
				s = 0, \>\>\> s\simeq -2\tau \eps
			\end{equation}
			The former is a stable attractor whereas the latter is an unstable repeller (see Fig.~\ref{fig:phaseSpace1D}a). 
			Remarkably, the critical condition $n=1$ for the Hawkes process corresponds to the condition of a transcritical bifurcation 
			(i.e., the repeller merges with the attractor; see Fig.~\ref{fig:phaseSpace1D}b) for the ``dynamical system" described by the Lagrange-Charpit equations.
			This picture is useful because it can be straightforwardly generalized to more general memory kernels $h(t)$, as shown later.
			
			Let us neglect the sub-leading contribution to obtain the general solution as
			\begin{equation}
				s = e^{-(1-n)(l-l_0)/\tau}, \>\>\> \log \tl{P}_{\mrss} \simeq \frac{n\nu_0}{\tau}\int dl~ s(l) + C
			\end{equation}
			with constants of integration $l_0$ and $C$.
			In the following, we set the initial ``time" (i.e., the initial point on the characteristic curve) as $l_0=0$. 
			We then obtain
			\begin{equation}
				\log \tl{P}_{\mrss} \simeq -\frac{n\nu_0 s}{1-n} + C 
			\end{equation}
			with constant of integration $C$. This constant is fixed by the condition of normalization of the PDF, 
			given by $\log \tl{P}_{\mrss} = 0$ for $s=0$, which imposes $C=0$. We thus obtain
			\begin{equation}
				\log \tl{Q}_{\mrss}(s) =  - \nu_0s + \log \tl{P}_{\mrss}(s) \simeq -\frac{\nu_0 s}{1-n},
				\label{yjnh23rb2}
			\end{equation}
			which is consistent with the asymptotic mean intensity in the steady state (see the long time limit of
			Eq.~\eqref{eq:avg_nu_single_expon_time_dependent}). 
		
			\paragraph{At criticality $n=1$.}
			For $n=1$, the lowest-order contribution in the Lagrange-Charpit equation is given by
			\begin{equation}
				\frac{ds}{dl} \simeq -\frac{s^2}{2\tau^2} \>\>\>
				\Longrightarrow \>\>\>
				s = \frac{2\tau^2}{l-l_0}.
			\end{equation}
			with constant of integration $l_0$. In the following, we set $l_0=0$ as the initial point on the characteristic curve. 
			We then obtain
			\begin{align}
				\log \tl{P}_{\mrss} = \frac{\nu_0}{\tau}\int dl ~s(l) + C \simeq -2\nu_0\tau \log |s| + C
			\end{align}
			with the constant of integration $C$. 
			The constant is an ``divergent" constant since it has to compensate the diverging logarithm to ensure that
			 $\log P_{\mrss}(s=0)=0$. This ``divergent" constant appears as a result of neglecting the ultra-violet (UV) 
			 cutoff for small $s$ (which corresponds to neglecting the exponential tail of the PDF of intensities). 
			By ignoring the divergent constant $C$, we obtain the intermediate asymptotics, 
			\begin{equation}
				P_{\mrss}(\nu) \propto \nu^{-1+2\nu_0\tau},	 
			\end{equation}
			which recovers the leading power law intermediate asymptotic 
			(\ref{eq:power-law_single_expon_steady}), which is a special case of the general solution (\ref{eq:main_finding_power-law_gen}).


	\subsection{Double exponential kernel}\label{sec:LinearStabilityOfLagrangeCharpitTwoExpon}
	
		We now consider the case where the memory function (\ref{yjtyhnwrgbqbq}) is made of $K=2$ exponential functions.
		Since the Laplace representation of the master equation is still a first-order partial differential equation, 
		its solution can be formally obtained by the method of characteristics (see Appendix.~\ref{sec:app:method_of_characterisics} for a short review). 
		Unfortunately, the time-dependent Lagrange Charpit equation cannot be exactly solved in explicit form anymore. 
		We therefore focus on the steady state solution of the master equation, with a special focus on the regime close 
		to the critical point. We develop the stability analysis of the Lagrange-Charpit equations 
		following the same approach as in Sec.~\ref{sec:single_expon_power-law_stability_analysis}. 

		Let us start from the Lagrange-Charpit equations, which are given by 
		\begin{subequations}
		\label{eq:LagrangeCharpit_2expon}
		\begin{align}
			\frac{ds_1}{dl} &= -e^{-(n_1s_1/\tau_1+n_2s_2/\tau_2)} + 1 - \frac{s_1}{\tau_1}, \\
			\frac{ds_2}{dl} &= -e^{-(n_1s_1/\tau_1+n_2s_2/\tau_2)} + 1 - \frac{s_2}{\tau_2}, \\
			\frac{d\Phi}{dl} &= -\nu_0 \left(e^{-(n_1s_1/\tau_1+n_2s_2/\tau_2)} - 1\right)~~~~~{\rm with}~~ \Phi := \log \tl{P}_{\mrss}
		\end{align}
		\end{subequations}
		and $l$ is the auxiliary ``time'' parameterising the position on the characteristic curve. 
		Let us develop the stability analysis around $s=0$ (i.e. for large $\nu$'s) for this pseudo dynamical system. 
		
		\paragraph{Sub-critical case $n<1$.}
			Assuming $n := n_1+n_2 < 1$, let us first consider the linearized dynamics of system (\ref{eq:LagrangeCharpit_2expon}) as
			\begin{equation}
				\frac{d\bm{s}}{dl} \simeq -\bm{H} \bm{s}, \>\>\> \frac{d\Phi}{dl} \simeq \nu_0\bm{K}\bm{s}
				\label{trhyr2hgbqb}
			\end{equation}
			with
			\begin{equation}
				\bm{s} :=	 	\begin{pmatrix}
									s_1\\
									s_2
								\end{pmatrix}, \>\>\>
				\bm{H} :=	 	\begin{pmatrix}
									\frac{1-n_1}{\tau_1}& \frac{-n_2}{\tau_2}\\
									\frac{-n_1}{\tau_1}& \frac{1-n_2}{\tau_2}
								\end{pmatrix}, \>\>\>
				\bm{K} :=	 	\begin{pmatrix}
									\frac{n_1}{\tau_1},  \frac{n_2}{\tau_2}
								\end{pmatrix}.
			\end{equation}

\begin{figure}
				\centering
				\includegraphics[width=100mm]{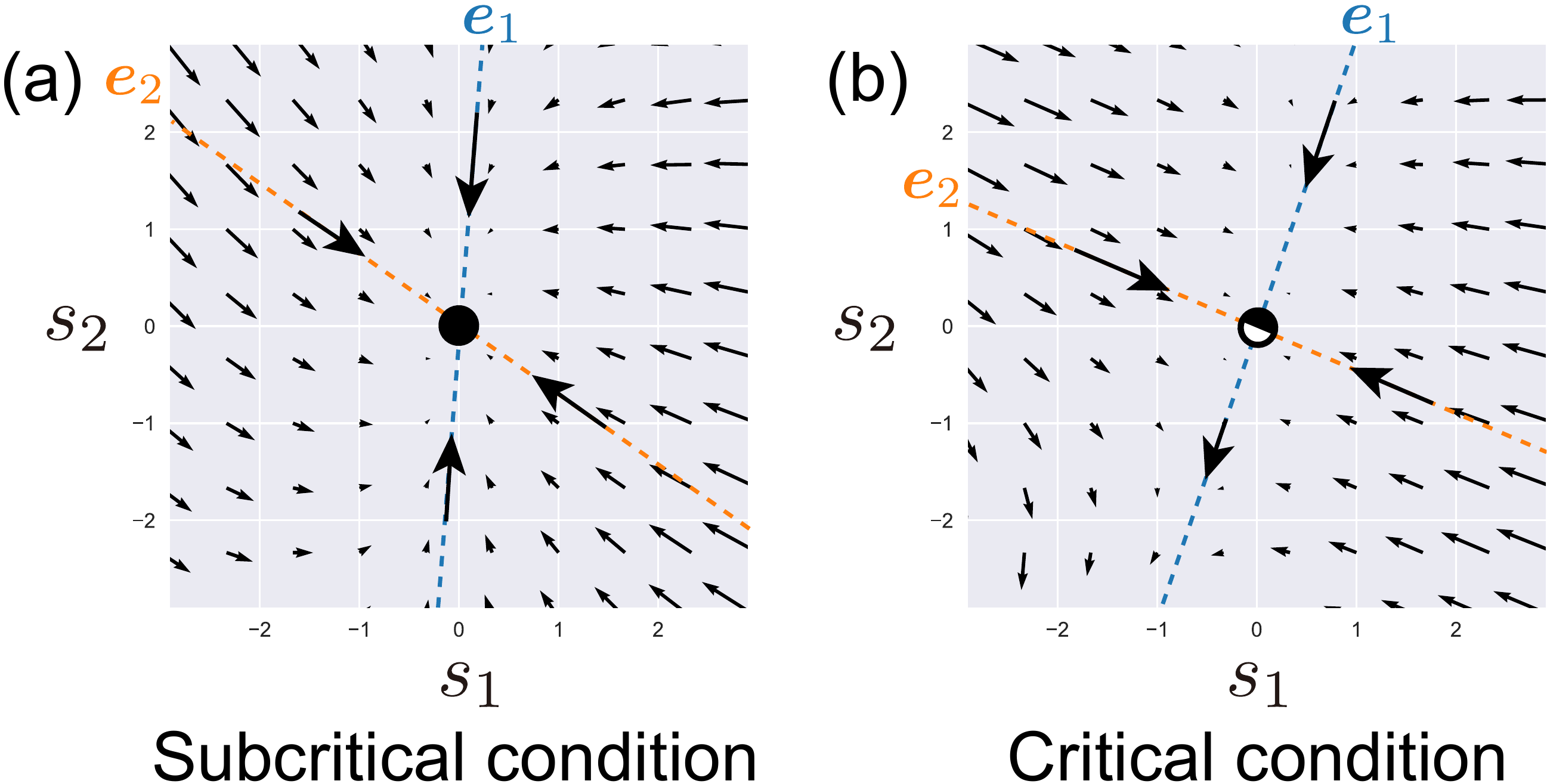}
				\caption{
							Qualitative representation of the Lagrange-Charpit equations in phase space. 
							By rewriting $d\bm{s}/dk := \bm{V}(\bm{s}) \simeq -\bm{H}\bm{s}$, the ``velocity" vector field $\bm{V}(\bm{s})$ 
							is plotted in the phase space $(s_1, s_2)$.
							(a) Subcritical case with $(\tau_1,\tau_2,n_1,n_2)=(1,3,0.3,0.1)$, showing that $\bm{s}=\bm{0}$ is a stable attractor. 
							(b) Critical case with $(\tau_1,\tau_2,n_1,n_2)=(1,3,0.3,0.7)$, showing that the $\bm{e}_1$ 
							direction is marginal in terms of the linear stability analysis
							(i.e., the repeller merges with the attractor, which corresponds to a transcritical bifurcation in dynamical systems). 
						}
				\label{fig:phaseSpace_twoExpon}
			\end{figure}

			Regarding this system as a dynamical system with the auxiliary ``time" $l$, its qualitative dynamics 
			can be illustrated by its phase space depicted in Fig.~\ref{fig:phaseSpace_twoExpon}.
			In the subcritical case $n<1$, the origin $\bm{s}=(0,0)$ is ``attractive" since all the eigenvalues of $\bm{H}$ 
			are positive (Fig.~\ref{fig:phaseSpace_twoExpon}a). 
			
			Let us introduce the eigenvalues $\lambda_1, \lambda_2$ and eigenvectors $\bm{e}_1, \bm{e}_2$ of $\bm{H}$,  such that
			\begin{equation}
				\bm{P}:=		 \begin{pmatrix}
									\bm{e}_1, \bm{e}_2
								\end{pmatrix}, \>\>\>
				\bm{P}^{-1}\bm{H}\bm{P} 
						= \begin{pmatrix}
							\lambda_1& 0 \\
							0& \lambda_2
						\end{pmatrix}.
			\end{equation}
			
			Because all eigenvalues are real (see Appendix.~\ref{app:sec:proof_eigenvalues_H_real} for the proof), 
			we denote $\lambda_1\leq \lambda_2$.  The determinant of $\bm{H}$ is given by
			\begin{equation}
				\det \bm{H} = \frac{1-n}{\tau_1\tau_2}.
			\end{equation}
			This means that the zero eigenvalue $\lambda_1=0$ appears at the critical point $n=1$. 
			Below the critical point $n<1$, all the eigenvalues are positive ($\lambda_1, \lambda_2>0$). 
			For $n<1$, the dynamics can be rewritten as
			\begin{equation}
				\frac{d}{dl}\bm{P}^{-1}\bm{s} =	-\begin{pmatrix}
							\lambda_1& 0 \\
							0& \lambda_2
						\end{pmatrix}
						\bm{P}^{-1}\bm{s} \>\>\>
				\Longrightarrow \>\>\>
				\bm{s}(l) = \bm{P}
						\begin{pmatrix}
							e^{-\lambda_1 (l-l_0)} \\
							e^{-\lambda_2 (l-l_0)}/C_1
						\end{pmatrix}
			\end{equation}
			with constants of integration $l_0$ and $C_1$. We can assume $l_0=0$ as the initial point of the characteristic 
			curve without loss of generality. Integrating the second equation in (\ref{trhyr2hgbqb}), we obtain    
			\begin{equation}
				\Phi = \nu_0\bm{K}\int \bm{s}(l)dl + C_2
					 = -\nu_0\bm{K}\bm{P}	\begin{pmatrix}
												1/\lambda_1& 0 \\
												0& 1/\lambda_2
											\end{pmatrix}
						\bm{P}^{-1}\bm{s} + C_2 = -\nu\bm{K}\bm{H}^{-1}\bm{s} + C_2. 
			\end{equation}
			The general solution is given by
			\begin{equation}
				\mathcal{H}(C_1) = C_2
			\end{equation}
			with a function $\mathcal{H}$ determined by the initial condition on the characteristic curve. 
			Let us introduce 
			\begin{equation}
				\bar{s} := \bm{P}^{-1}\bm{s} =
						\begin{pmatrix}
							\bar{s}_1\\
							\bar{s}_2
						\end{pmatrix} \>\>\>
				\Longrightarrow \>\>\>
				C_1 = \left(\bar{s}_1\right)^{\lambda_2/\lambda_1}\left(\bar{s}_2\right)^{-1} . 
			\end{equation}
			This means that the solution is given by the following form: 
			\begin{equation}
				\Phi(\bm{s}) = -\nu\bm{K}\bm{H}^{-1}\bm{s} + \mathcal{H}\left(\left(\bar{s}_1\right)^{\lambda_2/\lambda_1}\left(\bar{s}_2\right)^{-1}\right). 
			\end{equation}
			Because of the renormalization of the PDF, the following relation must hold 
			\begin{equation}
				\lim_{\bm{s}\to \bm{0}} \Phi(\bm{s}) = 0
			\end{equation}
			for any path in the $(s_1, s_2)$ space ending on the origin (limit $\bm{s}\to \bm{0}$). 
			Let us consider the specific limit such that $\bar{s}_1\to 0$ with $\bar{s}_2=x^{-1}(\bar{s}_1)^{\lambda_2/\lambda_1}$ 
			for an arbitrary positive $x$:
			\begin{equation}
				\lim_{\bar{s}_1\to 0} \Phi(\bm{s}) = \mathcal{H}(x).
			\end{equation}
			Since the left-hand side (LHS) is zero for any $x$, the function $\mathcal{H}(\cdot)$ must be identically zero. 
			With $\Phi := \log \tl{P}_{\mrss}$ as defined in (\ref{eq:LagrangeCharpit_2expon}), this leads to
			\begin{equation}
				\log \tl{P}_{\mrss} (\bm{s}) = -\nu\bm{K}\bm{H}^{-1}\bm{s}.
			\end{equation}
			By substituting with the special $\bm{s}=(s_1=s, s_2=s)^{\rm{T}}$, we obtain
			\begin{equation}
				\log \tl{Q}_{\mrss}(s) = -\nu_0 s + \Phi_{t=\infty} \left( s(1,1)^{\rm{T}} \right) \simeq -\frac{\nu_0}{1-n}s
				\label{temjkymku5jmn3}
			\end{equation}
			for small $s$, which recovers expression (\ref{yjnh23rb2}) derived above.

		\paragraph{Critical case $n=1$.}
			In this case, the eigenvalues and eigenvectors of $\bm{H}$ are given by
			\begin{equation}
				\lambda_1 = 0, \>\>\>
				\lambda_2 = \frac{n_1\tau_1+n_2\tau_2}{\tau_1\tau_2}, \>\>\>
				\bm{e}_1 = 	\begin{pmatrix}
							 \tau_1 \\
							 \tau_2
							\end{pmatrix}, \>\>\>
				\bm{e}_2 = 	\begin{pmatrix}
							 -n_2 \\
							 n_1
							\end{pmatrix}.
			\end{equation}
			This means that the eigenvalue matrix and its inverse matrix are respectively given by
			\begin{equation}
				\bm{P} = \begin{pmatrix}
							 \tau_1 & -n_2 \\
							 \tau_2 & n_1
							\end{pmatrix}, \>\>\>
				\bm{P}^{-1} =
							\frac{1}{\alpha}
							\begin{pmatrix}
							 n_1 & n_2 \\
							 -\tau_2 & \tau_1
							\end{pmatrix}, \>\>\>
				\alpha := \det \bm{P} = \tau_1n_1 + \tau_2n_2
				\label{teumk5im4jnw}
			\end{equation} 
			This value of $\alpha$ is the special case for two exponentials of the general definition (\ref{rhr2bg2}).
			Accordingly, let us introduce 
			\begin{equation}
				\bm{X} = (X,Y)^{\rm{T}} = \bm{P}^{-1}\bm{s}, \>\>\> \Longleftrightarrow \>\>\>
				X = \frac{n_1 s_1 + n_2 s_2}{\alpha}, \>\>\>
				Y = \frac{-\tau_2 s_1 + \tau_1 s_2}{\alpha}.
			\end{equation}
			We then obtain
			\begin{equation}
				\frac{dX}{dl} = 0, \>\>\> \frac{dY}{dl} = -\lambda_2 Y
			\end{equation}
			at the leading linear order in expansions in powers of $X$ and $Y$. Since the first linear term is zero 
			in the dynamics of $X$, corresponding to a transcritical bifurcation for the Lagrange-Charpit 
			equations~\eqref{eq:LagrangeCharpit_2expon}, we need to take into account the second order term in $X$, namely
			\begin{align}
				e^{-(n_1s_1/\tau_1+n_2s_2/\tau_2)} \simeq 1 - X + \frac{X^2}{2} + n_1n_2 \left(\frac{1}{\tau_1}-\frac{1}{\tau_2}\right)Y + O(XY, X^2Y, Y^2) 
			\end{align}
			where we have dropped terms of the order $Y^2$, $XY$ and $X^2Y$.
			We then obtain the dynamical equations at the transcritical bifurcation (see Fig.~\ref{fig:phaseSpace_twoExpon}b) to leading-order
			\begin{equation}
				\frac{dY}{dl} \simeq -\lambda_2 Y, \>\>\> \frac{dX}{dl} \simeq -\frac{X^2}{2\alpha} 
			\end{equation}
			whose solutions are given by
			\begin{equation}
				X(l) = \frac{2\alpha}{l-l_0}, \>\>\>
				Y(l) = C_1e^{-\lambda_2 (l-l_0)}
				\label{eq:asymptotic_speed_twoExpon}
			\end{equation}
			with constants of integration $l_0$ and $C_1$. 
			We can assume $l_0=0$ as the initial point on the characteristic curve. 
			Remarkably, only the contribution along the $X$ axis is dominant for the large $l$ limit (i.e., $|X|\gg |Y|$ for $l \to \infty$), which corresponds to the asymptotic limit $\bm{s}\to 0$. 
			We then obtain
			\begin{align}
				\Phi \simeq \nu_0 \int dl \left(\frac{n_1s_1(l)}{\tau_1} + \frac{n_2s_2(l)}{\tau_2}\right) \simeq -2\nu_0\alpha \log |X| + \frac{\nu_0n_1n_2}{\lambda_2}\left(\frac{1}{\tau_1}-\frac{1}{\tau_2}\right)Y + C_2
			\end{align}
			with constant of integration $C_2$. 
			The general solution is given by 
			\begin{equation}
				\mathcal{H}(C_1) = C_2 
			\end{equation}
			with a function $\mathcal{H}$, which is determined by the initial condition. 
			Considering that
			\begin{equation}
				C_1 = Y\exp\left[\frac{2\lambda_2 \alpha}{X}\right], 
			\end{equation}
			the solution is given by the following form: 
			\begin{equation}
				\Phi(\bm{s}) = -2\nu_0\alpha \log |X| + \frac{\nu_0n_1n_2}{\lambda_2}\left(\frac{1}{\tau_1}-\frac{1}{\tau_2}\right)Y + \mathcal{H}\left(Y\exp\left[\frac{2\lambda_2 \alpha}{X}\right]\right).
			\end{equation}
			Because we have neglected the UV cutoff for small $s$, there is an artificial divergent term $-2\nu_0\alpha \log |X|$ for small $X$. 
			Except for this divergent term, $\Phi(\bm{s})$ must be constant for $s\to 0$.
			The function $\mathcal{H}(\cdot )$ is thus constant because
			\begin{equation}
				\lim_{y\to 0}\left[\Phi(\bm{s}) + 2\nu_0\alpha \log |X|\right] = \mathcal{H}(Z) = \mbox{const.}
			\end{equation}
			with the choice of $X = 2\lambda_2 \alpha /\log(Z/Y)$ for any positive constant $Z$. 
			Therefore, we obtain the steady solution  
			\begin{equation}
				\log \tl{P}_{\mrss}(\bm{s}) \simeq -2\nu_0\alpha \log |X| + \frac{\nu_0n_1n_2}{\lambda_2}\left(\frac{1}{\tau_1}-\frac{1}{\tau_2}\right)Y
			\end{equation}
			for small $X$ and $Y$, by ignoring the UV cutoff and the constant contribution. 
			This recovers the power law formula of the intermediate asymptotics of the PDF of the Hawkes intensities:
			\begin{equation}
				\log \tl{Q}_{\mrss}(s) := -\nu_0s + \log \tl{P}_{\mrss}(s,s) \simeq - 2\nu_0 \alpha \log|s| \>\>\>(s\sim 0) \>\>\>
				\Longleftrightarrow \>\>\>
				P(\nu) \sim \nu^{-1+2\nu_0\alpha}\>\>\>(\nu\to +\infty),
				\label{yji,kiukm4j3w}
			\end{equation}
			with $\alpha = \tau_1n_1 + \tau_2n_2$ as defined in (\ref{teumk5im4jnw}).

			\paragraph{Numerical verification.}
				\begin{figure}
					\centering
					\includegraphics[width=180mm]{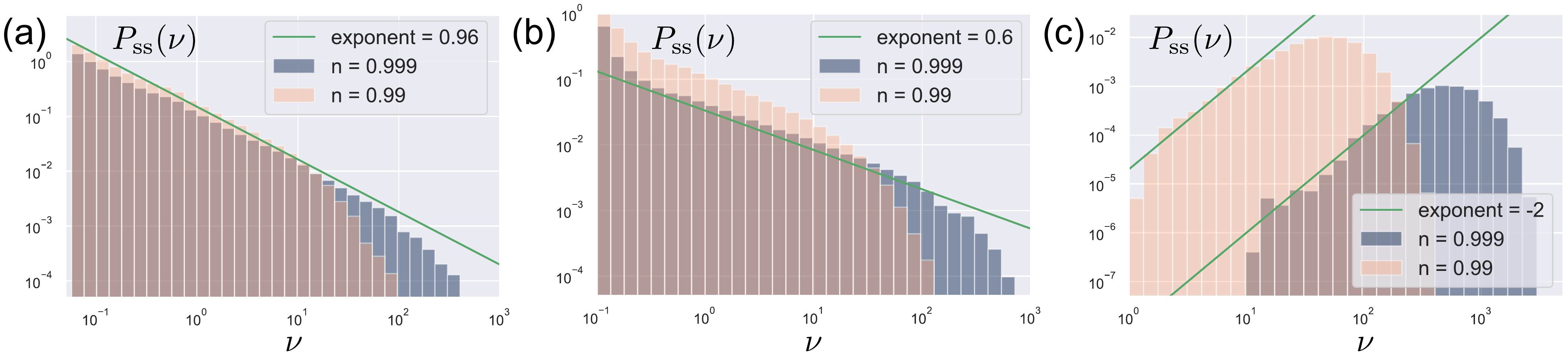}
					\caption{ 
								Numerical evaluation of the steady state PDF of the Hawkes intensity $\hnu$ 
								for the double exponential case $K=2$ (\ref{yjtyhnwrgbqbq}) with $(\tau_1,\tau_2)=(1,3)$,
								$(n_1,n_2)=(0.5,0.499)$ or $(n_1,n_2)=(0.5,0.49)$, near the critical point:
								(a) Background intensity $\nu_0=0.01$, leading to the power law exponent $0.96$.
								(b) $\nu_0=0.1$, leading to the power law exponent $0.6$. 
								(c) $\nu_0=0.75$, leading to the negative (i.e. growing PDF) power law exponent $-2.0$.
								For all simulations, the sampling time interval and total sampling time are $dt=0.001$ and $T_{\mathrm{tot}}=10000$ from the initial condition $\hz(0)=0$.
								The initial 10\% of the sample was discarded from the statistics. 
							}	
					\label{fig:Simulation_PDFs_DoubleExpon}
				\end{figure}
				
				We have numerically confirmed our theoretical prediction \eqref{yji,kiukm4j3w}, a special case of
				\eqref{eq:main_finding_power-law_gen} for a memory kernel with two exponentials, as shown in
				 Fig.~\ref{fig:Simulation_PDFs_DoubleExpon}. 
				The main properties are the same as those shown in Fig.~\ref{fig:Simulation_PDFs_SingleExpon}, 
				implying that our prediction is verified for memory kernels with one and two exponentials.


	\subsection{Discrete superposition of exponential kernels}
	
		We now consider the case where the memory kernel is the sum of an arbitrary finite number $K$ 
		of exponentials according to expression (\ref{yjtyhnwrgbqbq}). Our treatment follows the method 
		presented for the case $K=2$. 
		
		The corresponding Lagrange-Charpit equations read:
		\begin{equation}
			\frac{ds_k}{dl} = -e^{-\sum_{j=1}^Kn_js_j/\tau_j} + 1 - \frac{s_k}{\tau_k}, \>\>\> ~~~~~~
			\frac{d\Phi}{dl} = -\nu_0 \left(e^{-\sum_{j=1}^Kn_js_j/\tau_j} - 1\right).
			\label{eq:Lagrange_Charpit_eq_n_expon}
		\end{equation}
		The derivation of the PDF of the Hawkes intensities boils down to a 
		stability analysis of these equations around $s=0$ in the neighbourhood of the critical condition $n=1$. 

		\paragraph{Sub-critical case $n<1$.}
			We linearize the Lagrange-Charpit equations to obtain
			\begin{equation}
				\frac{d\bm{s}}{dl} \simeq -\bm{H}\bm{s}, \>\>\>
				\frac{d\Phi}{dl} \simeq \nu_0 \bm{K}\bm{s}
			\end{equation}
			with
			\begin{equation}
				\bm{H} :=	 	\begin{pmatrix}
									\frac{1-n_1}{\tau_1},& -\frac{n_2}{\tau_2},& \dots& -\frac{n_K}{\tau_K} \\
									-\frac{n_1}{\tau_1},& \frac{1-n_2}{\tau_2},& \dots& -\frac{n_K}{\tau_K} \\
									\vdots& \vdots& \ddots& \vdots \\
									-\frac{n_1}{\tau_1},& -\frac{n_2}{\tau_2},& \dots& \frac{1-n_K}{\tau_K}
								\end{pmatrix}, \>\>\>
				\bm{K} := \left(\frac{n_1}{\tau_1}, \dots, \frac{n_K}{\tau_K}\right). 
				\label{wrnhmnnh3}
			\end{equation}
			Considering that all eigenvalues  $\{\lambda_k\}_{k=1,\dots,K}$ of $\bm{H}$ are real 
			(see Appendix.~\ref{app:sec:proof_eigenvalues_H_real} for its proof), 
			we order them according to $\lambda_i<\lambda_j$ for $i<j$. We denote the corresponding eigenvectors 
			as $\{\bm{e}_k\}_{k=1,\dots,K}$. The matrix $\bm{H}$ can thus be diagonalised as follows
			\begin{equation}
				\bm{P}:= (\bm{e}_1,\dots, \bm{e}_K), \>\>\> 
				\bm{P}^{-1}\bm{H}\bm{P} = 	\begin{pmatrix}
												\lambda_1,& 0,& \dots& 0 \\
												0,& \lambda_2,& \dots& 0 \\
												\vdots& \vdots& \ddots& \vdots \\
												0,& 0,& \dots& \lambda_K 
											\end{pmatrix}.
			\end{equation}
			The critical case $n=1$ corresponds to the existence of a zero eigenvalue. 
			Therefore, at the critical point, the determinant of $\bm{H}$ is zero (see Appendix.~\ref{app:sec:proof_determinant_H} 
			for the derivation of the explicit form of its determinant):
			\begin{equation}
				\det \bm{H} = \frac{1-\sum_{k=1}^K n_k}{\prod_{k=1}^K\tau_k}= 0 \>\>\>\> \Longleftrightarrow \>\>\>\>
				n: =\sum_{k=1}^K n_k = 1.
			\end{equation}
			Following calculations similar those presented in to Sec.~\ref{sec:LinearStabilityOfLagrangeCharpitTwoExpon}, we obtain
			\begin{equation}
				\Phi(\bm{s}) \simeq -\nu_0 \bm{K}\bm{H}^{-1}\bm{s}
			\end{equation}
			where the inverse matrix $\bm{H}^{-1}$ is explicitly given in Appendix.~\ref{app:sec:inverse_matrix_H}. 
			We finally obtain
			\begin{equation}
				\log \tl{Q}_{\mrss}(s) = -\nu_0 s + \Phi(s(1,\dots,1)^T) = \frac{-\nu_0}{1-n}s,
				\label{uj4m4n32h}
			\end{equation}
			again recovering (\ref{temjkymku5jmn3}) and (\ref{yjnh23rb2}) derived above.

		\paragraph{Critical case $n=1$.} 
			At the critical point, the smallest eigenvalue of $\bm{H}$ is zero ($\lambda_1=0$). 
			By direct substitution, its corresponding eigenvector is
			\begin{equation}
				\bm{e}_1 = (\tau_1, \dots, \tau_K)^T~,
			\end{equation}
			as seen from
			\begin{equation}
				\bm{H}\bm{e}_1 = \begin{pmatrix}
									\frac{1-n_1}{\tau_1},& -\frac{n_2}{\tau_2},& \dots& -\frac{n_K}{\tau_K} \\
									-\frac{n_1}{\tau_1},& \frac{1-n_2}{\tau_2},& \dots& -\frac{n_K}{\tau_K} \\
									\vdots& \vdots& \ddots& \vdots \\
									-\frac{n_1}{\tau_1},& -\frac{n_2}{\tau_2},& \dots& \frac{1-n_K}{\tau_K}
								\end{pmatrix}
								\begin{pmatrix}
									\tau_1 \\
									\tau_2 \\
									\vdots \\
									\tau_K
								\end{pmatrix}
								=\begin{pmatrix}
									1-n \\
									1-n \\
									\vdots \\
									1-n
								\end{pmatrix}
								= \bm{0}    ~~~~~{\rm for}~n=1.
			\end{equation}
			We now introduce a new set of variables (i.e., representation based on the eigenvectors)
			\begin{equation}
				\bm{X} = \begin{pmatrix}
					X_1 \\
					X_2 \\
					\dots \\
					X_K
				\end{pmatrix} := \bm{P}^{-1}\bm{s}, \>\>\>\>~~~~~
				\bm{P}^{-1} =
				\begin{pmatrix}
					\bm{g}_1^T \\
					\bm{g}_2^T \\
					\dots \\
					\bm{g}_K^T
				\end{pmatrix}\bm{s}.
			\end{equation}
			The linearized Lagrange-Charpit equations are given by
			\begin{equation}
				\frac{dX_1}{dl} \simeq 0, \>\>\> ~~~~\frac{dX_j}{dl} \simeq -\lambda_j X_j~~~~{\rm for}~j\geq 2~.
			\end{equation} 
			Similarly to Eq.~\eqref{eq:asymptotic_speed_twoExpon}, 
			the leading-order contribution comes from the $X_1$ direction because $|X_1| \gg |X_j|$ for $j\geq 2$ in the asymptotic limit $l\to \infty$. 
			We thus neglect other contribution by assuming $X_j\sim 0$ for $j\geq 1$. 
			It is therefore necessary to take the second-order contribution along the $X_1$ direction,
			\begin{equation}
				e^{-\sum_{k=1}^{K}n_ks_k/\tau_k}-1 = -\sum_{k=1}^{K}\frac{n_ks_k}{\tau_k} 
				+ \frac{1}{2}\left(\sum_{k=1}^{K}\frac{n_ks_k}{\tau_k}\right)^2 + \dots.
			\end{equation}
			We note that $X_1$ is given by
			\begin{equation}
				X_1 = \bm{g}_1\cdot \bm{s} = \frac{1}{\alpha}\sum_{k=1}^K n_ks_k, 
				\>\>\>~~~~~ \bm{g}_1 = \left(\frac{n_1}{\alpha},\dots, \frac{n_K}{\alpha}\right)^T,
				\label{yjt4nh3g}
			\end{equation}
			where $\alpha := \sum_{k=1}^K\tau_kn_k$, which is a special case for a discrete sum of exponentials of the general definition (\ref{rhr2bg2}).
			
			Taking the derivative of (\ref{yjt4nh3g}) and using equation~\eqref{eq:Lagrange_Charpit_eq_n_expon}, we obtain
			\begin{align}
				\frac{dX_1}{dl} = \frac{1}{\alpha}\sum_{k=1}^Kn_k\frac{ds_k}{dl} = 0 - 
				\frac{1}{2\alpha}\left(\sum_{k=1}^{K}\frac{n_ks_k}{\tau_k}\right)^2 + \dots.
			\end{align}
			This means that $\bm{g}_1$ is a correct representation.  Note that the value of $\alpha$ given by (\ref{rhr2bg2})
			ensures consistency with the following identify: 
			\begin{equation}
				\bm{P}^{-1}\bm{P} = 
				\begin{pmatrix}
					\bm{g}_1^T \\
					\bm{g}_2^T \\
					\dots \\
					\bm{g}_K^T
				\end{pmatrix}
				\begin{pmatrix}
					\bm{e}_1, \bm{e}_2, \dots, \bm{e}_K
				\end{pmatrix}
				=
				\begin{pmatrix}
					n_1/\alpha,& n_2/\alpha,& \dots& n_K/\alpha \\
					\bigcirc,& \bigcirc,& \dots& \bigcirc \\
					\vdots& \vdots& \ddots& \vdots \\
					\bigcirc,& \bigcirc,& \dots& \bigcirc
				\end{pmatrix}
				\begin{pmatrix}
					\tau_1,& \bigcirc,& \dots& \bigcirc \\
					\tau_2,& \bigcirc,& \dots& \bigcirc \\
					\vdots & \vdots& \ddots& \vdots \\
					\tau_2,& \bigcirc,& \dots& \bigcirc
				\end{pmatrix}
				=
				\begin{pmatrix}
					1,& 0,& \dots& 0 \\
					0,& 1,& \dots& 0 \\
					\vdots& \vdots& \ddots& \vdots \\
					0,& 0,& \dots& 1
				\end{pmatrix},
			\end{equation}
			where $\bigcirc$ represents some unspecified value. 
			Since the contribution of $X_2, \dots X_K$ can be ignored for the description of the leading behavior 
			along $X_1$, let us set $X_2=X_3=\dots=X_K=0$, which leads
			\begin{equation}
				\bm{s} = \bm{P}\bm{X} \simeq (\bm{e}_1,\dots,\bm{e}_K)
				\begin{pmatrix}
					X_1\\
					0\\
					\vdots\\
					0
				\end{pmatrix}
				=X\bm{e}_1
				=\begin{pmatrix}
					X_1\tau_1\\
					X_1\tau_2\\
					\vdots\\
					X_1\tau_K
				\end{pmatrix}.
			\end{equation}
			
			We thus obtain the second-order contribution along the $X_1$ axis by ignoring nonlinear contribution from $X_2,\dots, X_K$: 
			\begin{equation}
				\frac{dX_1}{dl} \simeq -\frac{X_1^2}{2\alpha}.
			\end{equation}
			With calculations that follow step by step those in Sec.~\ref{sec:LinearStabilityOfLagrangeCharpitTwoExpon}, we obtain
			\begin{equation}
				\log \tl{Q}_{\mrss}(s) \simeq -2\nu_0 \alpha \log |s| \>\>\> \Longleftrightarrow \>\>\> P(\nu) \sim \nu^{-1+2\nu_0 \alpha}, 
				\>\>\>  {\rm with}~\alpha := \sum_{k=1}^Kn_k\tau_k. 
			\end{equation}
			This recovers the power law formula of the intermediate asymptotics of the PDF of the Hawkes intensities
			given by (\ref{eq:main_finding_power-law_gen}).


	\subsection{General case}
	
		We are now prepared to study the general case where the memory kernel of the Hawkes process is a 
		continuous superposition of exponential functions~\eqref{eq:continuous_decomposition_kernel}. 
		Introducing the steady state cumulant functional 
		\begin{equation}
			\Phi[s] := \log \tl{P}_{\mrss}[s],
		\end{equation}
		and from the master equation in its functional Laplace representation Eq.~\eqref{eq:master_gen_functional_Laplace}, 
		we obtain the following first-order functional differential equation in the steady state,
		\begin{equation}
			\int_0^\infty dx \left(e^{-\int_{0}^\infty dx' s(x')n(x')/x'} -1 + \frac{s(x)}{x}\right)\frac{\delta \Phi[s]}{\delta s(x)} = \nu_0\left(e^{-\int_{0}^\infty dx' s(x')n(x')/x'} -1\right).
			\label{eq:master_gen_functional_cumulant}
		\end{equation}
		The corresponding Lagrange-Charpit equations are the following partial-integro equations,
		\begin{equation}
			\frac{\partial s(l;x)}{\partial l} = 1 - e^{-\int_{0}^\infty dx' s(x')n(x')/x'} - \frac{s(x)}{\tau}, \>\>\>\>
			\frac{\partial \Phi(l)}{\partial l} = -\nu_0\left(e^{-\int_{0}^\infty dx' s(x')n(x')/x'} -1\right)\label{eq:LagrangeCharpitGeneral}
		\end{equation}
		where $l$ is the curvilinear parameter indexing the position along the characteristic curve.
		We now perform the stability analysis of this equation~\eqref{eq:LagrangeCharpitGeneral} in the neighbourhood of $s=0$ 
		close to the critical condition $n=1$. 
		
		\paragraph{Sub-critical case $n<1$.} 
			We linearize the Lagrange-Charpit equation~\eqref{eq:LagrangeCharpitGeneral} to obtain
			\begin{equation}
				\frac{\partial s(l;x)}{\partial l} = -\int_0^\infty dx' H(x,x')s(x'), \>\>\>\>
				\frac{\partial \Phi(l)}{\partial l} = \nu_0\int_0^{\infty} dx' K(x')s(x')
			\end{equation}
			with
			\begin{equation}
				H(x,x'):= \frac{\delta(x-x')-n(x')}{x'}, \>\>\>\> 
				K(x') := \frac{n(x')}{x'}. 
			\label{yhjuynbj2q}
			\end{equation}
			Let us introduce the eigenvalues $\lambda \geq \lambda_{\min}$ and eigenfunctions $e(x;\lambda)$, satisfying
			\begin{equation}
				\int_0^\infty dx' H(x,x')e(x';\lambda) = \lambda e(x;\lambda).
			\end{equation}
			Appendix ~\ref{app:sec:realEigenvalues_continuous} shows that all the eigenvalues are real.
			The inverse matrix of $H(x,x')$, denoted by $H^{-1}(x,x')$, can be explicitly obtained as shown in 
			Appendix~\ref{app:sec:inverseMatrix_continuous}. 
			Since the inverse matrix $H^{-1}(x,x')$ has a singularity at $n=1$, the critical condition of 
			this Hawkes process is given by $n=1$ as expected.
			Using calculations that are analogous to those in Sec.~\ref{sec:LinearStabilityOfLagrangeCharpitTwoExpon}, we obtain
			\begin{equation}
				\Phi[s] \simeq -\nu_0 \int_0^\infty dx \int_0^\infty dx' K(x)H^{-1}(x,x')s(x'),
			\end{equation}
			from which we state that
			\begin{equation}
				\log \tl{Q}_{\mrss}(s) = -\nu_0 s +\Phi[s\bm{1}(x)] = \frac{-\nu_0}{1-n}s.
			\end{equation}
			where $\bm{1}(x)$ is an indicator function defined by $\bm{1}(x) = 1$ for any $x$.
			This recovers \eqref{yjnh23rb2},  \eqref{temjkymku5jmn3} and \eqref{uj4m4n32h} derived above.	
						
		\paragraph{Critical case $n=1$.}
			At criticality, the smallest eigenvalue vanishes: $\lambda_{\min}=0$.
			Indeed, we obtain the zero eigenfunction
			\begin{equation}
				e(x;\lambda=0) = x,
			\end{equation}
			which can be checked by direct substitution:
			\begin{equation}
				\int_0^\infty dx H(x,x')e(x';\lambda=0) = \int_0^\infty dx \frac{\delta(x-x')-n(x')}{x'}x' = 1-n = 0~,~~{\rm for}~n=1.
			\end{equation}
			We now introduce a set of variables to obtain a new representation based on the eigenfunctions,
			\begin{equation}
				s(x) = \sum_{\lambda} e(x;\lambda)X(\lambda)\>\>\>
				\Longleftrightarrow \>\>\>
				X(\lambda) = \int_0^\infty dx e^{-1}(\lambda;x)s(x)
			\end{equation}
			with the inverse matrix $e^{-1}(\lambda;x)$ satisfying
			\begin{equation}
				\int_0^\infty dx e^{-1}(\lambda;x)e(x;\lambda') = \delta_{\lambda,\lambda'}.
			\end{equation}
			We assume the existence of the inverse matrix, which is equivalent to the assumption that the set of all eigenfunctions is complete.
			$H(x,x')$ can be diagonalized:
			\begin{equation}
				\int_0^\infty dx \int_0^\infty dx' e^{-1}(\lambda;x)H(x,x')e(x';\lambda')  = \lambda\delta_{\lambda,\lambda'}.
			\end{equation}
			We then obtain the linearized Lagrange-Charpit equations,
			\begin{equation}
				\frac{\partial X(\lambda)}{\partial l} \simeq -\lambda X(\lambda).
			\end{equation}
			The dominant contribution comes from the vanishing eigenvalue. We therefore focus on $X(0)$ by setting $X(\lambda)=0$ for $\lambda > 0$.
			We then form the expansion
			\begin{equation}
				e^{-\int_0^\infty dx' s(x')n(x')/x'} -1 = -\int_0^\infty dx' \frac{n(x')s(x')}{x'} + \frac{1}{2}\left(\int_0^\infty dx' \frac{n(x')s(x')}{x'}\right)^2 + \dots.
			\end{equation}
			The explicit representation of $x(0)$ is given by
			\begin{equation}
				X(\lambda=0) = \int_0^\infty dx e^{-1}(\lambda=0;x)s(x) = \frac{1}{\alpha}\int_0^\infty dx n(x)s(x), \>\>\>
				e^{-1}(\lambda=0;x) = \frac{n(x)}{\alpha},
				\label{en4h2g}
			\end{equation}
			where $\alpha$ is defined by expression (\ref{rhr2bg2}). Expression (\ref{en4h2g}) can be checked to be valid by direct substitution since,
			from Eq.~\eqref{eq:LagrangeCharpitGeneral}, we have
			\begin{equation}
				\frac{\partial X(0)}{\partial l} = \int_0^\infty dx e^{-1}(0;x)\frac{\partial s(x)}{\partial l} = 0 - \frac{1}{2\alpha}\left(\int_0^\infty dx\frac{n(x)s(x)}{x}\right)^2 + \dots,
			\end{equation}
			showing that the first-order contribution is actually null in this representation.  The parameter $\alpha$ (\ref{rhr2bg2})
			has the property to ensure the consistency with the following identity:
			\begin{equation}
				\int_0^\infty dx e^{-1}(\lambda=0;x)e(x;\lambda=0) = 
				\frac{1}{\alpha}\int_0^\infty dx n(x)x = \delta_{\lambda=0,\lambda'=0} = 1.
			\end{equation}
			Since we ignore the contribution from $X(\lambda)$ except for $\lambda=0$, let us set $X(\lambda)=0$ for $\lambda > 0$, which yields
			\begin{equation}
				s(x) = \sum_{\lambda} e(x;\lambda)X(\lambda) = e(x;0)X_0 = X_0x
			\end{equation}
			where we have written $X_0:= X(0)$.
			We then obtain the second-order contribution along the $X_0$ axis,
			\begin{equation}
				\frac{\partial X_0}{\partial l} \simeq -\frac{X_0^2}{2\alpha}
			\end{equation}
			From calculations mimicking those in Sec.~\ref{sec:LinearStabilityOfLagrangeCharpitTwoExpon}, we obtain
			\begin{equation}
				\log \tl{Q}_{\mrss}(s) \simeq -2\nu_0 \alpha \log |s| \>\>\> \Longleftrightarrow \>\>\> 
				P(\nu) \sim \nu^{-1+2\nu_0 \alpha}, \>\>\> \alpha := \int_0^\infty dx n(x)x. 
			\end{equation} 
			This recovers the power law formula of the intermediate asymptotics of the PDF of the Hawkes intensities
			given by (\ref{eq:main_finding_power-law_gen}).

\section{Discussion: Formal Relation to Quantum Field Theory}\label{sec:Discussion_quantumFieldTheory}
	We have formulated the non-Markovian Hawkes process as a classical field theory associated with stochastic excitation.
	While the formulation is entirely classical, it is interesting to point out its formal relationship with quantum field theory. 
	
	\subsection{Formal equivalence between the generalized Langevin equation and quantum field theory}
		We first discuss the formal relationship between the GLE and quantum field theory.
		In the beginning, let us study the case of the discrete-exponential-sum memory kernel~\eqref{eq:GLE_MarkovEmbedding_discrete}. 
		It is well-known that the Fokker-Planck/master equations have a structure that is quite similar to that of the Schr\"{o}dinger equation~\cite{QuantumMapForFP},
		and here we reformulate the Fokker-Planck/master equations according to this classical idea.  
		
		\subsubsection{Schr\"{o}dinger-like representation for the Fokker-Planck equation}
			After rewriting $v \to \Phi$, $u_k \to \phi_k$ and $|P_t\ra := \int d\Phi\prod_{k=1}^Kd\phi_k P_t(\Phi,\phi_1,\dots,\phi_K) | \Phi,\phi_1,\dots,\phi_K\ra$,
			the FP equation can be rewritten in a quantum-mechanics-like form: 
			\begin{equation}
				\frac{\partial}{\partial t}|P_t\ra = H_{\mathrm{orgn}}|P_t\ra, \>\>\> 
				H_{\mathrm{orgn}} := \sum_{k=1}^K\left[ \frac{-i}{M}\Pi\phi_k + i\pi_k\left(\frac{\phi_k}{\tau_k}+\kappa_k\Phi\right)- \frac{\kappa_kT}{\tau_k}\pi_k^2\right]
			\end{equation}
			with the ``momentum'' operators $\Pi:= - i\partial/\partial \Phi$ and $\pi_k:= -i\partial/\partial \phi_k$ satisfying commutative relations
			\begin{equation}
				[\Phi, \Pi] = i, \>\>\> [\phi_k, \pi_{k'}] = i\delta_{kk'}.
			\end{equation}
			While the Hamiltonian $H_{\mathrm{org}}$ is non-Hermitian (i.e., the evolution operator is not self-adjoint), this quantum-mechanical formulation is sometimes useful since analytical methods developed in quantum mechanics are available and can be formally transposed. 
			
			In particular, on the condition that the detailed balance is satisfied, there exists a mathematically better mapping~\cite{RiskenB}. 
			The steady solution of the FP is given by 
			\begin{equation}
				P_{\mrss}(\Phi,\phi_1,\dots,\phi_K)  \propto \exp\left[-\frac{1}{T}\left(\frac{M}{2}\Phi^2+\sum_{k=1}^K\frac{1}{2\kappa_k}\phi^2_k\right)\right].
			\end{equation}
			Here we make a transformation $|\psi_t\ra := P_{\mrss}^{-1/2}|P_t\ra$ to obtain 
			\begin{equation}
				-\frac{\partial}{\partial t}|\psi_t\ra = (H_H+H_A) |\psi_t\ra, \>\>\>
				H_H := \sum_{k=1}^K\left[\frac{\pi^2_k}{2m_k} +  \frac{m_k\omega^2_k}{2}\phi^2_k - \frac{\omega_k}{2}\right], \>\>\> H_A := i\sum_{k=1}^K \left[\kappa_k\Phi\pi_k - \frac{1}{M}\phi_k\Pi\right]
				\label{eq:FP_schrodinger_discrete}
			\end{equation}
			with $m_k:= \tau_k/(2T\kappa_k)$ and $\omega_k := 1/\tau_k$.
			Here,  $H_H$ and $H_A$ are Hermitian and anti-Hermitian operators, respectively (i.e., $H^\dagger_H=H_H$ and $H^\dagger_A=-H_A$). 
			Equation~\eqref{eq:FP_schrodinger_discrete} can be regarded as a non-Hermitian Schr\"{o}dinger equation based on the harmonic-oscillator Hamiltonian $H_H$.
			This decomposition is known to be mathematically useful in particular for the eigenfunction expansions~\cite{RiskenB}.
			
			We can introduce the creation and annihilation operators and their commutative relation
			\begin{equation}
				a_k^\dagger: = \sqrt{\frac{m_k\omega_k}{2}}\left(\phi_k - \frac{i}{m_k\omega_k}\pi_k \right), \>\>\>
				a_k: = \sqrt{\frac{m_k\omega_k}{2}}\left(\phi_k + \frac{i}{m_k\omega_k}\pi_k \right), \>\>\>
				[a_k,a_{k'}^{\dagger}] = \delta_{k,k'}
			\end{equation}
			to lead the Hamiltonian of the harmonic oscillators
			\begin{equation}
				H_H = \sum_{k=1}^K \omega_ka^{\dagger}_ka_k.
			\end{equation}
					
		\subsubsection{Schr\"{o}dinger-like representation for the field Fokker-Planck equation}\label{sec:quantumFieldTheoryGLE}
			This calculation can be generalized to the stochastic field theory of the Fokker-Planck equation. 
			Indeed, by introducing the field operator and the corresponding commutative relation
			\begin{equation}
				\pi(x):= -i\frac{\delta}{\delta \phi(x)}, \>\>\> [\phi(x), \pi(x')] = i\delta(x-x'),
				\label{wrth2rgq}
			\end{equation}
			we obtain the operator form of the FP field equation for the state vector $|P_t\ra := \int d\Phi\mathcal{D}\phi P_t(\Phi, \phi)|\Phi, \phi \ra$:
			\begin{equation}
				\frac{\partial}{\partial t}|P_t\ra = H_{\mathrm{orgn}}|P_t\ra, \>\>\> 
				H_{\mathrm{orgn}} := \int dx \left[ \frac{-i}{M}\Pi\phi(x) + i\pi(x)\left(\frac{\phi(x)}{x}+\kappa(x)\Phi\right)- \frac{\kappa(x)T}{x}\pi(x)^2\right].
			\end{equation}
			The FP field equation can be transformed into a non-Hermitian quantum-field theory composed of harmonic oscillators on the field:
			\begin{equation}
				-\frac{\partial}{\partial t}|\psi_t\ra = (H_H+H_A) |\psi_t\ra
			\end{equation}
			with Hermitian and anti-Hermitian operators 
			\begin{equation}
				H_H := \int dx \left[\frac{\pi^2(x)}{2m(x)} +  \frac{m(x)\omega^2(x)}{2}\phi^2(x) - \frac{\omega(x)}{2}\delta(0)\right], \>\>\> H_A := i\int dx \left[\kappa(x)\Phi\pi(x) - \frac{1}{M}\phi(x)\Pi\right]
			\end{equation}
			by defining
			\begin{equation}
				|\psi_t\ra := P_{\mrss}^{-1/2}|P_t\ra, \>\>\>
				P_{\mrss}(\Phi,\{\phi(x)\}_x)  \propto \exp\left[-\frac{1}{T}\left(\frac{M}{2}\Phi^2+\int dx\frac{1}{2\kappa(x)}\phi^2(x)\right)\right], \>\>\>
				m(x):= \frac{x}{2T\kappa(x)}, \>\>\>
				\omega(x) := \frac{1}{x}.
			\end{equation}
			While the divergent term $\delta(0)$ appears due to the singularity of the commutative relation (\ref{wrth2rgq}) at $x=x'$, this term is irrelevant to the physical observables. 
			Indeed, by introducing the field creation and annihilation operators
			\begin{equation}
				a^\dagger(x): = \sqrt{\frac{m(x)\omega(x)}{2}}\left(\phi(x) - \frac{i}{m(x)\omega(x)}\pi(x) \right), \>\>\>
				a(x): = \sqrt{\frac{m(x)\omega(x)}{2}}\left(\phi(x) + \frac{i}{m(x)\omega(x)}\pi(x) \right)
			\end{equation}
			satisfying the field commutative relation
			\begin{equation}
				[a(x), a^\dagger(x')] = \delta(x-x'),
			\end{equation}
			we obtain the field harmonic-oscillator representation
			\begin{equation}
				H_H = \int dx \omega(x)a^{\dagger}(x)a(x),
			\end{equation}
			where the divergent term $\delta(0)$ cancels out of the final results. 
			This technical procedure is essentially the same as that in quantum electrodynamics, 
			where renormalization of the energy is required to avoid divergence by removing the zero-point energy of harmonic oscillators.
			
			The formal correspondence between the FP field equation and non-Hermitian quantum field theory itself is not surprising.
			For example, it is known that a stochastic chemical reaction system, characterized by an SPDE, are formally equivalent to non-Hermitian quantum field theory (see Sec.~I E in Ref.~\cite{OdorRMP}).
			Since the formal relationship between classical and quantum mechanics is a recent hot topic in terms of non-Hermitian physics~\cite{Ashida2020}, 
			it might be interesting to further seek this mathematical relationship in understanding general non-Markovian processes.

	\subsection{Quantum-field-like representation for the Hawkes process}
		We can apply this formal procedure to the field master equation for the Hawkes process. Let us apply the functional Kramers-Moyal expansion to obtain
		\begin{equation}
			\frac{\partial P_t[z]}{\partial t} = \left[\int dx \frac{\delta }{\delta z(x)}\frac{z(x)}{x} + 
			\sum_{k=1}^\infty \frac{(-1)^k}{k!}\left( \int dx \frac{n(x)}{x}\frac{\delta }{\delta z(x)}\right)^k\left(\nu_0+\int z(x')dx'\right)\right]P_t[z],
		\end{equation}
		where we have used the functional Taylor expansion
		\begin{align}
			\left(\nu_0 + \int \left(z(x')-\frac{n(x')}{x'}\right)dx' \right)P_t\left[z-\frac{n}{x}\right] &= \sum_{k=0}^\infty \frac{1}{k!}\left( -\int dx \frac{n(x)}{x}\frac{\delta }{\delta z(x)}\right)^k\left(\nu_0+\int z(x')dx'\right)P_t[z] \notag \\
			&= \exp\left[ -\int dx \frac{n(x)}{x}\frac{\delta }{\delta z(x)}\right] \left(\nu_0+\int z(x')dx'\right)P_t[z].
		\end{align}
		We then introduce the field operators and the corresponding commutative relations,
		\begin{equation}
			\phi(x):= z(x), \>\>\> \pi(x):= -i\frac{\delta }{\delta \phi(x)}, \>\>\> [\phi(x), \pi(x')] = i\delta (x-x').
		\end{equation}
		We then obtain the Schr\"{o}dinger-like representation for the state vector $|P_t\ra := \int \mathcal{D}\phi P_t[\phi] | \phi \ra$
		\begin{equation}
			\frac{\partial}{\partial t}|P_t \ra = H |P_t \ra
		\end{equation}
		with non-Hermitian Hamiltonian $H$ defined by
		\begin{equation}
			H := \int \frac{dx}{x}\pi(x)\phi(x) + \left\{\exp\left[ \frac{1}{i}\int dx\frac{n(x)}{x}\pi(x)\right] -1\right\}\left(\nu_0+\int \phi(x')dx'\right).
		\end{equation}
		Since the Hamiltonian includes infinite-order ``momentum'' operators, the Hamiltonian is classified as a non-local operator, 
		which reflects trajectory jumps in the point processes. A similar non-local Hamiltonian representation for the master equation can be seen in Ref.~\cite{KleinertB} in the  
		context of path integral representations of L\'evy processes.
		In this sense, the field master equation can be formally regarded as a non-Hermitian quantum field theory without locality. 
		The exponential operator $T[y]:=\exp[-i\int dx y(x)\pi(x)]$ naturally appears because it is a translation operator: $T[y]P_t[z]=P_t[z-y]$.

		We note that the Hamiltonian reduces to a local operator if we can approximately truncate the Kramers-Moyal expansion up to the second order. 
		While the validity of such approximation is not obvious for this linear Hawkes process, we can actually formulate such a formal approximation 
		by generalizing the system size expansion for the field master equation in the case of nonlinear Hawkes processes~\cite{KanazawaSornetteFuture}.

\section{Conclusion}\label{sec:conclusion}

	We have presented an analytical framework of the Hawkes process for an arbitrary memory kernel, 
	based on the master equation governing the behavior of auxiliary field variables. We have derived systematically the corresponding 
	functional master equation for the auxiliary field variables. While the Hawkes point process is non-Markovian by construction,
	the introduction of auxiliary field variables provides a formulation in terms of linear stochastic partial differential equations
	that are Markovian. For the case of a memory kernel decaying as a single exponential, we presented the exact time-dependent 
	and steady state solutions for the probability density function (PDF) of the Hawkes intensities, using the Laplace representation of the 
	Master equation. For memory kernels represented as arbitrary sums of exponential (discrete and continuous sums), 
	we derived the asymptotic solutions of the Lagrange-Charpit equations for the hyperbolic master equations in the Laplace representation
	in the steady state, close to the critical point $n=1$ of the Hawkes process, where $n$ is the branching ratio.
	Our theory predicts a power law scaling of the PDF of the intensities in an intermediate asymptotics regime, which crosses over
	to an asymptotic exponential function beyond a characteristics intensity that diverges as the critical condition is
	approached ($n \to 1$). The exponent of the PDF is non-universal and a function of the background intensity $\nu_0$ of the Hawkes intensity and
	of the parameter $\alpha = n  \langle \tau \rangle$, where $\langle \tau \rangle$ is the first-order moment of the distribution
	of time scales of the memory function of the Hawkes process.
	We found that, the larger the memory $\langle \tau \rangle$,
	the larger the background intensity $\nu_0$ and the larger the branching ratio $n$, the smaller is the exponent $1-2\nu_0\alpha$
	of the PDF of Hawkes intensities.
	 	
	This work provides the basic analytical tools to analyse Hawkes processes from a different angle than hitherto developed
	and will be useful to study more general and complex models derived from the Hawkes process.  For instance, it is 
	straightforward to extend our treatment to the case where each event has a mark quantifying its impact or ``fertility'', 
	thus defining the more general Hawkes process with intensity $\hnu(t) = \nu_0 + n\sum_{i=1}^{\hN(t)} \hat{\rho}_i h(t-\htt_i)$
	with independent and identically distributed random numbers $\{\hat{\rho}_i\}_i$. Our framework is also well-suited
	to nonlinear generalisations of the Hawkes process, for instance with the intensity taking the form $\hnu(t) = g(\homg(t)) > 0$,
	where the auxiliary variable $\homg$ is given by $\homg(t) = \omg_0 + n\sum_{i=1}^{\hN(t)} \hat{\rho}_i h(t-\htt_i)$ and where
	the times $\{\htt_i\}_i$ of the events are determined from the intensity $\hnu(t)$. In this nonlinear version,
	the positivity of $\homg(t)$ and $\rho_i$ are not anymore required. 
	This nonlinear Hawkes process is more complex than the linear Hawkes process but
	our framework can be applied to derive its most important analytical properties~\cite{KanazawaSornetteFuture}.
	We note that such nonlinear Hawkes process include several models that have been proposed in the past, 
	with applications to explain the multifractal properties of earthquake seismicity \cite{SornetteMSA} and of financial volatility \cite{FiliSornette11}.

\begin{acknowledgements}
	This work was supported by the Japan Society for the Promotion of Science KAKENHI (Grand No.~16K16016 and No.~20H05526) and Intramural Research Promotion Program in the University of Tsukuba. 
	We are grateful for useful remarks on the manuscript provided by M. Schatz, A. Wehrli, S. Wheatley, H. Takayasu, and M. Takayasu.
\end{acknowledgements}


\appendix
\section{Explicit derivation of the master equations}

	\subsection{Derivation of Eqs.~{(\ref{eq:master_n_expon}, \ref{eq:master_n_expone_Laplace})}}\label{sec:master_eq_n_expon}
	
	Given the dynamical equations (\ref{eq:SDE_general_superposition_discrete}) for
	the excess intensities $\{\hz_k, k=1, ..., K\}$, which are short hand notations for the dynamics given by
	equation (\ref{jehygbqgb}), for an arbitrary function $f(\bm{\hz})$, its stochastic time evolution therefore reads
		\begin{equation}
			df(\bm{\hz}(t)) = f(\bm{\hz}(t+dt)) - f(\bm{\hz}(t))
							= 	\begin{cases}
									-\sum_{k=1}^K\frac{\hz_k}{\tau_k}\frac{\partial f(\bm{\hz})}{\partial \hz_k}dt & (\mbox{No jump during $[t,t+dt)$; probability} = 1-\hnu(t)dt) \\
									f(\bm{\hz}(t)+\bm{h}) - f(\bm{\hz}(t)) & (\mbox{Jump in $[t,t+dt)$; probability} = \hnu(t)dt)
								\end{cases}
			\label{rn3thn3hnwb}
		\end{equation}
		with jump size vector $\bm{h}$ and Hawkes intensity $\hnu$, defined by 
		\begin{equation}
			\bm{h} := \left(\frac{n_1}{\tau_1}, \frac{n_2}{\tau_2}, \dots, \frac{n_K}{\tau_K}\right)^{\mrT}, \>\>\>
			\hnu(t) := \nu_0 + \sum_{k=1}^K \hz_k(t). 
		\end{equation}
		Taking the ensemble average of both sides of (\ref{rn3thn3hnwb}) and after partial integration of the left-hand side, we get
		\begin{align}
			\int d\bm{z} f(\bm{z})\frac{\partial P_t(\bm{z})}{\partial t}dt = \int d\bm{z}\left[-\sum_{k=1}^K\frac{z_k}{\tau_k}\frac{\partial f(\bm{z})}{\partial z_k}dt + \left(\nu_0+\sum_{k=1}^K z_k\right)dt\left\{f(\bm{z}+\bm{h}) - f(\bm{z})\right\}\right]P_t(\bm{z}).
		\end{align}
		After partial integration of the right-hand side and making the change of variable $\bm{z}+\bm{h} \to \bm{z}$, we obtain
		\begin{equation}
			\int d\bm{z} f(\bm{z})\frac{\partial P_t(\bm{z})}{\partial t} = \int d\bm{z}\left[\sum_{k=1}^K\frac{\partial }{\partial z_k}\frac{z_k}{\tau_k}P(\bm{z}) + 
						\left\{ \nu_0+\sum_{k=1}^K (z_k-h_k)\right\}P(\bm{z}-\bm{h}) - 
						\left\{\nu_0+\sum_{k=1}^K z_k\right\} P(\bm{z})\right]f(\bm{z}).
		\end{equation}
		Since this is an identify for an arbitrary $f(\bm{z})$, we obtain Eq.~\eqref{eq:master_n_expon}. 
		
		We derive the corresponding Laplace representation~\eqref{eq:master_n_expone_Laplace} as follows: 
		Let us apply the Laplace transform to both sides of Eq.~\eqref{eq:master_n_expon},
		\begin{equation}
			\mathcal{L}_K\left[\frac{\partial P_t(\bm{z})}{\partial t}\right] = \mathcal{L}_K\left[\sum_{k=1}^K\frac{\partial }{\partial z_k}\frac{z_k}{\tau_k}P(\bm{z}) + 
						\left\{ \nu_0+\sum_{k=1}^K (z_k-h_k)\right\}P(\bm{z}-\bm{h}) - 
						\left\{\nu_0+\sum_{k=1}^K z_k\right\} P(\bm{z})\right]. \label{app:trans_master_n_expon_1}
		\end{equation}
		The left-hand side is given by
		\begin{equation}
			\mathcal{L}_K\left[\frac{\partial P_t(\bm{z})}{\partial t}\right] = \frac{\partial \tl{P}_t(\bm{s})}{\partial t}. 
		\end{equation}
		For the right-hand side, let us consider the following two relations: 
		\begin{align}
			\mathcal{L}_K\left[\frac{\partial }{\partial z_k}\frac{z_k}{\tau_k}P(\bm{z})\right] &= \int d\bm{z}e^{-\bm{s}\cdot \bm{z}}\frac{\partial }{\partial z_k}\frac{z_k}{\tau_k}P(\bm{z}) \notag\\
			&= \int \prod_{i | i\neq k}dz_i \int dz_k e^{-\bm{s}\cdot \bm{z}}\frac{\partial }{\partial z_k}\frac{z_k}{\tau_k}P(\bm{z})\notag \\
			&= \int \prod_{i | i\neq k}dz_i \left\{ \left[\frac{z_k}{\tau_k}P(\bm{z})\right]_{z_k=0}^{z_k=\infty} + s_k\int dz_k e^{-\bm{s}\cdot \bm{z}}\frac{z_k}{\tau_k}P(\bm{z})\right\} \notag \\
			&= s_k\int \prod_{i | i\neq k}dz_i \int dz_k e^{-\bm{s}\cdot \bm{z}}\frac{z_k}{\tau_k}P(\bm{z}) \notag \\
			&= s_k\int \prod_{i | i\neq k}dz_i \left(-\frac{1}{\tau_k}\frac{\partial}{\partial s_k}\right)\int dz_k e^{-\bm{s}\cdot \bm{z}}P(\bm{z}) \notag \\
			&= -\frac{s_k}{\tau_k}\frac{\partial }{\partial s_k} \tl{P}_t(\bm{s}),
		\end{align}
		where we have used the partial integration on the second line and have used the boundary condition~\eqref{eq:boundary_condition_n_expon} on the third line, and
		\begin{align}
			\mathcal{L}_K\left[\left\{ \nu_0+\sum_{k=1}^K (z_k-h_k)\right\}P(\bm{z}-\bm{h})\right] &= \int d\bm{z} e^{-\bm{s}\cdot \bm{z}}\left\{ \nu_0+\sum_{k=1}^K (z_k-h_k)\right\}P(\bm{z}-\bm{h}) \notag\\
			&= e^{-\bm{s}\cdot \bm{h}}\int d\bm{z} e^{-\bm{s}\cdot (\bm{z}-\bm{h})}\left\{ \nu_0+\sum_{k=1}^K (z_k-h_k)\right\}P(\bm{z}-\bm{h}) \notag \\
			&= e^{-\bm{s}\cdot \bm{h}}\int d\bm{z} e^{-\bm{s}\cdot \bm{z}}\left\{ \nu_0+\sum_{k=1}^K z_k\right\}P(\bm{z}) \notag \\
			&= e^{-\bm{s}\cdot \bm{h}}\left\{ \nu_0-\sum_{k=1}^K \frac{\partial }{\partial s_k}\right\}\int d\bm{z} e^{-\bm{s}\cdot \bm{z}}P(\bm{z}) \notag \\
			&= e^{-\bm{s}\cdot \bm{h}}\left\{ \nu_0-\sum_{k=1}^K \frac{\partial }{\partial s_k}\right\} \tl{P}_t(\bm{s}),
		\end{align}
		where we have applied the change of variable $\bm{z}-\bm{h} \to \bm{z}$ on the second line. 
		By applying these two relations to the right-hand side of Eq.~\eqref{app:trans_master_n_expon_1}, we obtain Eq.~\eqref{eq:master_n_expone_Laplace}

	\subsection{Derivation of Eq.~\eqref{eq:master_gen_functional}}\label{sec:master_eq_gen}
		
		The Hawkes intensity $\hnu$ is defined by 
		\begin{equation}
			\hnu(t) := \nu_0 + \int_0^\infty \hz(t,x)dx~,
		\end{equation}
		in terms of the continuous field of excess intensities $\{\hz(t,x)\}_{x\in (0,\infty)}$.		
		For an arbitrary functional $f[\hz]$, let us consider its stochastic time evolution:
		\begin{align}
			df[\hz] 	&= f[\{\hz(t+dt,x)\}_{x}] - f[\{\hz(t,x)\}_x] \notag\\
					&= 	\begin{cases}
									\displaystyle -dt\int_0^\infty dx \frac{\hz(t,x)}{x}\frac{\delta f[\hz]}{\delta \hz(x)} & (\mbox{No jump during $[t,t+dt)$; probability} = 1-\hnu_t dt) \\
									f[\hz+n/x] - f[\hz] & (\mbox{Jump in $[t,t+dt)$; probability} = \hnu_tdt)
								\end{cases},
			\label{grb2gb2b1v}
		\end{align}
		where we have used the the functional Taylor expansion
		\begin{equation}
			f[z+\eta] - f[z] = \sum_{k=0}^{\infty} \frac{1}{k!}\int dx_1\dots dx_k\frac{\delta^k f[z]}{\delta z(x_1)\dots \delta z(x_k)}\eta (x_1)\dots \eta(x_k)
		\end{equation}
		up to first-order. 
		Taking the ensemble average of both sides of (\ref{grb2gb2b1v}) yields
		\begin{align}
			\int \mathcal{D}z f[z]\frac{\partial P_t[z]}{\partial t}dt 
			= \int \mathcal{D}z \left[-\int_0^\infty dx \frac{z(x)}{x}\frac{\delta f[z]}{\delta z(x)}dt + \left(\nu_0+\int_0^\infty z(x)dx \right)dt\left\{f[z+n/x] - f[z]\right\}\right]P_t[z].
		\end{align}
		By partial integration and with the change of variable $z+n/x \to z$, we obtain
		\begin{align}
			\int \mathcal{D}z f[z]\frac{\partial P_t[z]}{\partial t} 
			= \int \mathcal{D}z \left[\int_0^\infty dx \frac{\delta }{\delta z}\frac{z}{x}P_t[z] + 
						\left\{ \nu_0+\int_0^\infty \left(z-\frac{n}{x}\right)dx\right\}P_t[z-n/x] - 
						\left\{\nu_0+\int_0^\infty zdx\right\} P[z]\right]f[z].
		\end{align}
		Since this is an identify for arbitrary $f[z]$, we obtain Eq.~\eqref{eq:master_gen_functional}.

\section{Derivations of the power law PDF of Hawkes intensities for the exponential memory kernel}\label{app:sec:various_derivation_power_law_single_expon}
	
	Here, we provide two different derivations of the power law PDF~\eqref{eq:power-law_single_expon_steady} of Hawkes intensities for 
	the exponential memory kernel~\eqref{def:single_expon_kernel}. 
	
	\subsection{Introduction of a UV cutoff.}
		
		We now investigate the steady solution of the master equation \eqref{eq:master_exp}
		for the probability density function (PDF) $P_t(z)$ of the excess intensity $\hz$ \eqref{hygqfqtgqb}, at the critical point $n=1$.  
		
		Let us introduce a UV cutoff $s_{\mathrm{uv}}$ to address the singularity at $s \to 0$ so that we can express
		\begin{equation}
			\tl{Q}_{\mrss}(s) \simeq \exp\left[-\frac{\nu_0}{\tau}\int_{s_{\mathrm{uv}}}^s \frac{sds}{e^{-s/\tau}-1+s/\tau}\right] 
			= \exp\left[-\nu_0(s-s_{\rm uv})-\nu_0\tau\log \left(\frac{e^{-s/\tau}-1+s/\tau}{e^{-s_{\rm uv}/\tau}-1+s_{\rm uv}/\tau}\right)\right]~.
		\end{equation}
		Recall that $\log \tl{Q}_{\mrss}(s) = -s\nu_0 + \log \tl{P}_{\mrss}(s)$ and $\tl{P}_{\mrss}(s)$ is the Laplace
		transform of the steady state $\tl{P}_{\mrss}(s):=\int_{0}^\infty d\nu e^{-s\nu}P_{\mrss}(z)$ of the master equation 
		(\ref{eq:master_exp}). The introduction of this UV cut-off $s_{\mathrm{uv}}$
		amounts to introducing a cut-off in the memory function $h(t)$ at large timescale (i.e., there exists 
		$t_{\mathrm{cut}}$ such that $h(t)$ is negligible for $t>t_{\mathrm{cut}}$).
		The validity of this approximation is confirmed by considering the time-dependent solution (see Sec.~\ref{htgb12fv1}).  
		At the critical point $n=1$, it has an asymptotic form for small $s_{\mathrm{uv}} < s \ll \tau$, 
		\begin{equation}
			\log \tl{Q}_{\mrss}(s)\simeq -\frac{\nu_0}{\tau}\int_{s_{\mathrm{uv}}}^s \frac{sds}{e^{-s/\tau}-1+s/\tau} \sim -\frac{\nu_0}{\tau}\int_{s_{\mathrm{uv}}}^s \frac{2\tau^2 ds}{s} = -2\nu_0 \tau\log \frac{s}{s_{\rm uv}},
		\end{equation}
		which implies the power-law relation for the tail distribution: 
		\begin{equation}
			P_{\mrss}(\nu) \propto  \nu^{-1+2\nu_0 \tau}
		\end{equation}
		for $0\ll \nu \ll \nu_{\max} := 1/s_{\rm uv}$.

	\subsection{Kramers-Moyal apparoch}
	
		We can derive relation~\eqref{eq:power-law_single_expon_steady} using the Kramers-Moyal expansion of the master equation~\eqref{eq:master_exp}. 
		Let us consider the expansion
		\begin{equation}
			(\nu_0+z-n/\tau)P_t(z-n/\tau) = \sum_{k=0}^\infty\frac{1}{k!}\left(\frac{-n}{\tau}\right)^k\frac{\partial^k}{\partial z^k}(\nu_0+z) P_t(z).
		\end{equation}
		By truncating the series at the second order, we obtain the Fokker-Planck equation at the critical point $n=1$ in the steady state: 
		\begin{equation}
			\left[-\frac{\nu_0}{\tau}\frac{\partial }{\partial z}+ \frac{1}{2\tau^2}\frac{\partial^2}{\partial z^2}(\nu_0+z) \right]P_{\mrss}(z) \simeq 0, 
		\end{equation}
		for $z\to \infty$. 
		We thus obtain an asymptotic formula
		\begin{equation}
			P_{\mrss}(z) \sim (\nu_0+z)^{-1+2\nu_0\tau}~,~~~~~{\rm for ~large}~z.
		\end{equation}
		This solution is consistent with the truncation of the Kramers-Moyal series at the second order, which consists
		in removing negligible higher order terms.
		For large $l\geq 3$, indeed, we obtain
		\begin{equation}
			\left|\frac{\partial^2}{\partial z^2}(\nu_0+z)P_{\mrss}(z)\right| \gg \left|\frac{\partial^l}{\partial z^l}(\nu_0+z)P_{\mrss}(z)\right|
			~,~~~~~{\rm for ~large}~z.
		\end{equation}

\section{Elementary summary of the method of characteristics}\label{sec:app:method_of_characterisics}

	The method of characteristics is a standard method to solve first-order PDEs. Here we focus on linear first-order PDEs
	that are relevant to the derivation of the PDF of Hawkes intensities.
	Let us consider the following PDE:
	\begin{equation}
		a(x,y,z)\frac{\partial z(x,y)}{\partial x} + b(x,y,z)\frac{\partial z(x,y)}{\partial y} = c(x,y,z). \label{eq:app_PDE_method_of_characteristics}
	\end{equation}
	According to the method of characteristics, we consider the corresponding Lagrange-Charpit equations:
	\begin{subequations}
	\begin{align}
		\frac{dx}{dl} &= -a(x,y,z) \\
		\frac{dy}{dl} &= -b(x,y,z) \\
		\frac{dz}{dl} &= -c(x,y,z) 
	\end{align}
	\end{subequations}
	with the parameter $l$ encoding the position along the characteristic curves.
	These equations are equivalent to an invariant form in terms of $l$
	\begin{equation}
		\frac{dx}{a(x,y,z)} = \frac{dy}{b(x,y,z)} = \frac{dz}{c(x,y,z)}.
	\end{equation}
	Let us write their formal solutions as $C_1 = F_1(x,y,z)$ and $C_2 = F_2(x,y,z)$ with constants of integration $C_1$ and $C_2$. 
	The general solution of the original PDE~\eqref{eq:app_PDE_method_of_characteristics} is given by
	\begin{equation}
		\phi\left(F_1(x,y,z), F_2(x,y,z) \right) = 0 
	\end{equation}
	with an arbitrary function $\phi(C_1, C_2)$, which is determined by the initial or boundary condition of the PDE~\eqref{eq:app_PDE_method_of_characteristics}. 
	This method can be readily generalized to systems with many variables.

\section{Analytical derivation of some main properties of $\mcF(s)$ (\ref{yi,m5i74jhnb2wg}) }\label{app:sec:mcF_characters_proof} 
	
	Here, we derive properties ($\alpha$1)-($\alpha$6) of $\mcF(s)$ (\ref{yi,m5i74jhnb2wg}).
	First, the following relations hold true:
	\begin{equation}
		s>\frac{\tau}{n}\log n \>\>\> \Longrightarrow \>\>\> \frac{d}{ds}\left(e^{-ns/\tau}-1+s/\tau\right) = -\frac{n}{\tau}e^{-ns/\tau} + \frac{1}{\tau} >0
	\end{equation}
	and 
	\begin{equation}
		\lim_{s\to \infty} \left(e^{-ns/\tau}-1+s/\tau\right) = \infty.
	\end{equation}
	These relations guarantee that there exists $s_0>0$ such that $e^{-ns/\tau}-1+s/\tau > 0$ for $s>s_0$. 
	Therefore, the integrand is positive-definite
	\begin{equation}
		\frac{1}{e^{-ns/\tau}-1+s/\tau} >0
	\end{equation}
	for $s > s_0$ by choosing an appropriate positive $s_0$. 
	Since the integrand is positive definite, the statement ($\alpha$1) is proved. 
	As a corollary of ($\alpha$1), the statement ($\alpha$2) is proved.  
		
	We next study $\mcF(s)$ in the sub-critical regime ($n<1$). 
	For $n<1$, the statement ($\alpha$3) is true because $e^{-ns/\tau}-1+s/\tau >0$ for any $s>0$ and $(\tau/n)\log n <0$. 
	The statement ($\alpha$4) is true because, for $0<s<s_0 \ll \tau/n$, we obtain
	\begin{equation}
		\mcF(s) \simeq \frac{\tau}{1-n}\log\frac{s}{s_0} \to -\infty
	\end{equation} 
	for $s\to +0$. The statement ($\alpha$5) is correct because
	\begin{equation}
		\mcF(s) \simeq \int_{c_0}^s\frac{ds'}{s'/\tau} + c_1 = \tau\log s + c_1 \to +\infty
	\end{equation}
	for $s\to +\infty$ with constants $c_0$ and $c_1$.
	As a corollary of ($\alpha$1), ($\alpha$4) and ($\alpha$5), the statement ($\alpha$6) is proved.

\section{Analytical derivation of the time-dependent solution~\eqref{eq:solution_time_dependent_single_expon}}

	\subsection{Consistency check 1: convergence to the steady solution.}\label{app:sec:convergence_to_steady}
		
		Let us check that the time-dependent solution~\eqref{eq:solution_time_dependent_single_expon} is consistent with the steady solution~\eqref{eq:exact_solution_single_expon_steady} for $n<1$.
		To prove this, it is sufficient to show that
		\begin{equation}
			\lim_{x\to \infty} \mathcal{H}(x) = 0. 
		\end{equation}
		Using ($\alpha$1), ($\alpha$4), and ($\alpha$5), we obtain  
		\begin{equation}
			\lim_{x\to \infty}S(x) =\lim_{x\to -\infty}\mathcal{F}^{-1}(x)=0~,
		\end{equation}
		and thus
		\begin{equation}
			\lim_{x\to \infty} \mathcal{H}(x) = \lim_{S\to 0}\left[\log\tl{P}_{t=0}\left(S\right)+\frac{\nu_0}{\tau}\int_0^{S} \frac{sds}{e^{-ns/\tau}-1+s/\tau}\right]=0.
		\end{equation}

	\subsection{Consistency check 2: relaxation dynamics of the average $\hnu(t)$ at finite times}\label{app:sec:average_nu_for_finite_time_single_expon}
		
		Let us now study the dynamics of the average intensity $\hnu(t)$ via the time-dependent formula~\eqref{eq:solution_time_dependent_single_expon}.  
		Below the critical point, we can use the renormalized expression~{\eqref{eq:F_transform_under_critical}}. 
		Then the integral $\mathcal{F}(s)$ can be asymptotically evaluated for small $0\leq s \ll \tau/n$:
		\begin{equation}
			\mathcal{F}(s) \simeq \frac{\tau}{1-n}\log s \to -\infty \>\>\>(s\to 0).\label{app:eq:transform_transient_finite_t_1}
		\end{equation}
		This means that the argument $x(t,s) = t-\mathcal{F}(s)$ shows the divergence 
		\begin{equation}
			x(t,s) = t-\mathcal{F}(s) \simeq t - \frac{\tau}{1-n}\log s \to +\infty \label{app:eq:transform_transient_finite_t_2}
		\end{equation}
		From Eq.~\eqref{app:eq:transform_transient_finite_t_1}, the inverse function shows the asymptotic behavior for large $x$:
		\begin{equation}
			S(x) = \mathcal{F}^{-1}(-x) \simeq \exp\left[-\frac{1-n}{\tau}x \right].\label{app:eq:transform_transient_finite_t_3}
		\end{equation}
		By substituting the relation~\eqref{app:eq:transform_transient_finite_t_2},
		we obtain 
		\begin{equation}
			S(x(t,s)) = \mathcal{F}^{-1}(-x(t,s)) \simeq \exp\left[-\frac{1-n}{\tau}\left(t-\frac{\tau}{1-n}\log s\right) \right] = se^{-(1-n)t/\tau}. 
		\end{equation}
		We now assume the initial condition $\hnu(0)=\nu_{\rm ini}$, or equivalently $\log \tl{P}_{t=0}(s)=-\nu_{\rm ini}s$. 
		From Eq.~\eqref{eq:solution_time_dependent_single_expon}, we thus obtain the relaxation dynamics for the tail $s\simeq 0$,
		\begin{equation}
			\log \tl{P}_t(s) \simeq -se^{-(1-n)t/\tau}\nu_{\rm{ini}}-\frac{\nu_0}{\tau}\int_{se^{-(1-n)t/\tau}}^s \frac{ sds}{e^{-ns/\tau}-1+s/\tau}
			\simeq -\left(\nu_{\rm ini}e^{-(1-n)t/\tau}+\frac{\nu_0(1-e^{-nt/\tau})}{1-n}\right)s,\label{eq:transient_finite_tail}
		\end{equation}
		which means that the average of $\hnu(t)$ is given by
		\begin{equation}
			\la \hnu(t) \ra = -\frac{d}{ds}\log \tl{P}_t(s)\bigg|_{s=0} = \nu_{\rm ini}e^{-(1-n)t/\tau} + \frac{\nu_0(1-e^{-nt/\tau})}{1-n}. 
		\end{equation}

	\subsection{Derivation of the asymptotic formula \eqref{eq:asymptotic_relaxation_single_expon}}
	
		Let us derive the asymptotic relaxation formula~\eqref{eq:asymptotic_relaxation_single_expon} for sufficiently large $t$, satisfying
		\begin{equation}
			t \gg \mathcal{F}(s)
		\end{equation}
		for a given $s$.
		Under such a condition, the asymptotic relation for the inverse function $\mathcal{F}^{-1}(-x(t,s))$ is available as Eq.~\eqref{app:eq:transform_transient_finite_t_3} with $x(t,s) = t - \mathcal{F}(s)$. 
		We then obtain
		\begin{equation}
			\mathcal{F}^{-1}(-x(t,s)) \simeq \exp\left[-\frac{1-n}{\tau}\left(t-\mathcal{F}_{\rm R}(s) - \frac{\tau}{1-n}\log s\right) \right] = s\exp\left[-\frac{1-n}{\tau}\left(t-\mathcal{F}_{\rm R}(s)\right)\right]
		\end{equation}
		for sufficiently large $t$. By substituting this into Eq.~\eqref{eq:solution_time_dependent_single_expon}, we obtain Eq.~\eqref{eq:asymptotic_relaxation_single_expon}.

\section{Proofs of mathematical properties of $\bm{H}$ (\ref{wrnhmnnh3})}
	
	Here, we summarize the proofs of the main mathematical properties of $\bm{H}$ (\ref{wrnhmnnh3}) for arbitrary values of $K$. 
	
	\subsection{Proof of that its eigenvalues are real}\label{app:sec:proof_eigenvalues_H_real}
		
		All eigenvalues of $\bm{H}$ are real numbers for the following reasons. $\bm{H}$ can be symmetrized as $\bar{\bm{H}}$, defined by
		\begin{align}
			\bar{\bm{H}} 
			:= \bm{A}\bm{H}\bm{A}^{-1} = 
							\begin{pmatrix}
								\frac{1-n_1}{\tau_1},& \sqrt{\frac{n_1n_2}{\tau_1\tau_2}},& \dots& \sqrt{\frac{n_1n_K}{\tau_1\tau_K}} \\
								\sqrt{\frac{n_2n_1}{\tau_2\tau_1}},& \frac{1-n_2}{\tau_2},& \dots& \sqrt{\frac{n_2n_K}{\tau_2\tau_K}} \\
								\vdots& \vdots& \ddots& \vdots \\
								\sqrt{\frac{n_Kn_1}{\tau_K\tau_1}},& \sqrt{\frac{n_Kn_2}{\tau_K\tau_2}},& \dots& \frac{1-n_K}{\tau_K}
							\end{pmatrix}, \>\>\>\>\>
			\bm{A} :=		\begin{pmatrix}
								\sqrt{\frac{n_1}{\tau_1}},& 0,& \dots& 0 \\
								0,& \sqrt{\frac{n_2}{\tau_2}},& \dots& 0 \\
								\vdots& \vdots& \ddots& \vdots \\
								0,& 0,& \dots& \sqrt{\frac{n_2}{\tau_2}}
							\end{pmatrix}. \>\>\>
		\end{align}
		Indeed, by representing all the matrices by their elements $\bar{\bm{H}}:=(\bar{H}_{ij})$, $\bm{H}:=(H_{ij})$, and $\bm{A}:= A_{ij}$, we obtain
		\begin{equation}
			\bar{H}_{ij} = \sum_{k,l}A_{ik}H_{kl}A^{-1}_{lj} = \sum_{k,l}\sqrt{\frac{n_i}{\tau_i}}\delta_{ik}\left(\frac{\delta_{kl}}{\tau_k}-\frac{n_l}{\tau_l}\right)\sqrt{\frac{\tau_j}{n_j}}\delta_{lj} = \frac{\delta_{ij}-\sqrt{n_in_j}}{\sqrt{\tau_i\tau_j}}.
		\end{equation}
			We therefore obtain
		\begin{equation}
			\bm{H}\bm{e}_i = \lambda_i \bm{e}_i \>\>\> \Longleftrightarrow \>\>\> \bar{\bm{H}}\left(\bm{A}\bm{e}_i\right) = \lambda_i \left(\bm{A}\bm{e}_i\right),
		\end{equation}
		implying that any eigenvalue of $\bm{H}$ is the same as that of $\bar{\bm{H}}$. 
		Because $\bar{\bm{H}}$ is a symmetric matrix, all the eigenvalues of $\bar{\bm{H}}$ are real. 
		Therefore, all the eigenvalues of $\bm{H}$ are also real.

	\subsection{Determinant}\label{app:sec:proof_determinant_H}
	
		Here, we derive the determinant $\det \bm{H}$ for arbitrary values of $K$. 
		Let us recall the following identities, showing the invariance of determinants:
		\begin{align}
			\det \bm{H} &=
					 	\det \begin{pmatrix}
							\bm{a}_1 \\
							\bm{a}_2 \\
							\vdots \\
							\bm{a}_j \\
							\vdots \\
							\bm{a}_K
						\end{pmatrix}
						=
					 	\det \begin{pmatrix}
							\bm{a}_1 \\
							\bm{a}_2 \\
							\vdots \\
							\bm{a}_j + c \bm{a}_k \\
							\vdots \\
							\bm{a}_K
						\end{pmatrix}.
		\end{align}
		This implies
		\begin{align}
			\det \bm{H} &=
					 	\det \begin{pmatrix}
							\bm{a}_1 \\
							\bm{a}_2 \\
							\bm{a}_3 \\
							\vdots \\
							\bm{a}_K
						\end{pmatrix}
						=
					 	\det \begin{pmatrix}
							\bm{a}_1 \\
							\bm{a}_2-\bm{a}_1 \\
							\bm{a}_3 \\
							\vdots \\
							\bm{a}_K
						\end{pmatrix}
						=
					 	\det \begin{pmatrix}
							\bm{a}_1 \\
							\bm{a}_2-\bm{a}_1 \\
							\bm{a}_3-\bm{a}_1 \\
							\vdots \\
							\bm{a}_K
						\end{pmatrix}
						=
						\dots
						=
					 	\det \begin{pmatrix}
							\bm{a}_1 \\
							\bm{a}_2-\bm{a}_1 \\
							\bm{a}_3-\bm{a}_1 \\
							\vdots \\
							\bm{a}_K-\bm{a}_1
						\end{pmatrix}
						:= 
					 	\det \begin{pmatrix}
							\bm{a}_1' \\
							\bm{a}_2' \\
							\bm{a}_3' \\
							\vdots \\
							\bm{a}_K'
						\end{pmatrix}
		\end{align}
		and 
		\begin{align}
						\det \bm{H} =
					 	\det \begin{pmatrix}
							\bm{a}_1' \\
							\bm{a}_2' \\
							\vdots \\
							\bm{a}_K'
						\end{pmatrix}
						=
					 	\det \begin{pmatrix}
							\bm{a}_1'+n_2\bm{a}_2' \\
							\bm{a}_2' \\
							\vdots \\
							\bm{a}_K'
						\end{pmatrix}
						=
					 	\det \begin{pmatrix}
							\bm{a}_1'+n_2\bm{a}_2'+n_3\bm{a}_3' \\
							\bm{a}_2' \\
							\vdots \\
							\bm{a}_K'
						\end{pmatrix}
						=
						\dots 
						=
						\det \begin{pmatrix}
							\bm{a}_1'+\sum_{k=2}^Kn_k\bm{a}_k' \\
							\bm{a}_2' \\
							\vdots \\
							\bm{a}_K'
						\end{pmatrix}
		\end{align}
		with constants $\{n_k\}_k$.
		Using these relations, the determinant of $\bm{H}$ is given by
		\begin{align}
			\det \bm{H} &=
					 	\det \begin{pmatrix}
										-n_1/\tau_1 + 1/\tau_1,& -n_2/\tau_2,& \dots,& -n_K/\tau_K \\
										-n_1/\tau_1,& -n_2/\tau_2 + 1/\tau_2,& \dots,& -n_K/\tau_K \\
										\vdots& \vdots& \ddots& \vdots \\
										-n_1/\tau_1,& -n_2/\tau_2,& \dots,& -n_K/\tau_K  + 1/\tau_K
						\end{pmatrix}
						\begin{matrix}
							\leftarrow \bm{a}_1 \\
							\leftarrow \bm{a}_2 \\
							\vdots \\
							\leftarrow \bm{a}_K \\
						\end{matrix}\notag\\
						&=
						\det \begin{pmatrix}
							(1-n_1)/\tau_1,& -n_2/\tau_2,& \dots& -n_K/\tau_K \\
							-1/\tau_1,& 1/\tau_2,& \dots& 0\\
							\vdots& \vdots& \ddots& \vdots\\
							-1/\tau_1,& 0,& \dots& 1/\tau_K
						\end{pmatrix}
						\begin{matrix}
							\leftarrow \bm{a}_1' &=\bm{a}_1 \\
							\leftarrow \bm{a}_2' &=\bm{a}_2&-&\bm{a}_1 \\
							\vdots \\
							\leftarrow \bm{a}_K' &=\bm{a}_K&-&\bm{a}_1 \\
						\end{matrix}\notag\\
						&=
						\det \begin{pmatrix}
							(1-\sum_{k=1}^Kn_k)/\tau_1,& 0,& \dots& 0 \\
							-1/\tau_1,& 1/\tau_2,& \dots& 0 \\
							\vdots& \vdots& \ddots& \vdots\\
							-1/\tau_1,& 0,& \dots& 1/\tau_K
						\end{pmatrix}
						\begin{matrix}
							\leftarrow \bm{a}_1'' &= \bm{a}_1' &+& \sum_{k=2}^Kn_k\bm{a}_k' \\
							\leftarrow \bm{a}_2'' &= \bm{a}_2' \\
							\vdots \\
							\leftarrow \bm{a}_K'' &= \bm{a}_K' \\
						\end{matrix}\notag\\
						&= \frac{1-\sum_{k=1}^Kn_k}{\tau_1\dots \tau_K}. 
		\end{align}

	\subsection{Inverse matrix}\label{app:sec:inverse_matrix_H}
		
		Here we derive the inverse matrix of $\bm{H}$ for arbitrary values of $K$.
		The inverse matrix is derived from the method of row reduction:
		\begin{align}
			&\left(
				\begin{array}{cccc|cccc}
					-n_1/\tau_1 + 1/\tau_1,& -n_2/\tau_2,& \dots& -n_K/\tau_K& 1,& 0,&\dots,& 0 \\
					-n_1/\tau_1,& -n_2/\tau_2 + 1/\tau_2,& \dots& -n_K/\tau_K& 0,& 1,&\dots,& 0 \\
					\vdots& \vdots& \ddots & \vdots& \vdots& \vdots & \ddots& \vdots \\
					-n_1/\tau_1,& -n_2/\tau_2,& \dots,& -n_K/\tau_K  + 1/\tau_K& 0,& 0,&\dots,& 1
				\end{array}
			\right)
						\begin{matrix}
							\leftarrow \bm{b}_1 \\
							\leftarrow \bm{b}_2 \\
							\vdots \\
							\leftarrow \bm{b}_K \\
						\end{matrix}\notag\\
			\to
			&\left(
				\begin{array}{cccc|cccc}
					(1-n_1)/\tau_1,& -n_2/\tau_2,& \dots& -n_K/\tau_K& 1,& 0,&\dots& 0 \\
					-1/\tau_1,& 1/\tau_2,& \dots& 0& -1,& 1,&\dots& 0 \\
					\vdots& \vdots& \ddots& \vdots& \vdots& \vdots& \ddots& \vdots\\
					-1/\tau_1,& 0,& \dots& 1/\tau_K& -1,& 0,&\dots& 1
				\end{array}
			\right)
						\begin{matrix}
							\leftarrow \bm{b}_1'&=\bm{b}_1& \\
							\leftarrow \bm{b}_2'&=\bm{b}_2&-&\bm{b}_1 \\
							\vdots \\
							\leftarrow \bm{b}_K'&=\bm{b}_K&-&\bm{b}_1 \\
						\end{matrix}\notag\\
			\to
			&\left(
				\begin{array}{cccc|cccc}
					(1-n)/\tau_1,& 0,& \dots& 0& 1-\sum_{k=2}^Kn_k,& n_2,&\dots& n_K \\
					-1/\tau_1,& 1/\tau_2,& \dots& 0& -1,& 1,&\dots& 0 \\
					\vdots& \vdots& \ddots& \vdots& \vdots& \vdots& \ddots& \vdots\\
					-1/\tau_1,& 0,& \dots& 1/\tau_K& -1,& 0,&\dots& 1
				\end{array}
			\right)
						\begin{matrix}
							\leftarrow \bm{b}_1'' &= \bm{b}_1' &+& \sum_{k=2}^Kn_k\bm{b}_k' \\
							\leftarrow \bm{b}_2'' &= \bm{b}_2' \\
							\vdots \\
							\leftarrow \bm{b}_K'' &= \bm{b}_K' \\
						\end{matrix}\notag\\
			\to
			&\left(
				\begin{array}{cccc|cccc}
					1,& 0,& \dots& 0& \tau_1+\tau_1n_1/(1-n),& \tau_1n_2/(1-n),&\dots& \tau_1n_K/(1-n) \\
					-\tau_2/\tau_1,& 1,& \dots& 0& -\tau_2,& \tau_2,&\dots& 0 \\
					\vdots& \vdots& \ddots& \vdots& \vdots& \vdots& \ddots& \vdots\\
					-\tau_K/\tau_1,& 0,& \dots& 1& -\tau_K,& 0,&\dots& \tau_K
				\end{array}
			\right)
						\begin{matrix}
							\leftarrow \bm{b}_1''' &= \frac{\tau_1}{(1-n)}\bm{b}_1'' \\
							\leftarrow \bm{b}_2''' &= \tau_2\bm{b}_2'' \\
							\vdots \\
							\leftarrow \bm{b}_K''' &= \tau_K\bm{b}_K'' \\
						\end{matrix}\notag\\
			\to
			&\left(
				\begin{array}{cccc|cccc}
					1,& 0,& \dots& 0& \tau_1+\tau_1n_1/(1-n),& \tau_1n_2/(1-n),&\dots& \tau_1n_K/(1-n) \\
					0,& 1,& \dots& 0& \tau_2n_1/(1-n),& \tau_2 + \tau_2n_2/(1-n),&\dots& \tau_2n_K/(1-n) \\
					\vdots& \vdots& \ddots& \vdots& \vdots& \vdots& \ddots& \vdots\\
					0,& 0,& \dots& 1& \tau_Kn_1/(1-n),& \tau_Kn_2/(1-n),&\dots& \tau_K + \tau_Kn_K/(1-n)
				\end{array}
			\right)
						\begin{matrix}
							\leftarrow \bm{b}_1'''' &= \bm{b}_1''' \\
							\leftarrow \bm{b}_2'''' &= \bm{b}_2''' &+& \frac{\tau_2}{\tau_1}\bm{b}_1''' \\
							\vdots \\
							\leftarrow \bm{b}_K'''' &= \bm{b}_K''' &+& \frac{\tau_2}{\tau_1}\bm{b}_1'''\\
						\end{matrix}
		\end{align}
		which implies 
		\begin{align}
			\bm{H}^{-1} = 
			\begin{pmatrix}
				\tau_1+\tau_1n_1/(1-n),& \tau_1n_2/(1-n),&\dots& \tau_1n_K/(1-n) \\
				\tau_2n_1/(1-n),& \tau_2 + \tau_2n_2/(1-n),&\dots& \tau_2n_K/(1-n) \\
				\vdots& \vdots& \ddots& \vdots\\
				\tau_Kn_1/(1-n),& \tau_Kn_2/(1-n),&\dots& \tau_K + \tau_Kn_K/(1-n)
			\end{pmatrix}
		\end{align}
		or equivalently
		\begin{equation}
			H^{-1}_{ij}=\tau_i \delta_{ij} + \frac{\tau_i n_j}{1-n}
		\end{equation}
		in the representation by matrix elements. 
		As a check of the above calculation, we can directly confirm the following relation, defining the inverse matrix:
		\begin{equation}
			\bm{H}\bm{H}^{-1}=\bm{I} \>\>\> \Longleftrightarrow \>\>\> 
			\sum_{j=1}^K H_{ij}H^{-1}_{jk}=\sum_{j=1}^K \left(-\frac{n_j}{\tau_j}+\frac{1}{\tau_j}\delta_{ij}\right)\left(\tau_j \delta_{jk} + \frac{\tau_j n_k}{1-n}\right) = \delta_{ik}. 
		\end{equation}
		The inverse matrix has a singularity at $n=1$, which corresponds to the critical regime of the Hawkes process.

\section{Proofs of the mathematical properties of $\bm{H}(x,x')$ (\ref{yhjuynbj2q})} 

	\subsection{Proof that the eigenvalues are real}\label{app:sec:realEigenvalues_continuous}
	
		Considering the analogy to the eigenvalue problem of the finite-dimensional matrix $\bm{H}$ (\ref{wrnhmnnh3}), 
		it is obvious by taking the continuous limit that all the eigenvalues of $H(x,x')$ are positive.
		As an appendix, we remark that all eigenvalues can be proved real for $H(x,x')$ by making some specific assumptions. 
		For example, let us assume that $n(x)$ has a finite cutoff, such that
		\begin{equation}
			n(x) = 	\begin{cases}
							\tl{n}(x) & (x<\tau^*) \\
							0 & (x\geq \tau^*)
						\end{cases}
		\end{equation}
		with a positive continuous function $\tl{n}(x)> 0$ and a cutoff $\tau^* > 0$. 
		In this case, the eigenvalue problem for $H(x,x')$ can be rewritten as 
		\begin{equation}
			\int_0^\infty dx' H(x,x')e(x';\lambda) = \int_0^{\tau^*} H(x,x')e(x';\lambda) = \lambda e(x;\lambda)
			\label{app:eq:eigenvalue_integral_equation}
		\end{equation}
		While $H(x,x')$ itself is not a symmetric kernel, $H(x,x')$ can be symmetrized by introducing
		\begin{equation}
			\bar{H}(x,x') := \frac{\delta (x-x')-\sqrt{n(x)n(x')}}{\sqrt{x x'}},
		\end{equation}
		such that 
		\begin{equation}
			\int_0^{\tau^*} dx'\bar{H}(x,x') \bar{e}(x;\lambda) = \lambda \bar{e}(x;\lambda), \>\>\>
			\bar{e}(x;\lambda) := \sqrt{\frac{n(x)}{x}}e(x;\lambda)
		\end{equation}
		or equivalently, 
		\begin{equation}
			\int_0^{\tau^*} dx'\sqrt{\frac{n(x)n(x')}{xx'}} \bar{e}(x;\lambda) = (1/x-\lambda) \bar{e}(x;\lambda)
			\label{app:eq:eigenvalue_integral_equation2}
		\end{equation}
		This implies that all the eigenvalues of $H(x,x')$ are identical to those of $\bar{H}(x,x')$.
		Since Eq.~\eqref{app:eq:eigenvalue_integral_equation2} is a homogeneous Fredholm integral equation of the second kind with a continuous and symmetric kernel $\sqrt{n(x)n(x')/(xx')}$ and with a finite interval $[0,\tau^*]$, 	
		all the eigenvalues of $\bar{H}(x,x')$ are real according to the Hilbert-Schmidt theory~\cite{ArfkenBook}. 
		Therefore, all the eigenvalues of $H(x,x')$ are also real.

	\subsection{Inverse matrix}\label{app:sec:inverseMatrix_continuous}
	
		The inverse matrix of $H(x,x')$ is given by
		\begin{equation}
			H^{-1}(x,x') := x\left\{ \delta (x-x') + \frac{n(x')}{1-n}\right\}.
		\end{equation}
		Indeed, we verify that
		\begin{equation}
			\int_0^{\infty}dx' H(x,x')H^{-1}(x',x'') = \int_0^\infty dx' \frac{\delta(x-x')-n(x')}{x'}x'\left\{ \delta (x'-x'') + \frac{n(x'')}{1-n}\right\} = \delta(x-x'').
		\end{equation}
		The inverse matrix has a singularity at $n=1$, corresponding to the critical regime of the Hawkes process.

\end{document}